\documentclass[12pt]{article}
\pdfoutput=1
\usepackage{sectsty}
\usepackage[toc,page]{appendix}
\usepackage{mathrsfs}
\allsectionsfont{\sffamily}
\usepackage[nosort]{cite}

\usepackage[pdftex]{graphicx}
\usepackage[font={scriptsize,singlespacing}]{caption}
\usepackage{epstopdf}

\usepackage{amsmath}
\usepackage{amssymb}
\usepackage{subfigure}
\usepackage{hyperref}
\usepackage{url}
\usepackage{xcolor}
\usepackage{color}
\definecolor{amaranth}{rgb}{0.9, 0.17, 0.31}
\definecolor{purple(munsell)}{rgb}{0.62, 0.0, 0.77}
\definecolor{americanrose}{rgb}{1.0, 0.01, 0.24}
\definecolor{palatinateblue}{rgb}{0.15, 0.23, 0.89}
\definecolor{royalblue(web)}{rgb}{0.25, 0.41, 0.88}
\definecolor{hanpurple}{rgb}{0.32, 0.09, 0.98}
\definecolor{beaublue}{rgb}{0.74, 0.83, 0.9}
\definecolor{carminered}{rgb}{1.0, 0.0, 0.22}
\definecolor{brightpink}{rgb}{1.0, 0.0, 0.5}
\definecolor{vividviolet}{rgb}{0.62, 0.0, 1.0}
\hypersetup{ linktoc=all,
    colorlinks, linkcolor={palatinateblue},
    citecolor={brightpink}, urlcolor={amaranth}}

\def\sideremark#1{\ifvmode\leavevmode\fi\vadjust{\vbox to0pt{\vss
 \hbox to 0pt{\hskip\hsize\hskip1em
 \vbox{\hsize2cm\tiny\raggedright\pretolerance10000
 \noindent #1\hfill}\hss}\vbox to8pt{\vfil}\vss}}}%
\newcommand{\bo}{\raise-1mm\hbox{\Large$\Box$}}

\newcommand{\f}[2]{\frac{#1}{#2}}

\newcommand{\y}{\gamma}
\newcommand{\bd}{\boldsymbol}

\newcommand{\be}{\begin{equation}}
\newcommand{\ee}{\end{equation}}
\newcommand{\bea}{\begin{eqnarray}}
\newcommand{\eea}{\end{eqnarray}}

\def\brcurs{{\mbox{$\resizebox{.09in}{.08in}{\includegraphics[trim= 1em 0 14em 0,clip]{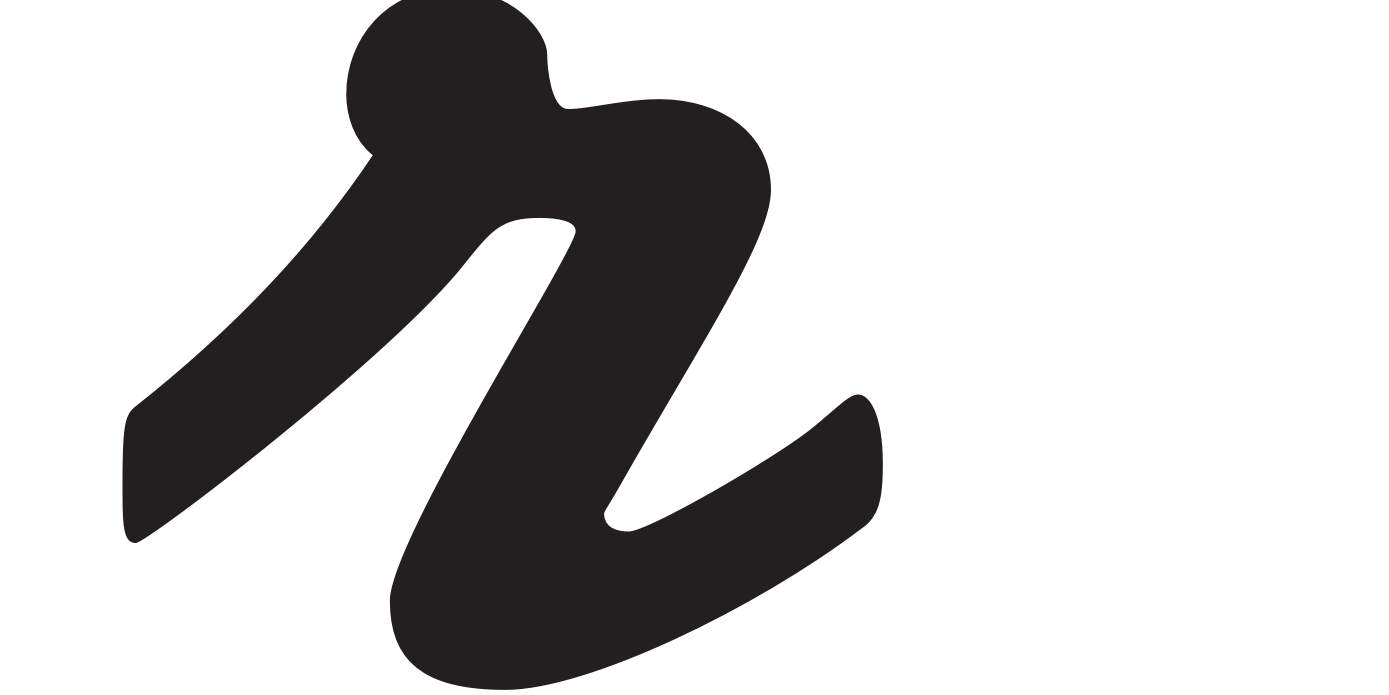}}$}}}

\def\hrcurs{{\mbox{$\hat \brcurs$}}}

\begin{document}
\thispagestyle{empty}
\begin{center}

\null \vskip-1truecm \vskip2truecm
{\LARGE{\bf \textsf{Stationary Worldline Power Distributions }}}

\vskip1truecm
\textbf{\textsf{Michael R.R. Good, Maksat Temirkhan, Thomas Oikonomou}}\\
{\footnotesize\textsf{Department of Physics, School of Science and Technology, Nazarbayev University, Astana, Kazakhstan}\\
{\tt Email: michael.good@nu.edu.kz, maksat.temirkhan@nu.edu.kz,
thomas.oikonomou@nu.edu.kz }}\\




\end{center}
\vskip1truecm \centerline{\textsf{ABSTRACT}} \baselineskip=15pt

\medskip

A worldline with a time-independent spectrum is called  stationary.  Such worldlines are arguably the most simple motions in physics. Barring the trivially static motion, the non-trivial worldlines are uniformly accelerated.  As such, a point charge moving along a stationary worldline will emit constant radiative power. The angular distribution, maximum angle scaling and Thomas precession of this power is found for all stationary worldlines including those with torsion and hypertorsion.


\vskip0.4truecm
\hrule
\tableofcontents

\section{Introduction}


\subsection*{Overview}
This paper is concerned with the properties of the radiation from a relativistically uniformly accelerated charge. The general formula for an arbitrary trajectory of the total radiated power and its angular distribution are specified to uniform accelerated motion with torsion and hypertorsion.  Schwinger \cite{Schwinger:1949ym} first calculated the angular distribution of power for point charges undergoing linear and circular acceleration for arbitrary dynamics and specification to uniformly accelerated circular motion.  Letaw \cite{Letaw:1980yv} found three previously unknown classes of uniformly accelerated worldlines.  This paper extends Schwinger's initial specification of circular motion to include the power distributions for these three fundamental cases of uniform acceleration.

\subsection*{Motivation}
Stationary background systems based on these new worldlines are of interest because the worldlines are trajectories of time-like Killing vector fields \cite{Letaw:1980yv, Rosu:1999ad, LetawJ}. The less-known trajectories (3, 4, 5; see below) are interesting in their own right because they are simple uniformly accelerated motions.  Their excitation spectra are connected to the question of coordinate dependence of thermodynamics and quantum field theory in flat spacetime (i.e. they may yield insights into the Unruh effect \cite{Unruh:1976db}, with crucial dependence on study of accelerating world lines).  

There is broad motivation for investigating the light from these fundamental motions: connections between quantum theory and gravity have been made (e.g. consider relativistic superfluidity \cite{Xiong:2014oga} or geometric creation of quantum vortexes \cite{Good:2014iua}) by pursuing the effects of the influence of quantum fields under external conditions (external effects like the Davies-Fulling effect\footnote{Consider the scale invariant accelerated solution in \cite{Good:2013lca} and the spin-statistics derivation from it \cite{Good:2012cp}.} \cite{Davies:1976hi}, the Hawking effect \cite{Hawking1975}, and the Parker effect \cite{Parker:1968mv}, i.e. respectively, moving mirrors, black holes, and expanding cosmologies). This program \cite{Parkerbook} continues to lead to progress that demonstrate strong correspondences \footnote{Example solutions are explored in the recent black hole-moving mirror correspondence contained in \cite{Good:2016oey}, temperature without a horizon in \cite{Good:2016horizonless}, and a black hole birth cry and death gasp in \cite{Good:2015nja}.} and simplifying physics \footnote{ The Kerr black hole has a temperature $2\pi T = g - k$ where $k = m\Omega^2$ is the spring constant and $g= (4m)^{-1}$ is the Schwarzschild surface gravity \cite{Good:2014uja}. }.  

Accelerated trajectories in flat spacetime and motion in curved spacetime are joined via the Equivalence Principle.  The thermodynamics `addendum' made to the Equivalence Principle, (e.g. Unruh effect), has been immensely helpful for further understanding the gravitational influence on quantum fields\footnote{Hawking's effect as a gravitational phenomena is further understood as a result of the quantum field phenomena of the Unruh effect.}. These motions are interesting and simple, and they deserve more attention particularly because of the fundamentally important and wide-reaching nature of uniform acceleration. 

\subsection*{Naming Convention} 

With an emphasis on torsion \cite{Letaw:1980yv}, the worldlines themselves can be called (e.g. a shortened use of the convention introduced by Rosu \cite{Rosu:1999ad}):
\begin{enumerate}
	\item Nulltor \hskip1.0cm	2. Ultrator  \hskip1.2cm 3. Parator \hskip1.0cm 4. Infrator  \hskip1.0cm 5. Hypertor
\end{enumerate}
 The worldline plots are in Appendix \ref{Appendix:Worldline} where we have also included their four-vector position parametrization $x^\mu(s)$ in proper time \cite{Letaw:1980yv}. 
 
Projected in space, the worldlines take the forms: 

\begin{enumerate}
	\item Line \hskip1.0cm	2. Circle  \hskip1.2cm 3. Cusp \hskip1.0cm 4. Catenary  \hskip1.0cm 5. Spiral
\end{enumerate}
where we have ignored the zero acceleration solution, $\kappa=0$, (the inertial path). These spatial projections are in Appendix \ref{Appendix:Spatial}, with the assigned axis convention. The main new results of this paper will be focused on Infrator and Hypertor.  

\subsection*{Curvature Invariants}
The three curvature invariants are, $\kappa$, $\tau$, and $\nu$, (curvature, torsion, and hypertorsion, respectively). As mentioned above, the stationary world lines separate naturally into five classes according to the values of the curvature invariants, if we exclude case when particle  at the rest (calling the motions by their projections in space):
\begin{itemize}
  \item Line :   $\kappa \neq 0$, $\nu=\tau=0$;
  \item Circle : $|\kappa|<|\tau|$, $\nu=0$;
  \item Cusp :   $|\kappa|=|\tau|$, $\nu=0$;
  \item Catenary : $|\kappa|>|\tau|$, $\nu=0$;
  \item Spiral : $\nu \neq 0$
\end{itemize}

\subsection*{Power and Acceleration}\label{sec:acc_basics}

Using the the relativistic covariant form of Larmor's formula, an accelerated point charge has power, 
\be\label{Power1} P = \frac{2}{3} q^2 \alpha^2, \ee
where $\alpha$  is the proper acceleration (as we will see in Eq.~(\ref{properacc})) and $q$ is the charge.  A constant proper acceleration yields constant power. The proper acceleration is the scalar invariant magnitude 
\be \alpha^2 \equiv -a_{\mu}a^{\mu}, \ee
of the four-acceleration, $a^\mu$,
\be a^{\mu} = \f{d^2 x^{\mu}}{d\tau^2},\ee
most easily expressed (see e.g. \cite{ Rindler:1960zz, Proceedings}), in two nice forms:
\begin{subequations}\label{properacc}
\bea \alpha^2
	&=& \y^4a^2 + \y^6(\bd{v}\cdot \bd{a})^2,\\
  &=& \gamma^6a^2 - \y^6(\bd{v} \times \bd{a})^2 \;. \eea
 \end{subequations}
In the case of straight-line motion, parallel vectors $\bd{v}\times \bd{a} = \bd{0}$ yield $\alpha^2 = \y^6 a^2$. In the case of circular motion, the three-acceleration vector is perpendicular to the three-velocity vector, $\bd{v}\cdot \bd{a} = 0$, and one obtains $\alpha^2 = \gamma^4 a^2$. The acceleration ratio, $A_\textrm{WL} \equiv \alpha^2/a^2$, is given for each worldline in Appendix \ref{Appendix:Acceleration}. 

In the instantaneous rest frame of the accelerating particle, the proper acceleration, $\alpha$, is the acceleration as measured by a hand-held accelerometer.  It is what is `felt' and is the \textit{proper}-ty of the particle, a Lorentz scalar invariant in all frames. For a stationary worldline the Lorentz scalar remains constant and the proper acceleration, $\alpha = \kappa$ is the curvature, the simplest of the curvature invariants for stationary worldlines (the others being torsion, $\tau$, and hypertorsion, $\nu$ \cite{Letaw:1980yv}). The power is consequently,
\be P = \frac{2}{3} q^2 \kappa^2. \ee

\subsection*{Vacuum Spectra}
Stationary world lines are special in part because each have a unique spectrum, e.g. Parator is exactly calculable and distinctly non-Planckian.  Each unique motion is a stationary world line solution of the Frenet equations when the curvature invariants of proper acceleration and proper angular velocity are constant.  There is a dearth of papers that have treated these worldlines over the last four decades\footnote{A more recent treatment discussing the difficulties in finding a characterization of the notion of higher order time derivatives of constant proper $n-$acceleration itself as a Lorentz invariant statement and its relationship to the Frenet-Serret formalism is given by Pons and Palol \cite{Josep}.}, for instance the exactly solvable motion has been treated in only a handful of papers: 
\cite{Rosu:1999ad, Padmanabhan:1983ub,Takagi:1986kn,Audretsch:1995iw,Sriramkumar:1999nw,Leinaas:2000mh,Rosu:2005iu,Louko:2006zv,Obadia:2007qf,Russo:2009yd}. 

As an illustration of the importance of investigating these worldlines, it is good to remark that the vacuum spectra differ among the stationary worldlines. For instance, two can be calculated exactly: the Nulltor and Parator motions which are both one parameter motions ($\kappa$ only).  The vacuum fluctuation spectrum for the zero torsion case \cite{Letaw:1980yv}, is the same Planckian distribution as found by Unruh \cite{Unruh:1976db} for uniform acceleration radiation, and scales (with $\omega/\kappa \gg 1$):  
\be \frac{1}{e^{2\pi \omega/\kappa}-1} \approx e^{-2\pi \omega/\kappa}, \quad \textrm{with} \quad T = \frac{\kappa}{2\pi} = 0.159\kappa.  \ee
The spectrum for the paratorsional worldline is Letawian, with a hotter temperature (by a factor of $\pi/\sqrt{3} = 1.81$), scaling:
\be e^{-\sqrt{12} \omega/\kappa}, \quad \textrm{with} \quad T = \frac{\kappa}{\sqrt{12}} = 0.289\kappa. \ee
The Nulltor and Parator cases both have uniform acceleration of $\kappa$, with an exactly analytic spectrum,  confirmed in, e.g. \cite{Proceedings}, as first found in Letaw \cite{Letaw:1980yv}. 

The current work is a continuation of \cite{Proceedings} which focused exclusively on Parator.  This work contains the new power distributions of both Infrator and Hypertor which are harder because they are not single parameter systems and their spectra are not analytic despite being time-independent.  We have included the distribution of Parator in this work for completeness.

\section{Calculating the Power Distributions}
For a charge in relativistic motion the power distribution has a well-known (e.g. Jackson \cite{DavidJackson}), result (using the notation of Griffiths \cite{Griffiths}) with the intricate vector algebra details in Appendix \ref{Appendix:Angular},
\be\label{AngDistr}  \frac{dP}{d\Omega} = \frac{q^2}{4\pi} \frac{|\hrcurs \times (\bd{u} \times \bd{a})|^2}{(\hrcurs \cdot \bd{u})^5}. \ee
Two types of relativistic effects are present: (1) the spatial relationship between the velocity and acceleration and (2) the transformation between the instantaneous rest frame of the particle to the observer.\footnote{The denominator terms constitutes this dominant effect for ultra-relativistic motion.} The relevant derivatives for each worldline are given in Appendix \ref{Appendix:Derivatives}.  
We define the power distribution function $I_{\textrm{WL}}(\theta,\phi)$,
\be  
\frac{dP}{d\Omega} \equiv  \frac{2}{3} q^2 \kappa^2 I_{\textrm{WL}}(\theta,\phi), 
\ee
so that for any particular worldline, the function $I_\mathrm{WL}$ satisfies the relation,
\be \label{unitone} 
\int_0^{2\pi} \int_0^\pi I_{\textrm{WL}} \; \textrm{sin}\theta d\theta d\phi =1.
\ee
The total radiating power is therefore given by the constant,
\be
P = \frac{2}{3} q^2 \kappa^2. 
\ee
Computing the power distribution, Eq.~(\ref{AngDistr}), requires straightforward but involved vector algebra.  We have done this work and the results for each class of motion are in Appendix \ref{Appendix:Power}, where we include the simplified exact algebraic answers, the main results of this paper.  Furthermore, we resolve two dimensional polar plots $\phi = 0$, and 3D plots of the shape of the light for various speeds in Appendix \ref{Appendix:Polar}. 
\subsection*{Normalized Power Distribution}
The normalized power distribution, $I_{\textrm{WL}}(\theta,\phi)$, is the power distribution without the coefficient $\frac{2}{3}q^2 \kappa^2$ and satisfies Eq.~(\ref{unitone}).  Below are the results of the normalized power distributions for Infrator and Hypertor motions:
\begin{eqnarray}
I_{\textrm{infra}}=\frac{3}{8 \pi H(\theta,\phi)^5} \left[C_2 \, F(\theta,\phi)^2+2\, C_1 F(\theta,\phi)G(\theta,\phi)-\frac{1}{\gamma ^2} G(\theta,\phi)^2\right]\,\,, 
\end{eqnarray}
\begin{eqnarray}
I_\mathrm{hyper}=\frac{3}{8\pi Q(\theta,\phi)^5} \left[K_1 Q(\theta,\phi)^2+2\, K_2 Q(\theta,\phi)P(\theta,\phi) -\frac{1}{\gamma^2}P(\theta,\phi)^2\right]\,\,.
\end{eqnarray}
While ostensibly similar in form, the angular distribution functions are carefully defined with unique scalings and their detailed definitions and form of the coefficients can be found in Appendix \ref{Appendix:Power} .

\newpage

\section{Scaling at High Speeds}
%
The angle $\theta_{\textrm{max}}$ at which maximum radiation is emitted occurs when
\be \frac{d}{d\theta} \Big[\frac{dP}{d\Omega}\Big] = 0. \ee
Calculation of derivative for all world lines was done with respect to the angle $\theta$ in the plane $\phi = 0$, except hypertor, since its  $\phi$ range is between $0$ and $2\pi$. After numerical and graphical analysis we find the solution angles and expand them for the high speed limit, 
\begin{eqnarray}
\theta_{\textrm{max}}^{\textrm{nulltor}} = \frac{1}{2\gamma}+ \mathcal{O}\left(\frac{1}{\gamma^2}\right)^{3/2}\,,\quad
\end{eqnarray}
\be \label{maxangle}\theta^{\textrm{parator}}_{\textrm{max}} = 
\sqrt{\frac{2}{\gamma}}  + \mathcal{O}\left(\frac{1}{\gamma}\right)^{3/2}, \ee

For the case of Nulltor and Parator motions, the answer does not depend on curvature invariants, and we will always get the same maximum angle, as it was demonstrated in \cite{Proceedings}. For the case of Infrator and Hypertor motions 
we removed the constant variables $\kappa, \tau$, and $\nu$, because they do not affect the final scaling results with respect to $\gamma$.  If we substitute values for the curvature invariants the same series by order  gamma with different numerical coefficients result. For this reason, the maximum angle for  Infrator and Hypertor motions below are shown to leading order in gamma,

\be \label{maxangle1}\theta^{\textrm{infrator}}_{\textrm{max}} \sim 
\frac{1 }{\gamma}+ \mathcal{O}\left(\frac{1}{\gamma^2}\right)^{3/2}, \ee

\be \label{maxangle1}\theta^{\textrm{hypertor}}_{\textrm{max}} \sim 
\frac{1}{\gamma}+ \mathcal{O}\left(\frac{1}{\gamma^2}\right)^{3/2}. \ee

The case of Ultrator is trivial as the radiation is always maximally beamed along $\theta_{\textrm{max}}^{\textrm{ultrator}}=0$ for a stationary worldline with circular spatial projection, as $\phi$ ranges from 0 to $2\pi$. These expressions demonstrate the maximum angle scaling for the aforementioned trajectories, placing emphasis on the $\gamma^{-1/2}$ scaling of Parator.  The slow scaling (relative to rectilinear motion) here explains the physics behind the delayed collimation of radiation at high speeds that can be seen in the polar plot, $\phi = 0$, in Appendix \ref{Appendix:Polar}.  The outlier case of Parator is also corroborated by the different scaling in the simple calculation of Thomas precession for the worldlines given in the next section.  The $\gamma^{-1}$ result for Infrator and Hypertor corresponds to the usual beaming that occurs in rectilinear motion at high speeds.  

\section{Thomas precession}\label{sec:Thomas}
 Five  types of Thomas precession in all case of stationary motions with approximations are found. Thomas precession is important as a relativistic correction and a kinematic effect in the flat spacetime of special relativity.  Its importance is underscored by its algebraic origin as a result of the non-commutativity of Lorentz transformations,  and its accounting of relativistic time dilation between the electron and nucleus of an atom, giving the right correction to the spin-orbit interaction in quantum mechanics. Using the cross product definition of proper acceleration Eq. (\ref{properacc}) and the definition of Thomas precession, \be(\gamma+1)^2\omega_T^2 = \gamma^4(\bd{v}\times\bd{a})^2, \ee and the fact that stationary worldlines have constant acceleration, $\alpha = \kappa$, one finds:
\be \gamma^2(\gamma+1)^2 \omega_T^2  = \gamma^6 a^2 - \kappa^2. \ee
We choose the sign for the square root using the standard convention in the the canonical expression $+\omega_T \sim \bd{a}\times\bd{v}$, i.e. the right-hand rule in the use of the cross product that results in a positive sign overall. Applying this to the worldlines give the Thomas precession and the lowest order term in $\beta$  for Nulltor and Ultrator motions, respectively: 
\be \omega_T = 0, \ee

\be \omega_T = \frac{\kappa \beta}{\gamma+1} \approx \frac{\kappa \beta}{2}, \ee
and for Parator and Infrator motions, respectively
\be \omega_T =  \frac{\kappa}{\gamma}\left(\frac{\gamma-1}{\gamma+1}\right)\approx \frac{\kappa \beta^2}{4},\ee
\be \omega_T = \frac{\tau}{\gamma+1} \approx \frac{\tau}{2} = \frac{\kappa \beta_R}{2}. \ee
Finally,  precession for Hypertor motion is found. As its approximation looks complicated, for brevity two quantities ($a_N$ and $a_H$)  are introduced:
\be \omega_T=\frac{\kappa}{\gamma(\gamma+1)}\sqrt{ \left(\frac{a_H}{a_N}\right)^2-1}. \ee
where $a_N=\kappa/\gamma^3$ is the Nulltor acceleration and  the quantity $a_H$ is the magnitude of the 3-acceleration of Hypertor:
\begin{equation}
a_H \equiv \frac{\gamma_{min}^2}{\gamma^2} \frac{R}{\Delta}\sqrt{\frac{R_{+}^2}{\gamma^2 }+R^2 v_{min}^2}.
\end{equation} 
The variable  $\gamma_{min}$ and all others are described in Appendix \ref{Appendix:Power}.

\section{Conclusion}\label{sec:conclusions} 

We have found the power distributions for light emitted by a point charge undergoing uniform acceleration for all five classes of stationary worldlines, including those with torsion and hypertorsion. After computing their proper accelerations and confirming that all of the stationary motions have values equal to $\kappa$ (as they should be!), we have calculated acceleration ratios, minimum velocities and confirmed the results with constant power emission.  

We have graphically described the emitted radiation with different speeds, starting from rest to near light speed,  in two and three dimensional space. After numerical and graphical analysis we found  the maximum angle at which radiation is emitted. Additionally, in the Appendices  of this work we have  included worldline plots of all systems  with  the four  functions  $x^{\mu}(s)$, specifying the coordinates of each point $s$ on the curve and special projections with their general trajectory equations including curvature, torsion and hypertorsion. Finally, we have derived Thomas precession for all  classes of stationary worldlines. 

It may be possible to use these results to act as benchmark for investigating non-stationary trajectories and their power distributions.  Non-stationary trajectories have been investigated in a different context (that of the Unruh effect) in Obadia-Milgrom \cite{Obadia:2007qf}.  For instance, there are several fundamental non-uniformly accelerated motions that may emit interesting shapes of radiation to be distinguished from the stationary cases, e.g. the black hole collapse trajectory \cite{MG14one,MG14two,Good:2016LECOSPA} or the eternal thermal boundary acceleration \cite{Carlitz:1986nh}.  Other non-uniformly accelerated trajectories like those related to black hole remnants \cite{Proex,GTC,universe,MG15}, and particularly those that are of recent interest with respect to entropy and energy emission \cite{paper1,paper2,paper3} can be explored via the power distribution method in this work. 

Uniform accelerated motion was crucial for construction of the equivalence principle in general relativity.  Motivated by the fact that investigations into uniform acceleration has yielded insights into gravity, thermodynamics and quantum field theory, we hope that this study may be an incipient direction that can lead toward further insights into the basic radiation emitted from fundamental uniformly accelerated motion (in electrodynamics\footnote{For an interesting and recent paper on paradoxes involving the electric field of a uniformly accelerating charge see \cite{DavidG} and references therein.} or otherwise). 

These solutions are exact results in classical electrodynamics and may be a first step toward a better understanding (or disentangling) related radiation effects like Letaw's vacuum spectra, Unruh's uniform acceleration radiation with torsion, or Davies-Fulling moving mirror radiation.  

Further work along this direction includes the computation of radiation reaction for each of the five classes of stationary worldline motions. This can be interesting to address because results may be used to contrast with the consensus example that Nulltor radiates without radiation reaction (a nice book on uniformly accelerated point charges including much discussion of this example was written by Lyle \cite{Lyle}).  Other extensions include investigation into the intensity spectrum distribution, and maximum intensity measure\cite{Griffiths}.

\section*{Acknowledgment}

MG thanks Rafael Sorkin for clarifying a point about integral parameter independence and Harret Rosu for discussions on the semicubical parabola. Funding from state-targeted program ``Center of Excellence for Fundamental and Applied Physics" (BR05236454) by the Ministry of Education and Science of the Republic of Kazakhstan is acknowledged. Funding also provided by the ORAU FY2018-SGP-1-STMM Faculty Development Competitive Research Grant No. 090118FD5350 at Nazarbayev University. 
\newpage

\begin{appendices}


\section{Computational Details}

\subsection{Details to find Angular Distribution}
\label{Appendix:Angular}

The necessary tools for determining the stationary worldline power distributions, 
\be \frac{dP}{d\Omega} = \frac{q^2}{4\pi} \frac{|\hrcurs \times (\bd{u} \times \bd{a})|^2}{(\hrcurs \cdot \bd{u})^5}, \ee
are shown in the following.  Here 
\be \hrcurs \times(\bd{u} \times \bd{a}) = (\hrcurs \cdot \bd{a})\bd{u} - (\hrcurs \cdot \bd{u})\bd{a}, \ee
so that
\be  |\hrcurs \times (\bd{u} \times \bd{a})|^2 = (\hrcurs \cdot \bd{a})^2 u^2 - 2 (\bd{u}\cdot \bd{a})(\hrcurs \cdot \bd{a})(\hrcurs \cdot \bd{u}) + (\hrcurs \cdot \bd{u})^2 a^2. \ee
Using the definitions, including $\brcurs \equiv \bd{r} - \bd{r}'$, (the vector from a source point $\bd{r}'$ to a field point $\bd{r}$), where the source point is our origin, $\bd{r}'=0$,
\be \hrcurs \equiv \frac{\bd{r} - \bd{r}'}{|\bd{r} - \bd{r}'|} = \bd{\hat{r}} \equiv \sin\theta\cos\phi \bd{\hat{x}} + \sin\theta\sin\phi \bd{\hat{y}} + \cos\theta \bd{\hat{z}},\ee
\be \bd{u} \equiv \hrcurs - \bd{v}, \ee
one finds the terms in the order that they appear,
\be \hrcurs \cdot \bd{a} = a_x \sin\theta\cos\phi + a_y\sin\theta\sin\phi + a_z \cos\theta, \ee
\be u^2 = (\hrcurs - \bd{v})^2 = 1 - 2( v_x \sin\theta\cos\phi + v_y\sin\theta\sin\phi + v_z \cos\theta) + v^2,\ee
\be \bd{u}\cdot \bd{a} = (\hrcurs - \bd{v})\cdot \bd{a} = a_x \sin\theta \cos\phi + a_y\sin\theta\sin\phi + a_z \cos\theta -v_x a_x -v_y a_y - v_z a_z, \ee
\be \hrcurs \cdot \bd{u} = \hrcurs^2 - \hrcurs\cdot \bd{v} = 1- v_x \sin\theta\cos\phi -v_y\sin\theta\sin\phi - v_z \cos\theta, \ee
\be a^2 = a_x^2 + a_y^2 + a_z^2, \quad v^2 = v_x^2 + v_y^2 + v_z^2. \ee

\newpage
\subsection{Derivatives of the Stationary Worldlines}
\label{Appendix:Derivatives}
\subsection*{Nulltor}
The rectilinear values for the component velocities and accelerations are,
\be a_x = 0, \quad a_y = 0, \quad a_z = -\frac{\kappa }{\gamma ^3},\ee
\be v_x = 0, \quad v_y = 0, \quad v_z = \beta.\ee
Here $\gamma$ is the usual Lorentz factor and $\beta$ is the speed.
\subsection*{Ultrator}

The values for the component velocities and accelerations are (at proper time moment $s = 0$, axis chosen such that $v_z = v \hat{z}$ and $a_x=a \hat{x}$), using $\beta = \kappa/\tau$,
\be a_x = -\frac{\kappa }{\gamma ^2}, \quad a_y = 0, \quad a_z = 0,\ee
\be v_x = 0, \quad v_y = 0, \quad v_z = \beta.\ee
Notice $\beta$ is constant in this stationary worldline case. 

\subsection*{Parator}
The values for the component velocities and accelerations are,
\be a_x = -\frac{(\gamma -2) \kappa }{\gamma ^3}, \quad a_y = 0, \quad a_z = \frac{\sqrt{2\gamma -2} \kappa }{\gamma ^3},\ee
\be v_x = \gamma^{-1}\sqrt{2(\gamma-1)}, \quad v_y = 0, \quad v_z = 1-\gamma^{-1}.\ee
The main difference here is a single parameter $\kappa = \tau$ for the motion, while $\gamma$ is the usual Lorentz factor. 
\subsection*{Infrator}
The values for the component velocities and accelerations are,
\be a_x =\frac{\sigma ^2 }{\kappa}\text{sech}^3(s \sigma ) , \quad a_y = 0, \quad a_z = - \frac{\sigma ^2 \tau  }{\kappa^2}\tanh (s \sigma ) \text{sech}^2(s \sigma ),\ee
\be v_x = \tanh (s \sigma ), \quad v_y =0 , \quad v_z = v_R \text{sech}(s \sigma ),\ee
\\
where $s=\frac{1}{2 \sigma }\textrm{ln} \left(2 \frac{\gamma ^2}{\gamma_R ^2} -2\frac{\gamma }{\gamma_R } \sqrt{\frac{\gamma ^2}{\gamma_R ^2}-1}-1\right)$.



\subsection*{Hypertor}
The hypertor values with proper time 
 for the component velocities and accelerations are,
\begin{multline}
 a_x = \frac{R R_{+} }{\Delta }\text{sech}^3(R_{+} s)\,\,, \qquad a_y = \frac{R  v_{\textrm{min}  } }{\Delta } \text{sech}^2(R_{+} s) [ R_{+} \sin (R_{-} s) \tanh (R_{+} s)-R_{-}\cos (R_{-} s)]\,\,,\\  a_z =- \frac{R v_{\textrm{min} }  }{\Delta }\text{sech}^2(R_{+} s)[R_{+} \cos (R_{-} s) \tanh (R_{+} s)+R_{-} \sin (R_{-} s)]\,\,,\\
 \end{multline}
\be v_x = \tanh \left(R_+ s\right), \quad v_y = v_{\textrm{min}} \sin \left(R_- s\right) \text{sech} \left(R_+ s\right), \quad v_z = v_{\textrm{min}} \cos \left(R_- s\right) \text{sech} \left(R_+ s\right),\ee

where  $s=-\frac{1}{R_+}\cosh ^{-1}\left(\frac{\gamma}{\gamma_{\textrm{min}}}\right)$ .


\subsection{Acceleration Ratios}\label{Appendix:Acceleration}
The acceleration ratio is defined simply as $A^2 \equiv \alpha^2/a^2$. 
\subsection*{Nulltor}
\be A^2_{\textrm{line}} = \gamma^6. \ee
\subsection*{Ultrator}
\be A^2_{\textrm{circle}} = \gamma^4, \quad v \equiv \kappa/\tau. \ee
\subsection*{Parator}
\be A^2_{\textrm{cusp}} = \frac{\gamma^6}{\gamma^2 - 2\gamma + 2}, \quad \kappa = \tau. \ee
\subsection*{Infrator}
\be A^2_{\textrm{cat}} = \frac{\gamma^6}{1+ \gamma^2v_R^2}, \quad v_R \equiv \tau/\kappa. \ee

\subsection*{Hypertor}
\be A^2_{\textrm{hyper}} = \frac{\gamma ^4  \kappa ^3 \tau}{R^2 \gamma _{\textrm{min} }^4 v_{\textrm{min}} \left(R^2 v_{\textrm{min} }^2+\frac{R_+^2}{\gamma ^2}\right)}\,\,,  \ee
\be \gamma_{{\textrm{min}}} \equiv (1-v_{\textrm{min}}^2)^{-1/2} , \qquad v_{\textrm{min}} \equiv \frac{\kappa \tau}{\Delta^2} \,\,,\ee
\be  \Delta^2 \equiv \frac{1}{2}( R^2 +\kappa^2 + \tau^2 + \nu^2 ) \,\,,
\ee
 \be R^2 \equiv R_+^2 + R_-^2 \,\,,\ee
 \be  R_{\pm}^2 \equiv {\sqrt{a^2+b^2} \pm a}\,\,,\ee
 \be a=\frac{1}{2} \left(\kappa ^2-\nu ^2-\tau ^2\right) \,\,,\qquad b= \kappa  \nu\,\,.\ee

 
  

 \newpage

\subsection{Compilation of Power Distributions}\label{Appendix:Power}
\subsection*{Nulltor}
\be \label{LineHS} I_{\textrm{null}} = \frac{3}{8\pi \gamma^6} \frac{\textrm{sin}^2\theta}{(1-\beta\textrm{cos}\theta)^5}. \ee
\subsection*{Ultrator}
\be I_{\textrm{ultra}} = \frac{3}{8\pi} \frac{(1-\beta\textrm{cos}\theta)^2-(1-\beta^2)\textrm{sin}^2\theta\textrm{cos}^2\phi}{\gamma^4(1-\beta\textrm{cos}\theta)^5}. \ee
\subsection*{Parator}
\be\label{even_func} I_{\textrm{para}} \equiv 
\frac{\lambda_1
+\lambda_2 \textrm{cos}{\phi} 
+\lambda_3\textrm{cos}^2\phi }{ 
\left( \textrm{cos}\phi + \lambda_4 \right){}^5},\ee
%
%
\begin{subequations}
\begin{eqnarray}
\lambda_0&=&-\frac{3}{8\pi \gamma\sqrt{2\gamma _1}^5 \textrm{sin}^5\theta},\\
\lambda_1&=&\lambda_0
\left[2 + \gamma  \gamma _2 + \gamma _1 \textrm{cos}\theta 
\left(\gamma_3 \textrm{cos}\theta-2 \gamma_2\right)\right],\\
\lambda_2&=& -2\lambda_0 \sqrt{2\gamma _1}\textrm{sin}\theta \left(\gamma _1-\gamma _2 \textrm{cos}\theta\right),\\
\lambda_3&=& 2\lambda_0 \gamma_{3/2}\,\textrm{sin}^2\theta,\\
\lambda_4&=& \frac{\gamma _1 \textrm{cos}\theta - \gamma}{\sqrt{2\gamma _1}\textrm{sin}\theta},
\end{eqnarray}
\end{subequations}
\be \gamma_n \equiv \gamma-n \ee
\be n=1,2,3, 3/2 \ee

\subsection*{Infrator}
\begin{eqnarray}
I_{\textrm{infra}}=\frac{3}{8 \pi H(\theta,\phi)^5} \left[C_2 \, F(\theta,\phi)^2+2\, C_1 F(\theta,\phi)G(\theta,\phi)-\frac{1}{\gamma ^2} G(\theta,\phi)^2\right]\,\,, 
\end{eqnarray}

\begin{subequations}
\begin{eqnarray}
 G(\theta,\phi)&=&\frac{\sin (\theta ) \, F(\theta,\phi)}{H(\theta,\phi)}\,\left( \cos (\phi )-v_R \, \sin (\phi )\, \sinh (\alpha)\right)\,\,,\\
F(\theta,\phi)&=&(1-v_R^2)\,\text{sech}^3(\alpha)\, H(\theta,\phi)\,\,,\\
H(\theta,\phi)&=&1-v_R \, \sin (\theta ) \, \sin (\phi )\, \text{sech}(\alpha)-\,\sin (\theta )\, \cos (\phi ) \, \tanh (\alpha ) \,\,, 
\end{eqnarray}
\end{subequations}

\begin{subequations}
\begin{eqnarray}
C_1&=&(1-v_R^2)\tanh (\alpha)\,\,, \\
C_2&=& v_R^2 \sinh ^2(\alpha)+1 \,\,,
\end{eqnarray}
\end{subequations}
\begin{subequations}
\begin{eqnarray}
 \alpha&=& \frac{1}{2}\, \textrm{ln}(2\,\xi \,\eta-1)\,\,, \\
 \eta &=&\sqrt{\xi^2-1}+\xi\,\,, \\
 \xi &=&\gamma \sqrt{1-v_R^2} \,\,.
\end{eqnarray}
\end{subequations}

\subsection*{Hypertor}

\begin{eqnarray}
I_\mathrm{hyper}=\frac{3}{8\pi Q(\theta,\phi)^5} \left[K_1 Q(\theta,\phi)^2+2\, K_2 Q(\theta,\phi)P(\theta,\phi) -\frac{1}{\gamma^2}P(\theta,\phi)^2\right]\,\,,
\end{eqnarray}

\begin{subequations}
\begin{eqnarray}
Q(\theta,\phi)&=&1 - A_1\,\cos(\theta) + A_2\, \sin(\theta)\cos(\phi) -  A_3\,\sin(\theta) \sin(\phi)\,\,,\\
P(\theta,\phi)&=& B_3\,\cos(\theta) + B_2\, \sin(\theta)\cos(\phi) + B_1\,\sin(\theta) \sin(\phi)\,\,,
\end{eqnarray}
\end{subequations}

\begin{subequations}
\begin{eqnarray}
A_1&=& v_{\textrm{min}} b\cos(\Omega_\gamma)\,\,,\\
A_2&=&\sqrt{1-b^2}\,\,,\\
A_3&=& v_{\textrm{min}}b\sin(\Omega_\gamma)\,\,,
\end{eqnarray}
\end{subequations}

\begin{subequations}
\begin{eqnarray}
B_1&=& b(A_2A_3D_1-A_1D_2)\,\,,\\
B_2 &=& D_1 b^3 \,\,,\\
B_3 &=& b(A_3D_2+A_2A_1D_1) \,\,,
\end{eqnarray}
\end{subequations}

\begin{subequations}
\begin{eqnarray}
K_1&=& B_2^2+B_1^2+B_3^2\,\,,\\
K_2&=& -A_2 B_2+A_3B_1+A_1B_3\,\,,
\end{eqnarray}
\end{subequations}

\begin{subequations}
\begin{eqnarray}
D_1 &=& \frac{R\,R_{+}}{\kappa \Delta}\,\,, \\
D_2 &=& \frac{R\,R_{-}}{\kappa \Delta}\,\,,
\end{eqnarray}
\end{subequations}

\begin{subequations}
\begin{eqnarray}
\Omega_\gamma&=& \frac{R_{-}}{R_{+}}\cosh^{-1}(\frac{1}{b}) \,\,,\,\,\,\,\,\,\, 
b=\frac{\gamma_{\textrm{min}}}{\gamma}\,\,,\,\,\,\,\,\,v_{\textrm{min}}=\frac{\kappa\,\tau}{\Delta^2} \,\,,\,\,\,\, \gamma_{\textrm{min}}= \frac{1}{\sqrt{1-v_\textrm{min}^2}} \,\,,
\end{eqnarray}
\end{subequations}

\begin{subequations}
\begin{eqnarray}
R^2=R_{+}^2 + R_{-}^2\,\,,
\end{eqnarray}
\end{subequations}

\begin{subequations}
\begin{eqnarray}
R_{+}^2&=& \frac{1}{2}(\kappa^2-\nu^2-\tau^2) +\sqrt{\kappa^2\,\nu^2+\frac{1}{4}(\kappa^2-\nu^2-\tau^2)^2}  \,\,,\\
R_{-}^2&=& -\frac{1}{2}(\kappa^2-\nu^2-\tau^2) +\sqrt{\kappa^2\,\nu^2+\frac{1}{4}(\kappa^2-\nu^2-\tau^2)^2} \,\,.
\end{eqnarray}
\end{subequations}

\newpage

\section{Plots}

\subsection{Worldline Plots}\label{Appendix:Worldline}

\subsection*{Nulltor}
\be x^\mu (s) = \kappa^{-1} \left(\sinh(\kappa s), 0, 0, \cosh(\kappa s)\right). \ee
\begin{figure}[ht]
\begin{center}
{\rotatebox{0}{\includegraphics[width=1.0in]{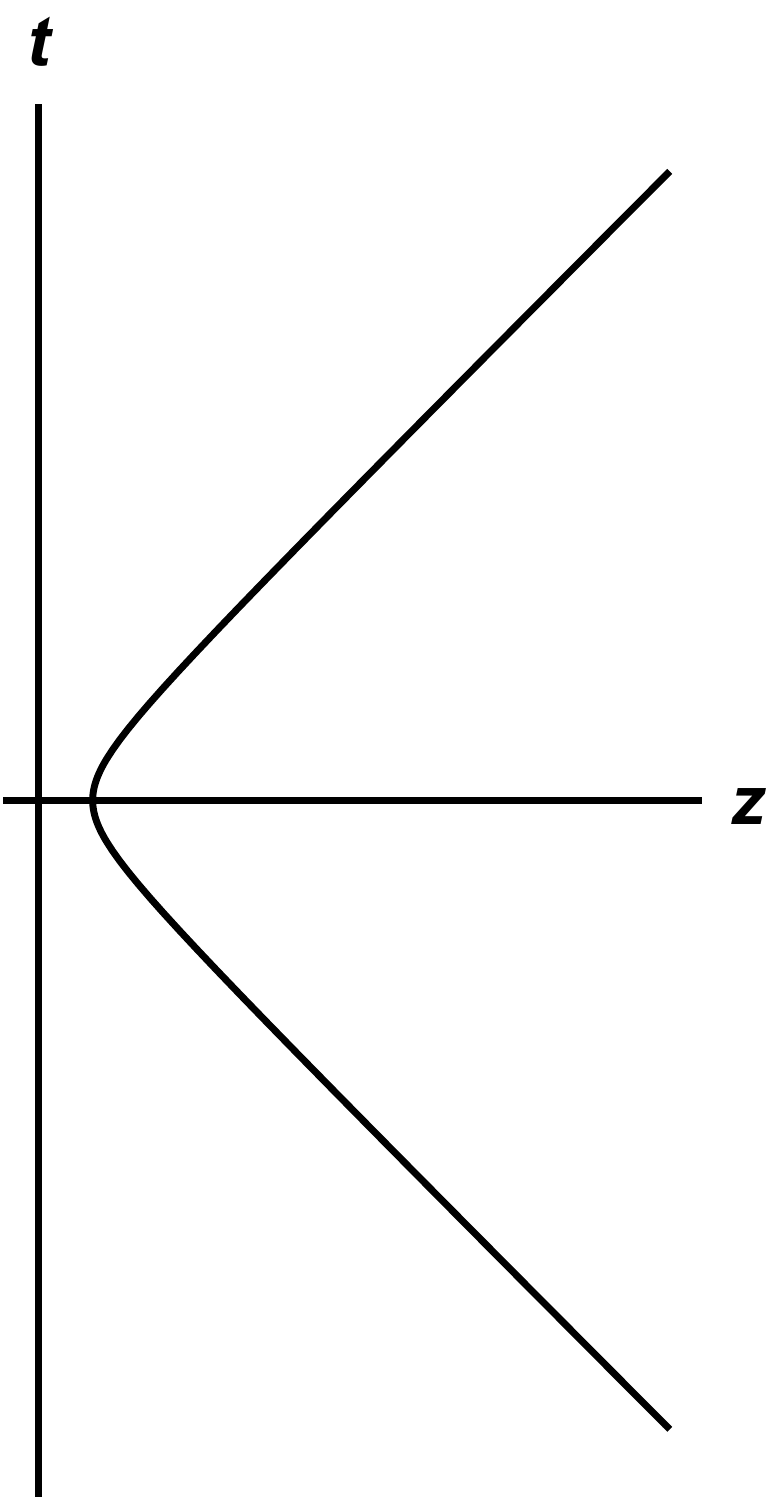}}} 
{\caption{\label{fig:WL_Line} The hyperbola in a spacetime diagram.  The vertical axis is time, the horizontal axis is space.  This is the uniformly accelerated rectilinear worldline.  This is usual straightline motion with constant proper acceleration.  Here $s$ is proper time and $\kappa$ is the proper acceleration.  It is a one parameter motion defined by $\kappa$ only.  }} 
\end{center}
\end{figure}  

\subsection*{Ultrator}
\be x^\mu (s) = \rho^{-2} \left(\tau \,\rho \,s, \kappa \cos\rho s, 0, \kappa \sin \rho s\right). \ee
\begin{figure}[ht]
\begin{center}
\includegraphics[width=2.0in]{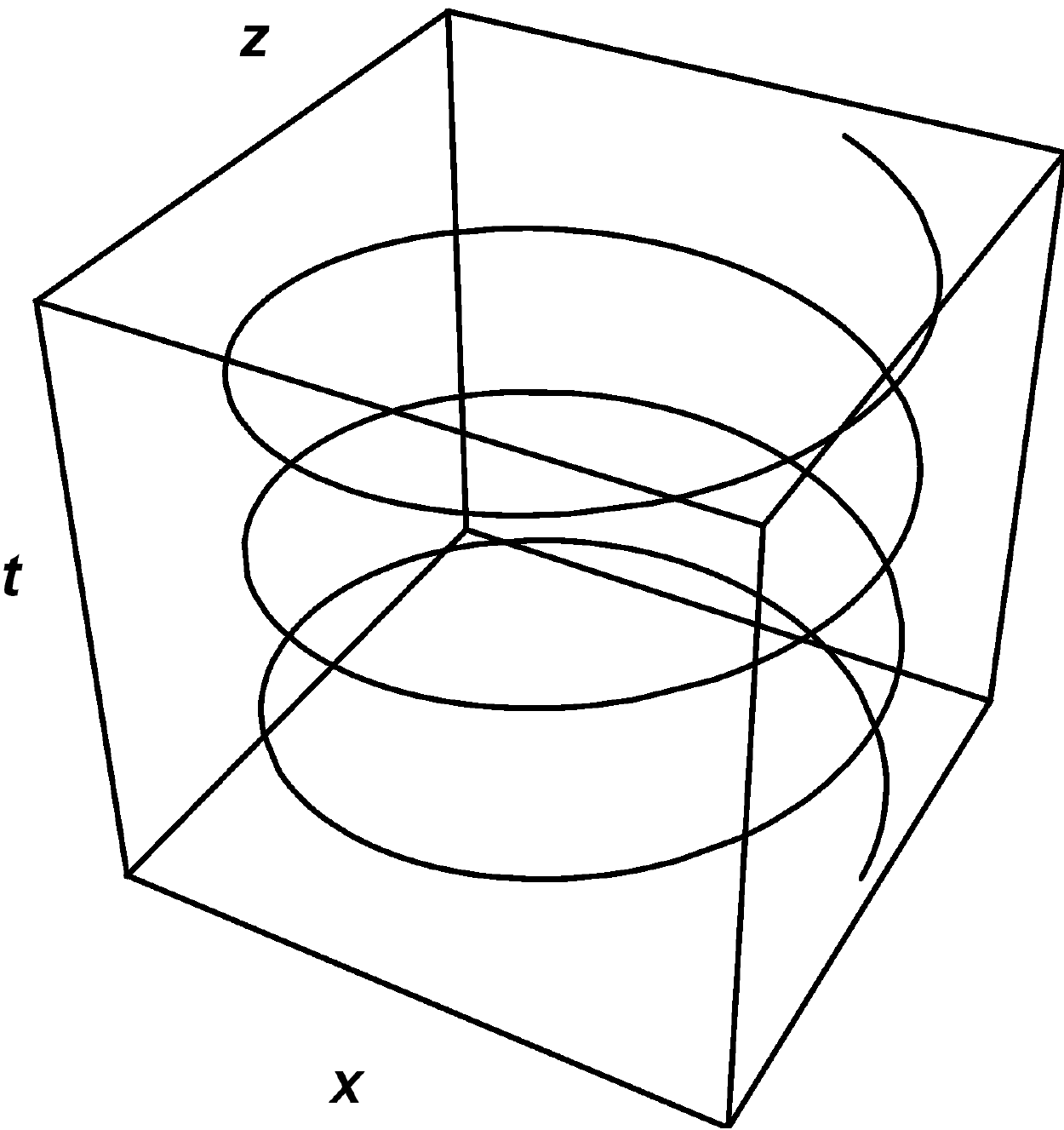}
%
{\caption{\label{fig:WL_Circle} The helix worldline for familiar circular motion.  This is a 3D parametric plot of the uniformly accelerated worldline that is spatially projected as circular motion.  Here $s$ is proper time, $\rho^2 = \tau^2 - \kappa^2$, $\tau$ is the torsion, and $\kappa$ is the proper acceleration.  It is a two-parameter system defined by $\kappa$ and $\tau$.  }} 
\end{center}
\end{figure}  

\newpage 

\subsection*{Parator}
\be x^\mu (s) = \left(s + \frac{1}{6} \kappa^2 s^3, \frac{1}{2} \kappa s^2, 0, \frac{1}{6} \kappa^2 s^3\right). \ee
\begin{figure}[ht]
\begin{center}
\includegraphics[width=2.0in]{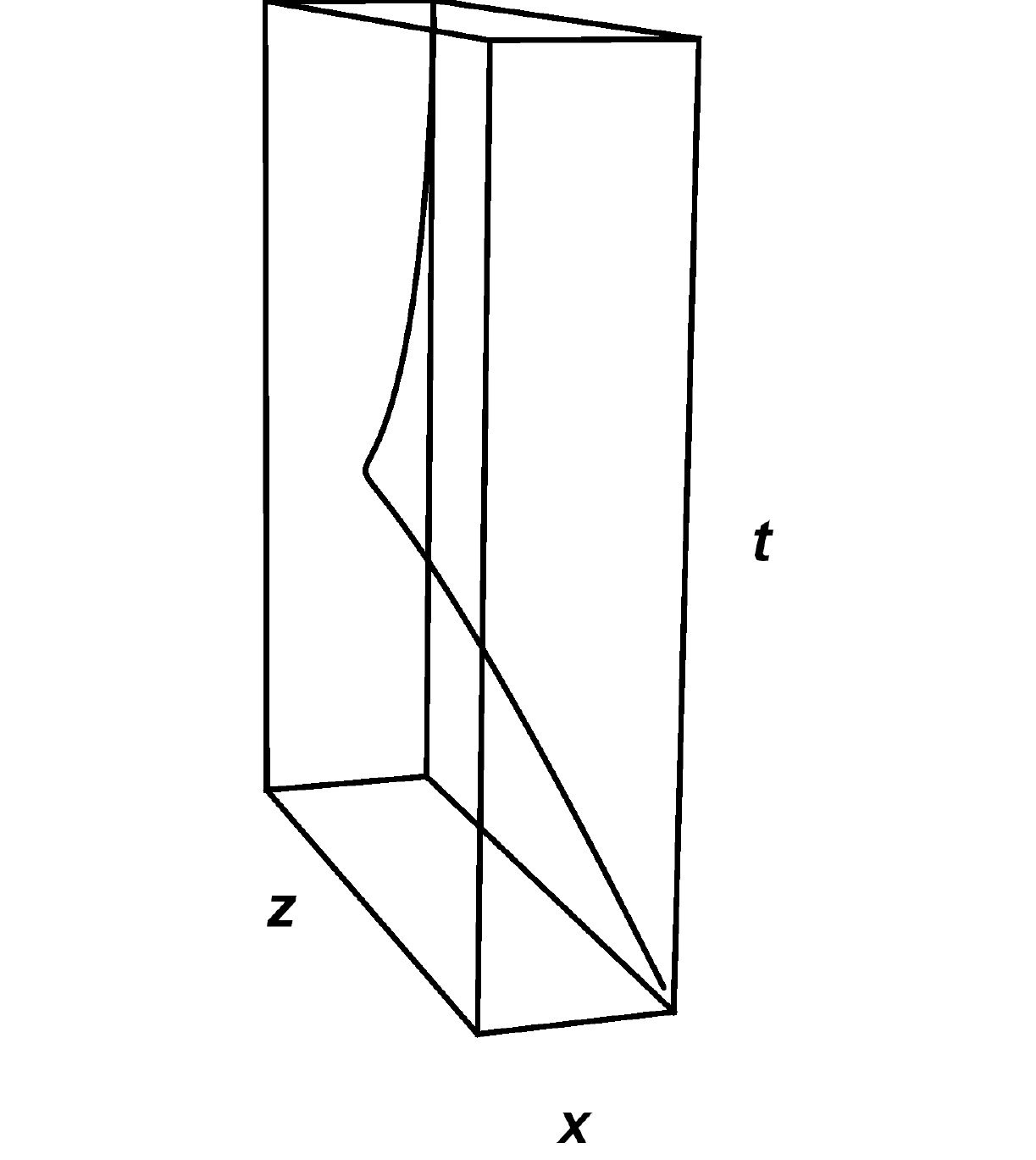}
{\caption{\label{fig:WL_Cusp} The cusp world line which is spatially a semi-cubic parabola plotted in a 3D parametric plot.  It is a one-parameter system defined by $\kappa = \tau$. Here $s$ is proper time and the system moves in two dimensions of space.  
}} 
\end{center}
\end{figure}  

\subsection*{Infrator}
\be x^\mu (s) = \sigma^{-2} \left(\kappa \sinh (s \sigma ), \kappa \cosh (s \sigma ), 0, s \tau \sigma\right). \ee
\begin{figure}[ht]
\begin{center}
\includegraphics[width=2.0in]{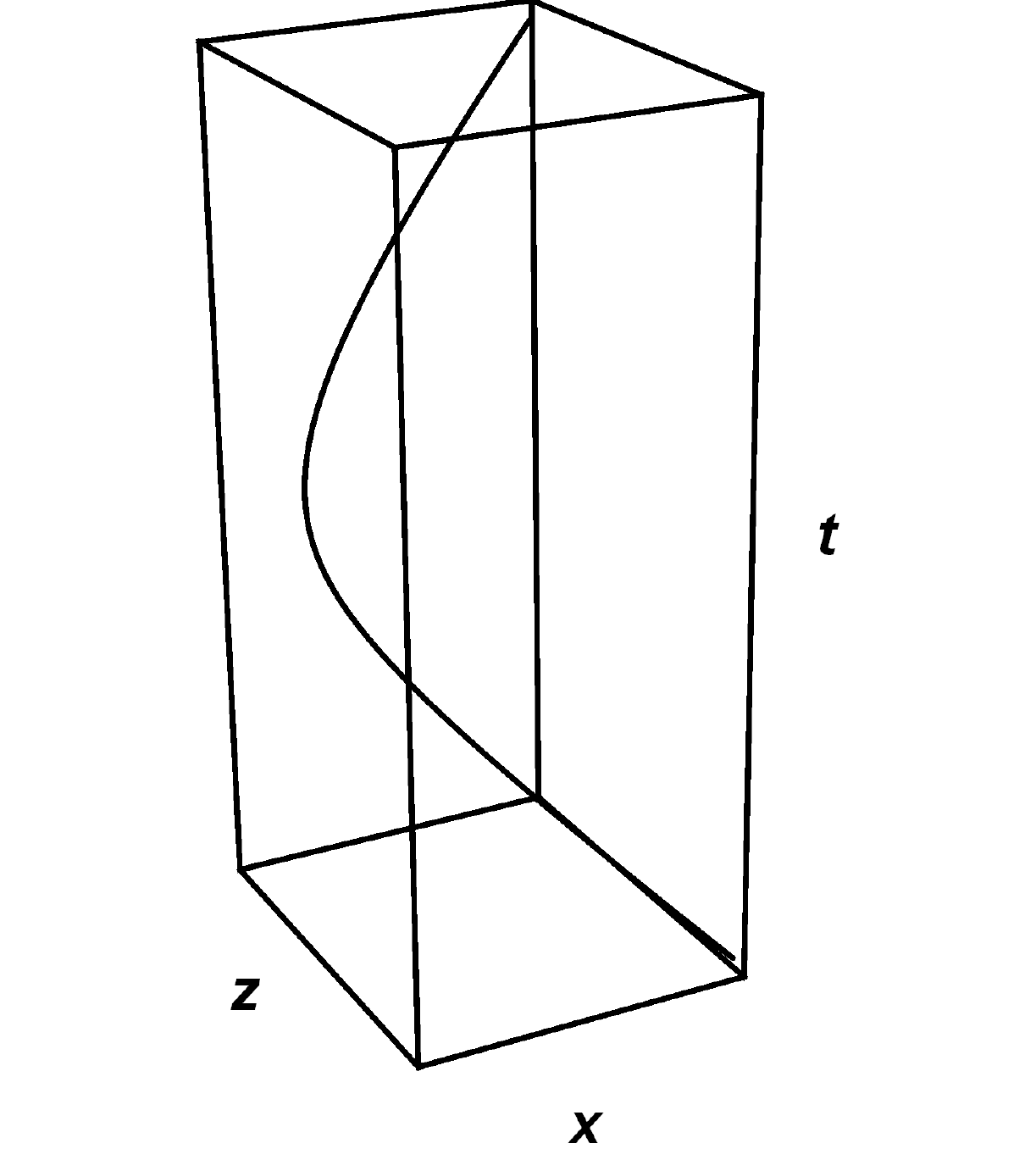}
{\caption{\label{fig:WL_Cat1} The infrator worldline (spatial catenary) plotted in a 3D parametric plot. Here $s$ is proper time, $\sigma^2 = \kappa^2- \tau^2$, and $\tau<\kappa$.  This is a two-parameter system in two dimensions of space.  Its projected spatial shape is the familiar hanging chain. } } 
\end{center}
\end{figure}  

\subsection*{Hypertor}
The Hypertor worldline is 3+1 dimensional, and so cannot be plotted on a 3D plot.  Its worldline function is written down explicitly in Letaw (1980): \cite{Letaw:1980yv}.

\newpage


\subsection{Spatial Projections}\label{Appendix:Spatial}

\subsection*{Nulltor: Line}
For the zero torsional worldline whose spatial projection is just a line, its direction has been assigned to the z-axis in this work.   

\subsection*{Ultrator: Circle}
\be z=\pm\sqrt{\frac{\kappa^2}{(\tau^2-\kappa^2)^2}-x^2} \ee
\begin{figure}[ht]
\begin{center}
\includegraphics[width=2.0in]{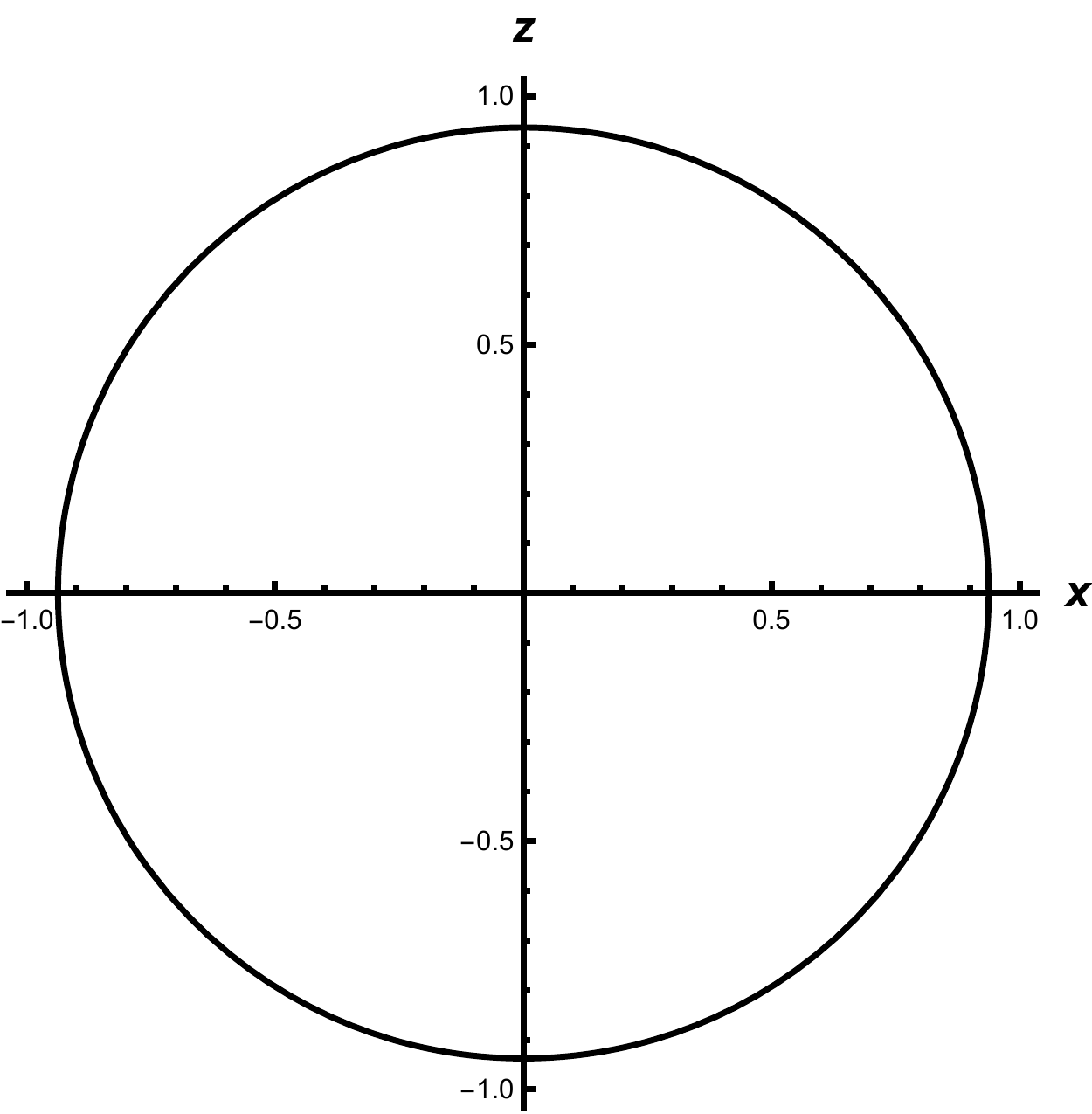}
{\caption{\label{fig:WL_Circ} The spatial Ultrator plot with invariants $\tau > \kappa$ for $\tau = 1$ and  $\kappa = 3/5$, with radius $R=\frac{\kappa}{\rho^2}=15/16 $.}} 
\end{center}
\end{figure}  

\subsection*{Parator: Cusp}
\be z = \frac{\sqrt{2 \kappa}}{3} x^{3/2}\ee
\begin{figure}[ht]
\begin{center}
\includegraphics[width=2.0in]{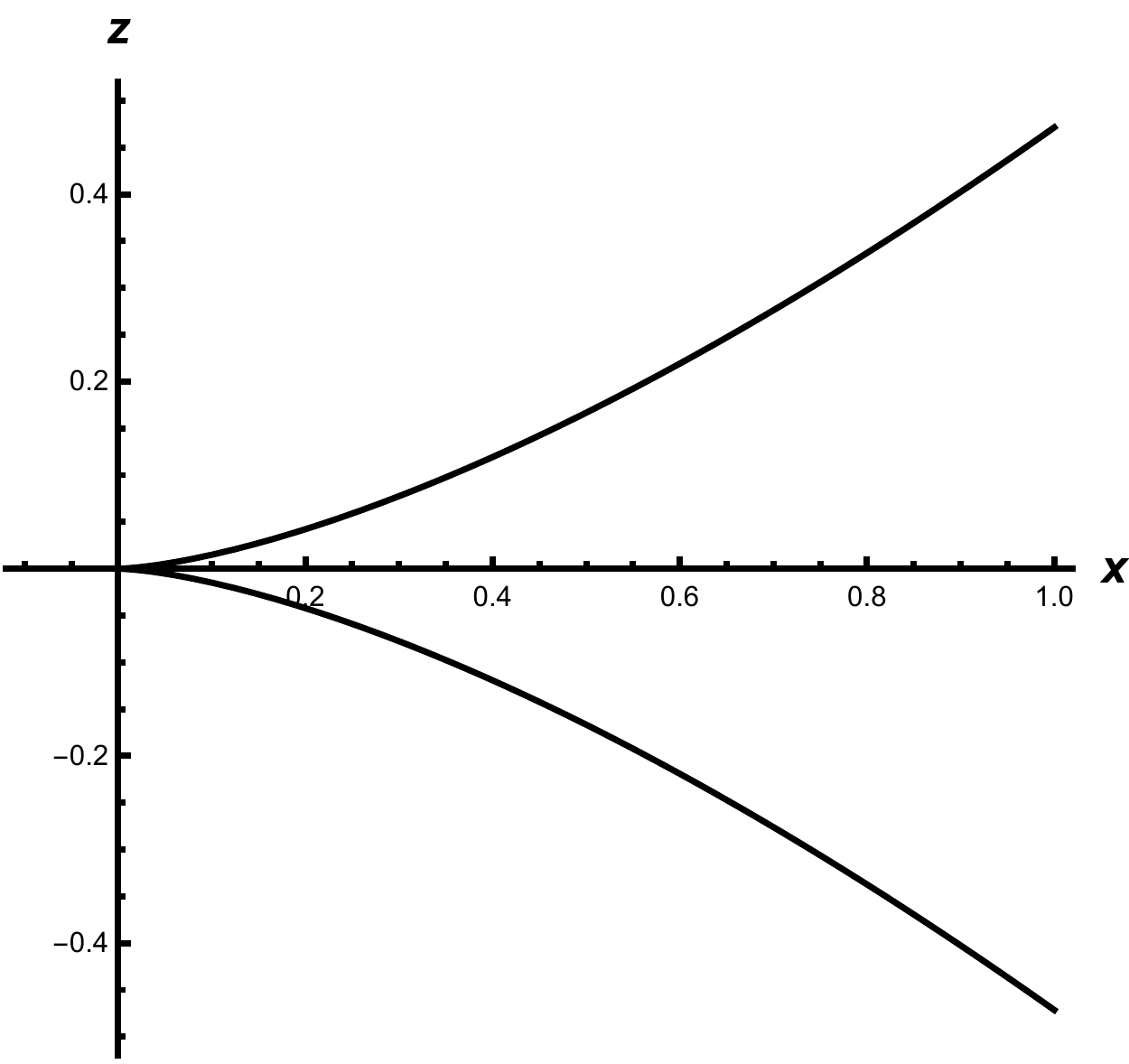}
{\caption{\label{fig:WL_Cusp1} The spatial Parator plot with invariants $\kappa =\tau= 1$.}} 
\end{center}
\end{figure}

\subsection*{Infrator: Catenary}
\be z= \frac{\tau}{\kappa^2-\tau^2}\cosh^{-1}\left[ \frac{x(\kappa^2-\tau^2)}{\kappa}\right]  \ee
\begin{figure}[ht]
\begin{center}
\includegraphics[width=2.0in]{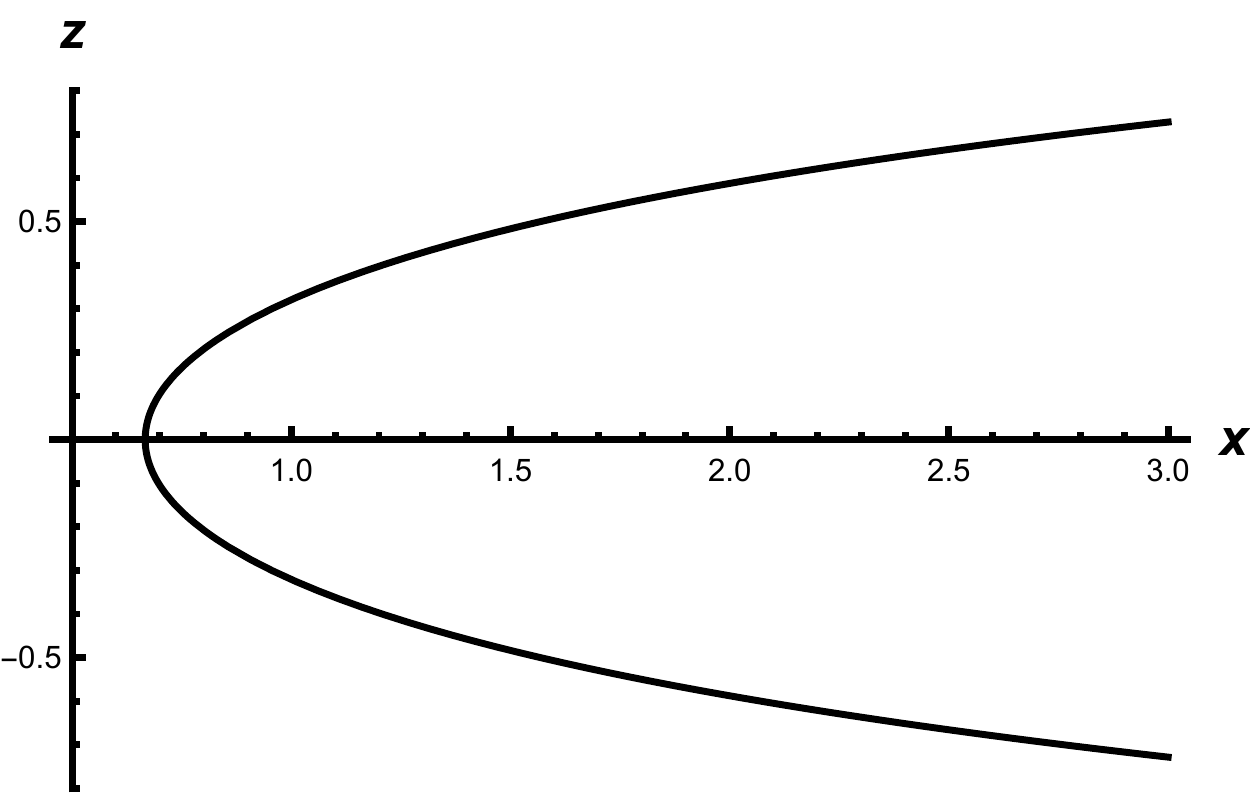}
{\caption{\label{fig:WL_Cat3} Two dimensional catenary  plot with invariants $\tau < \kappa$ for $\tau = 1$,  $\kappa = 2$ of the Infrator motion.}} 
\end{center}
\end{figure}  

\subsection*{Hypertor: Helix with Variable Pitch}
Parametric equations:
\be x = \frac{\Delta}{RR_{+}} \cosh (R_{+}s ),\ee
\be y = \frac{\kappa \tau}{R\Delta R_{-}}\cos(R_{-}s),\ee
\be z = \frac{\kappa \tau}{R\Delta R_{-}}\sin(R_{-}s). \ee
where variable $\Delta$ and others are shown in Appendix \ref{Appendix:Acceleration}. 

\begin{figure}[ht]
\begin{center}
\includegraphics[width=2.0in]{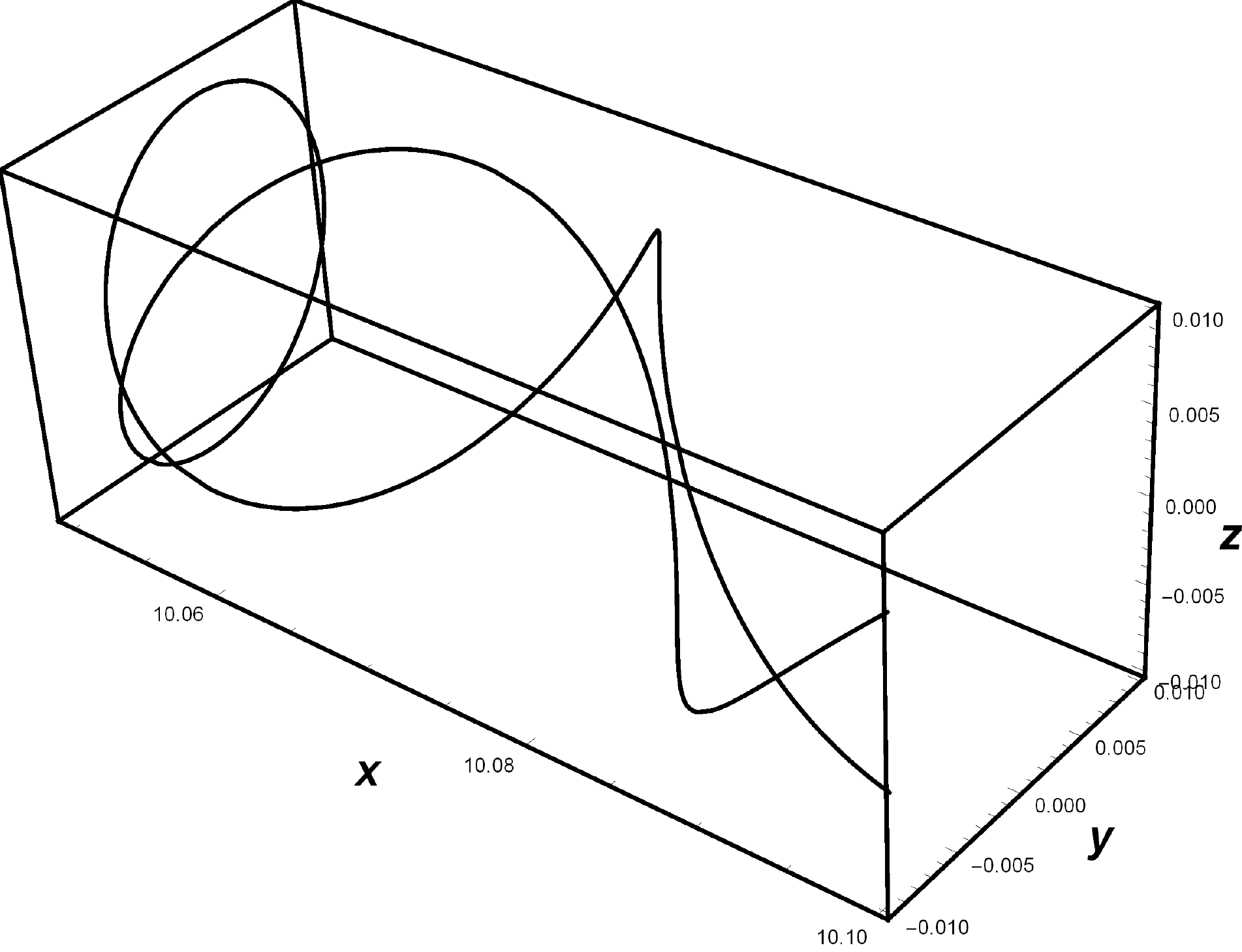}
{\caption{\label{fig:WL_hyper} Spatial Projection of the Hypertorsional worldline with $\tau = 10$, $\nu = \kappa = 1$.  Proper time ranges from $-1$ to $1$.}} 
\end{center}
\end{figure}  

\newpage

\subsection{Polar Distribution Plots}\label{Appendix:Polar}
\subsection*{Nulltor}
\begin{figure}[ht]
\begin{center}
\mbox{
\subfigure{\includegraphics[width=1.2in]{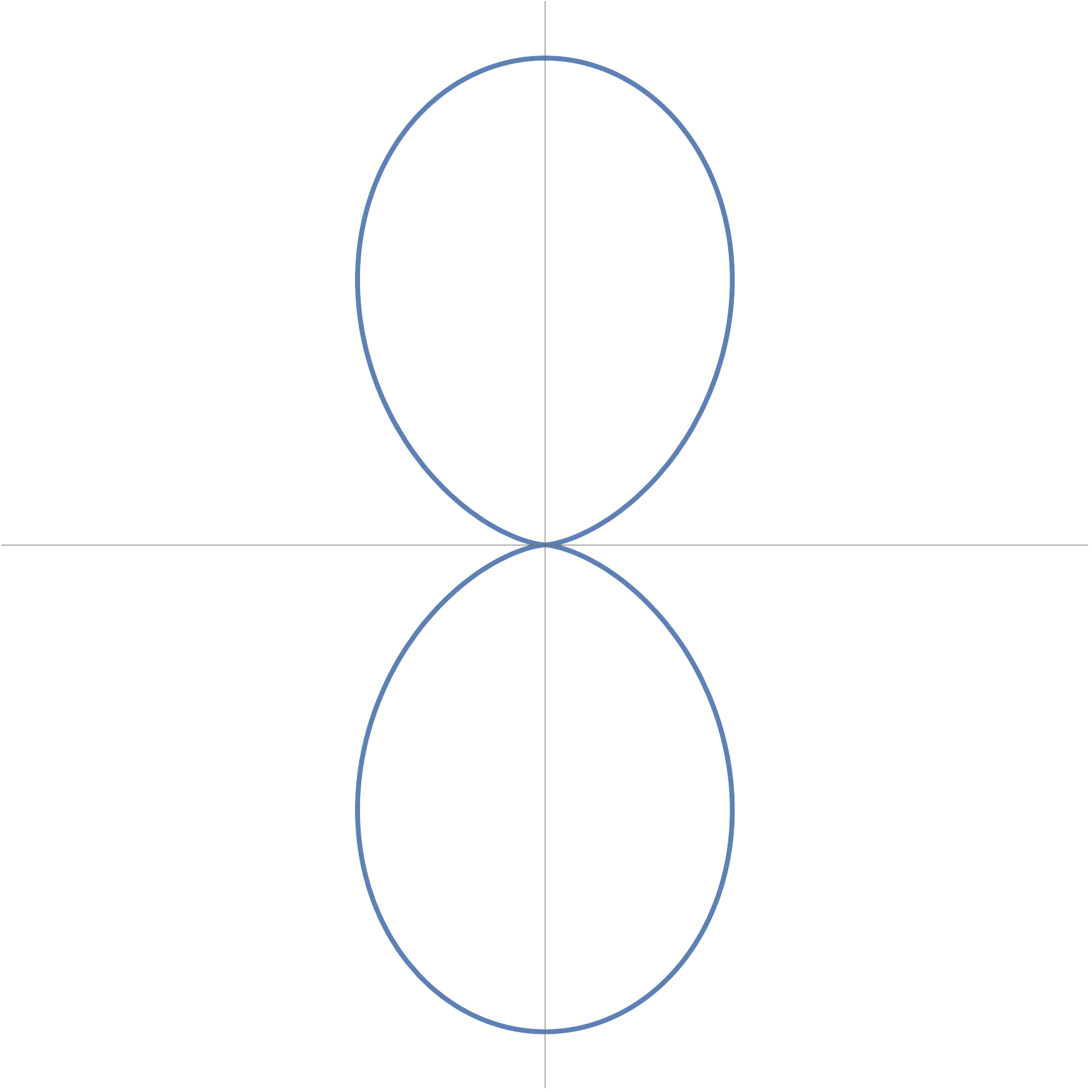}}\quad
\subfigure{\includegraphics[width=1.2in]{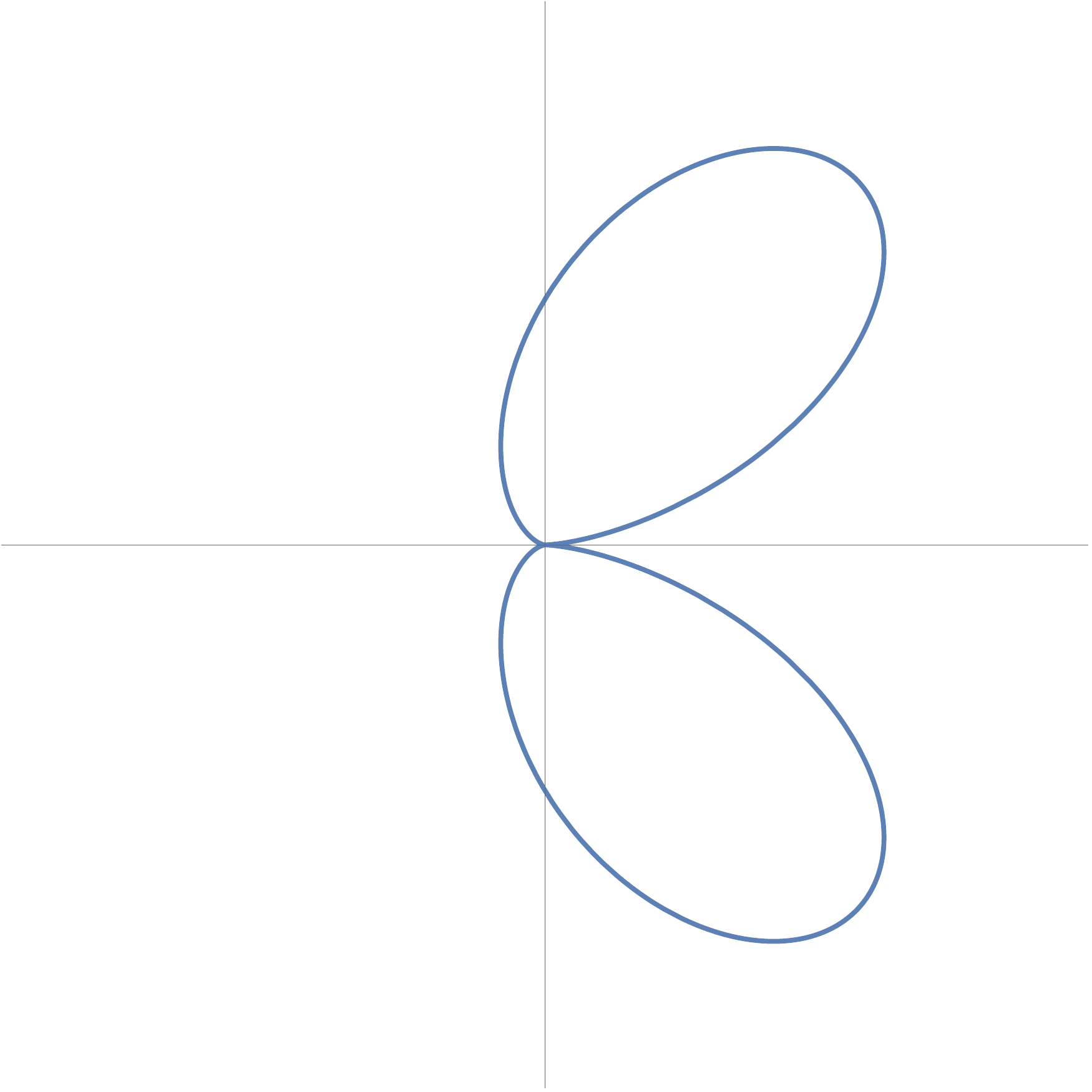}}\quad
\subfigure{\includegraphics[width=1.2in]{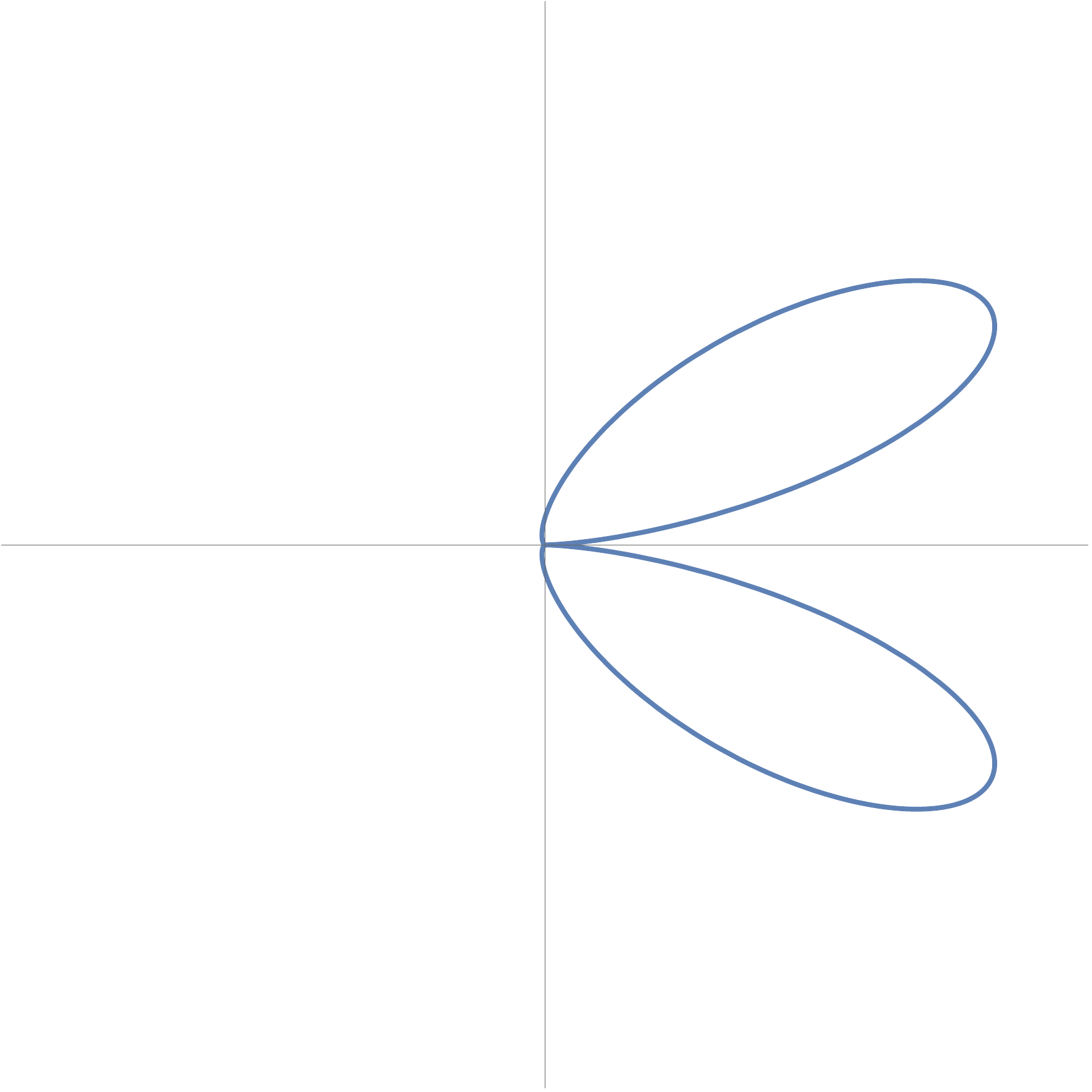}}\quad
\subfigure{\includegraphics[width=1.2in]{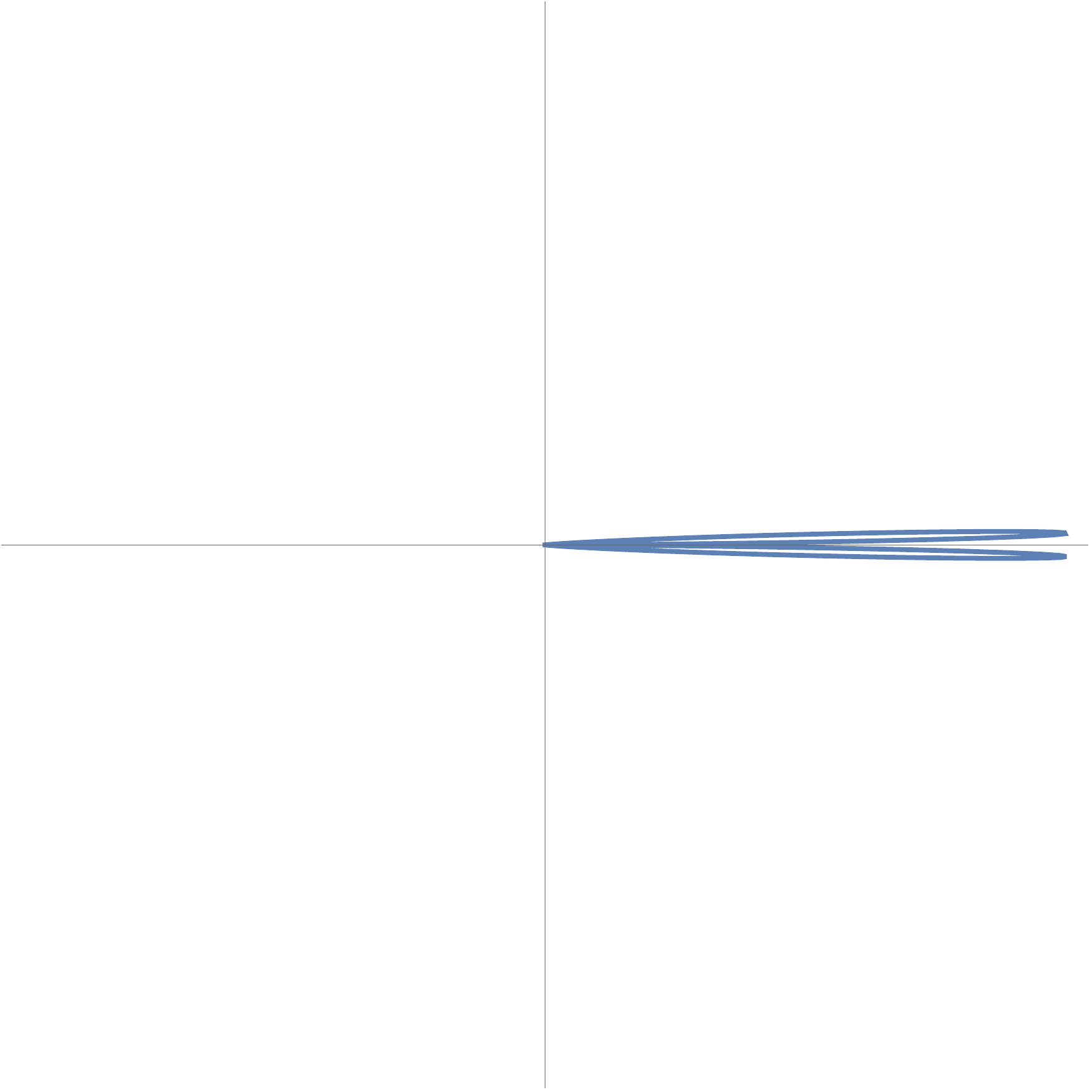} }}
\caption{\label{fig3} Rectilinear Angular Distribution; A polar plot with different velocities for the particle, $\beta = 0.001, 0.333, 0.666, 0.999$. In the first figure the particle has a very small velocity. In  other figures particle moving with  $33\%$ ,$66\%$, $99\%$ speed of light.  The vertical axes is $x$ and the horizontal axis is $z$. The electron moves forward along the $z$ straight line direction.  This is the usual textbook distribution of braking radiation with forward beaming around the axis of motion (no actual radiation in the precise direction of motion, just around it).   } 
\end{center}
\end{figure}  
\subsection*{Ultrator}

\begin{figure}[ht]
\begin{center}
\mbox{
\subfigure{\includegraphics[width=1.2in]{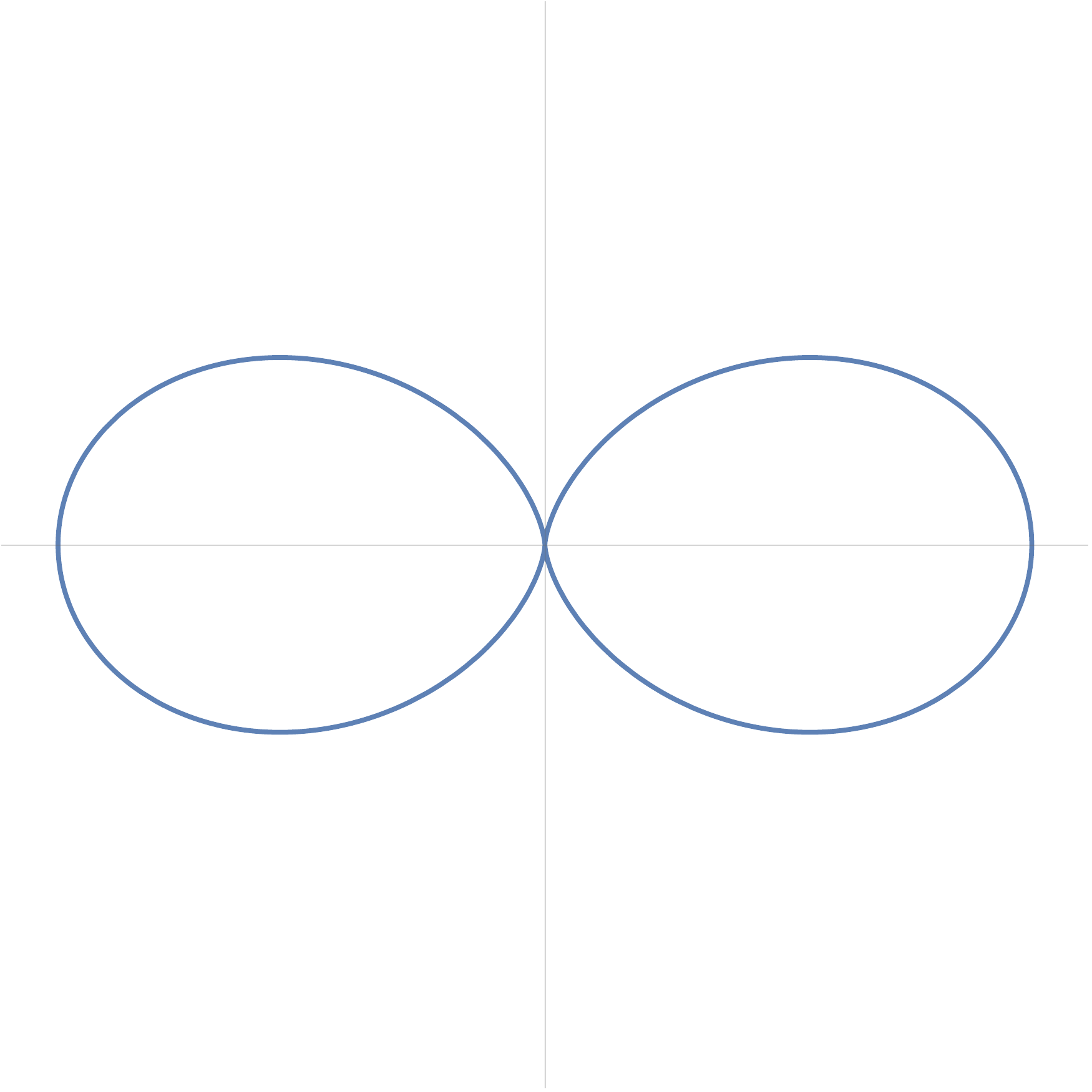}}\quad
\subfigure{\includegraphics[width=1.2in]{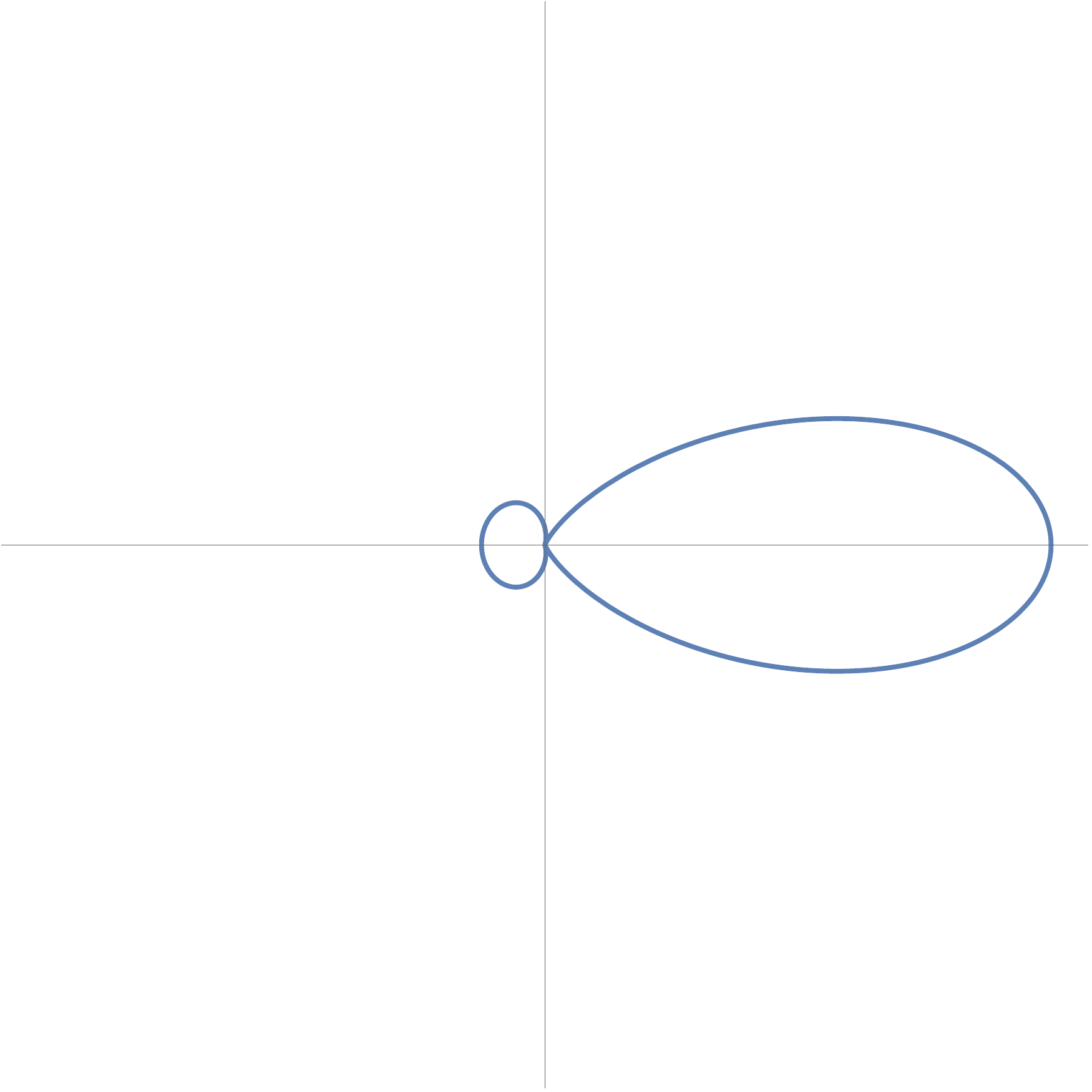}}\quad
\subfigure{\includegraphics[width=1.2in]{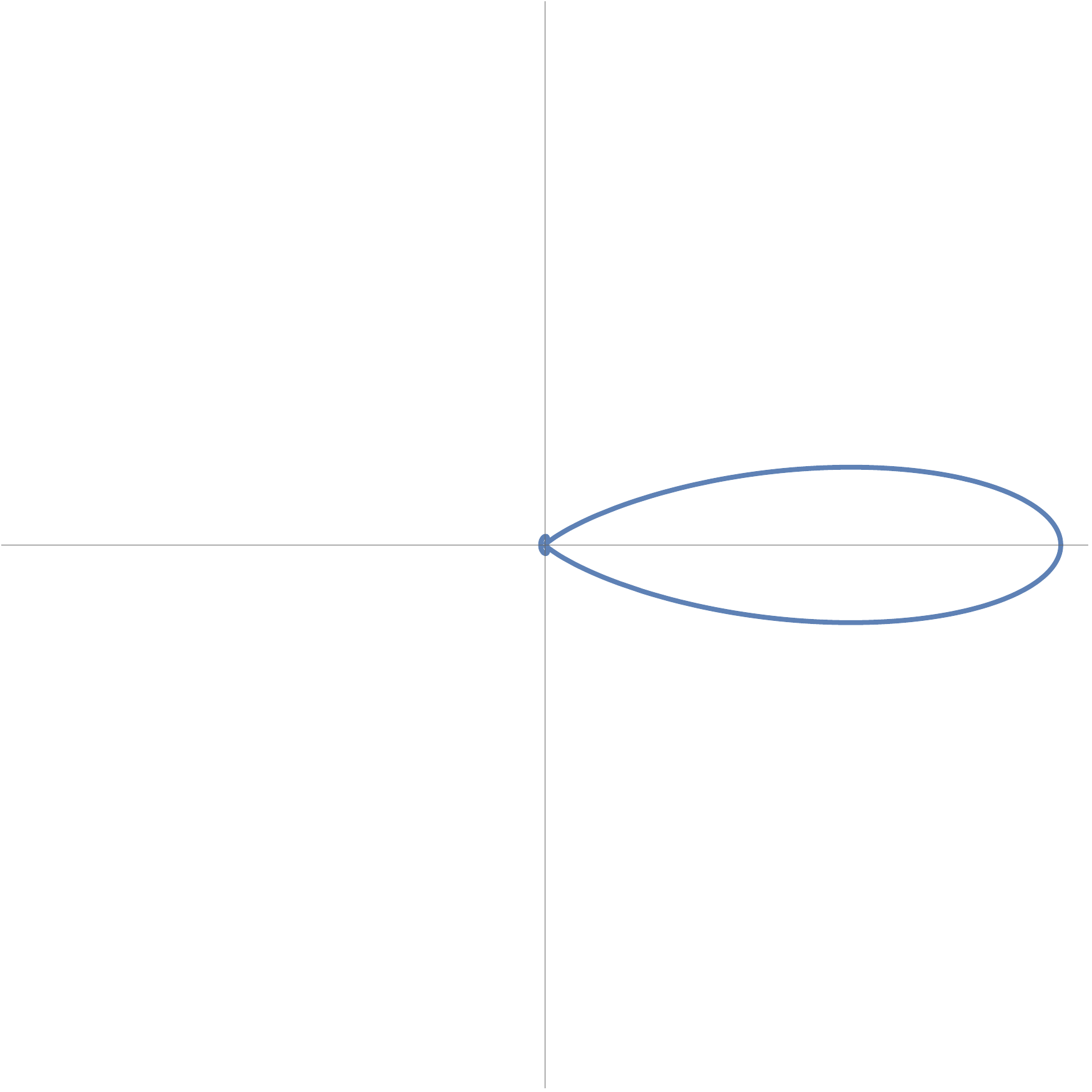}}\quad
\subfigure{\includegraphics[width=1.2in]{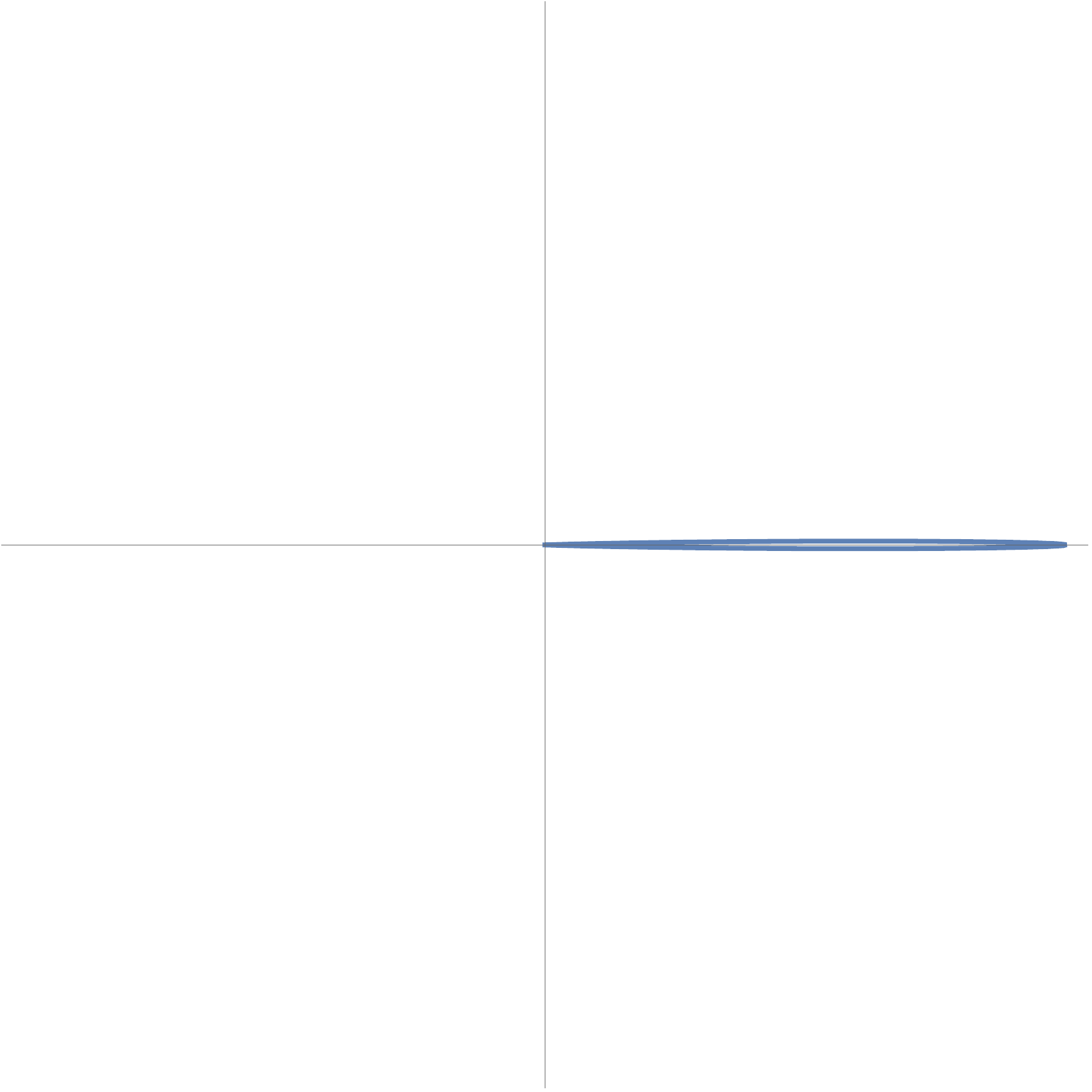} }}
\caption{\label{fig4a} Synchrotron Angular Distribution; A polar plot with $\beta = 0.001, 0.333, 0.666, 0.999$. In the first figure  particle  moving with constant velocity along a circle. In the second figure, the particle is moving with  $33\%$ speed of light, in the third and fourth figures, we have $66\%$ and $99\%$ speed of light respectively. The vertical axes is $x$ and the horizontal axis is $z$. The electron moves circularly in the horizontal $z$-$y$ plane. Notice each graph is a different worldline as the shape remains the same along a constant $\beta$ stationary worldline. This is the usual textbook example of synchrotron radiation, where the radiation is as a `train headlight', always directed in the forward motion, quite distinct from braking radiation. } 
\end{center}
\end{figure}  
\newpage
\subsection*{Parator}
\begin{figure}[ht]
\begin{center}
\mbox{
\subfigure{\includegraphics[width=1.2in]{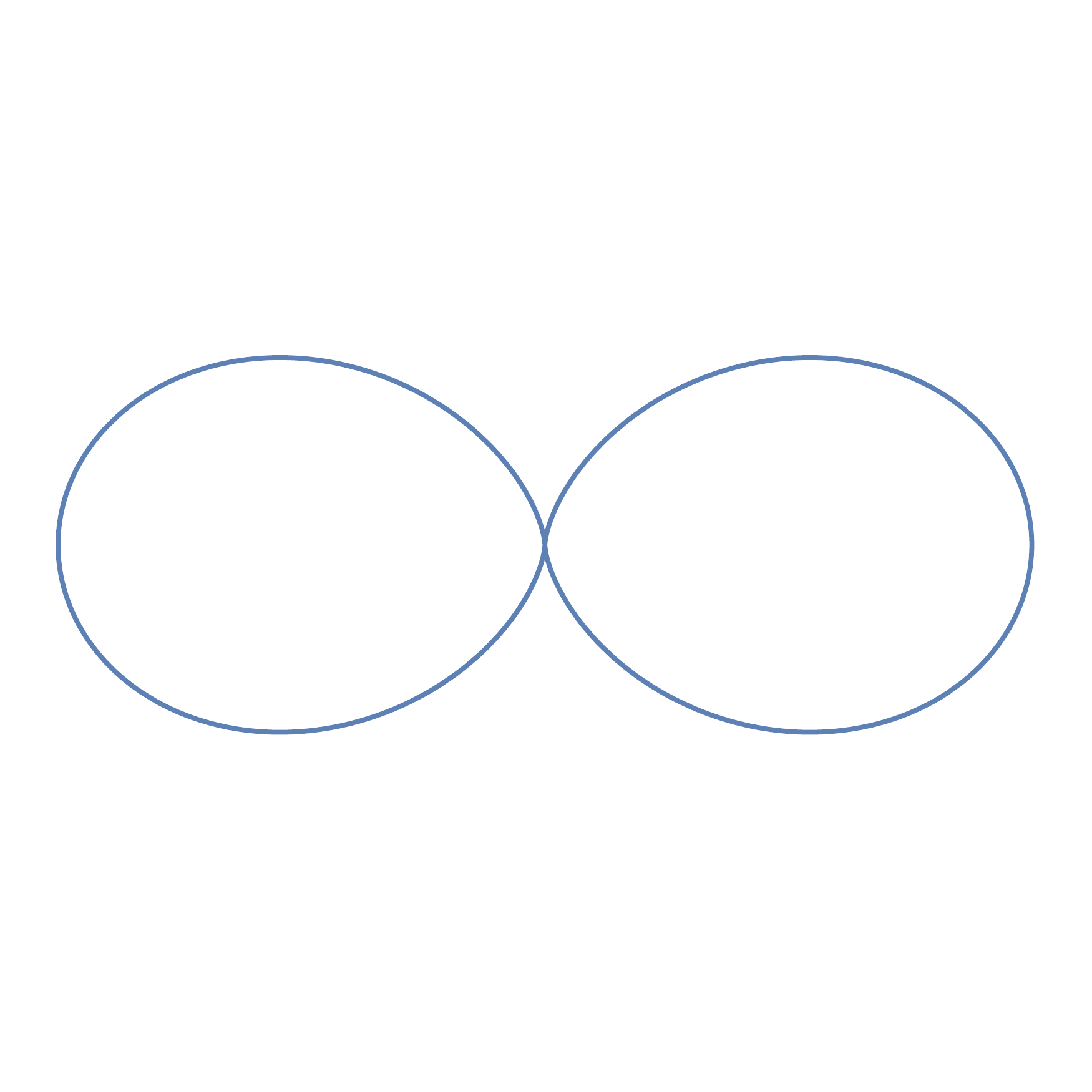}}\quad
\subfigure{\includegraphics[width=1.2in]{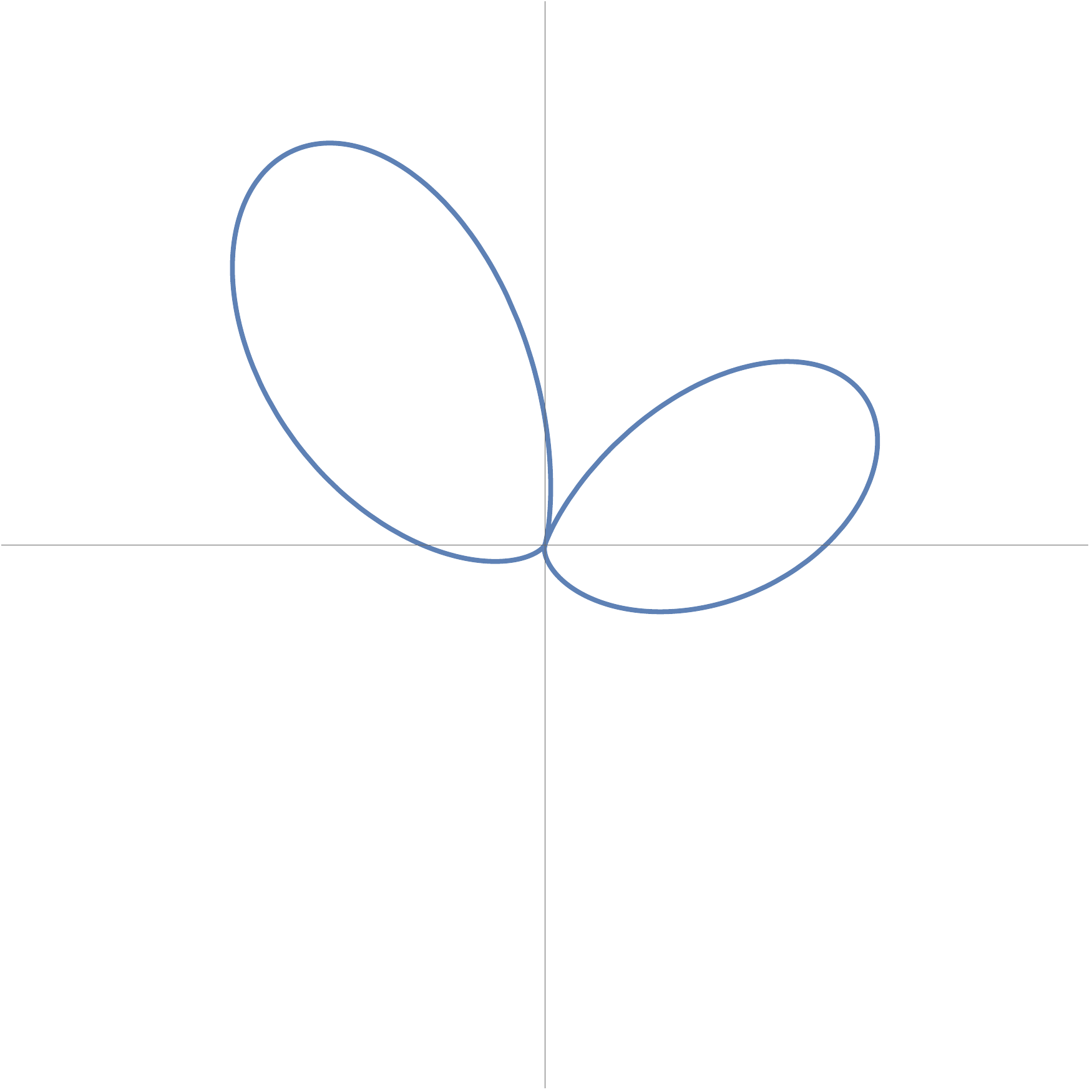}}\quad
\subfigure{\includegraphics[width=1.2in]{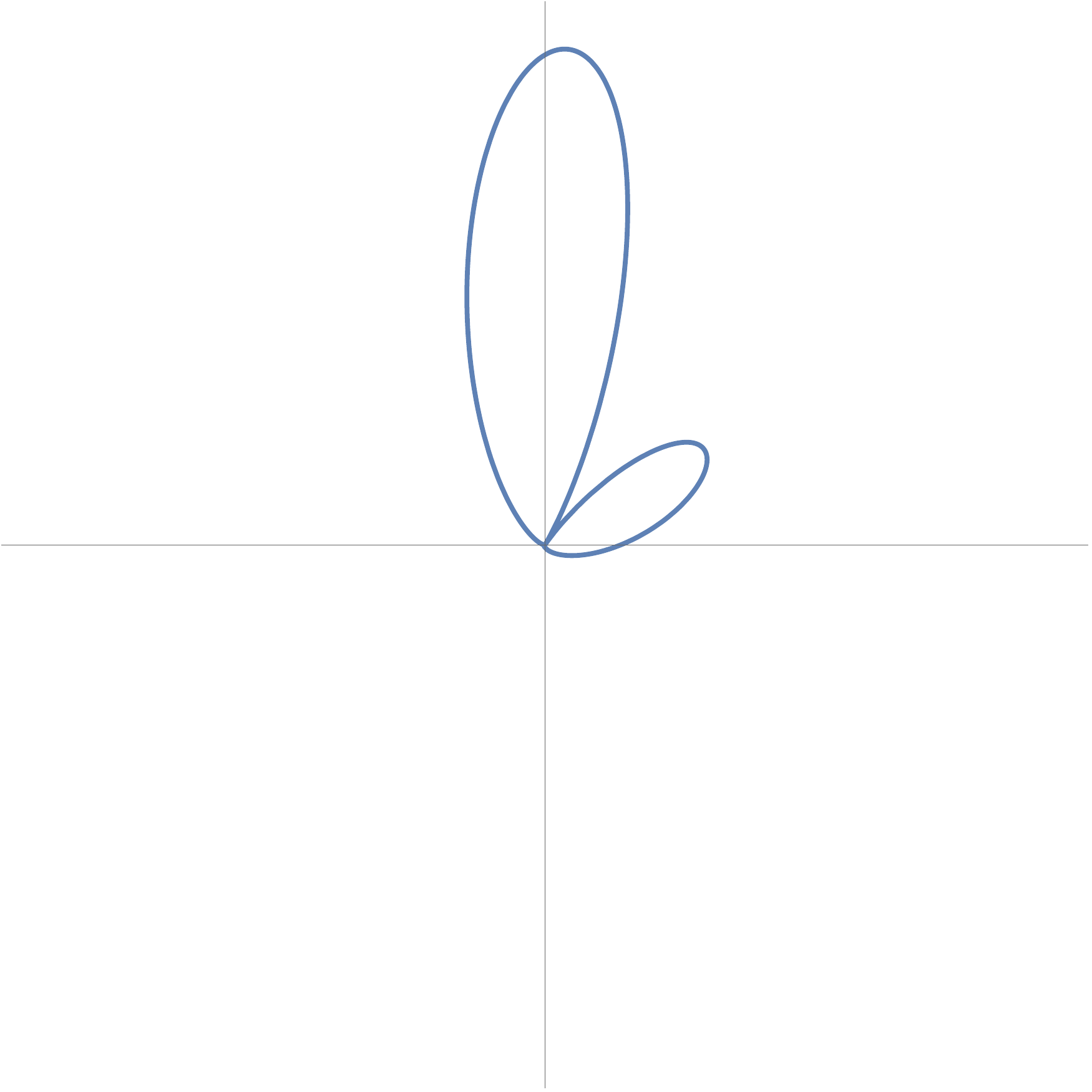}}\quad
\subfigure{\includegraphics[width=1.2in]{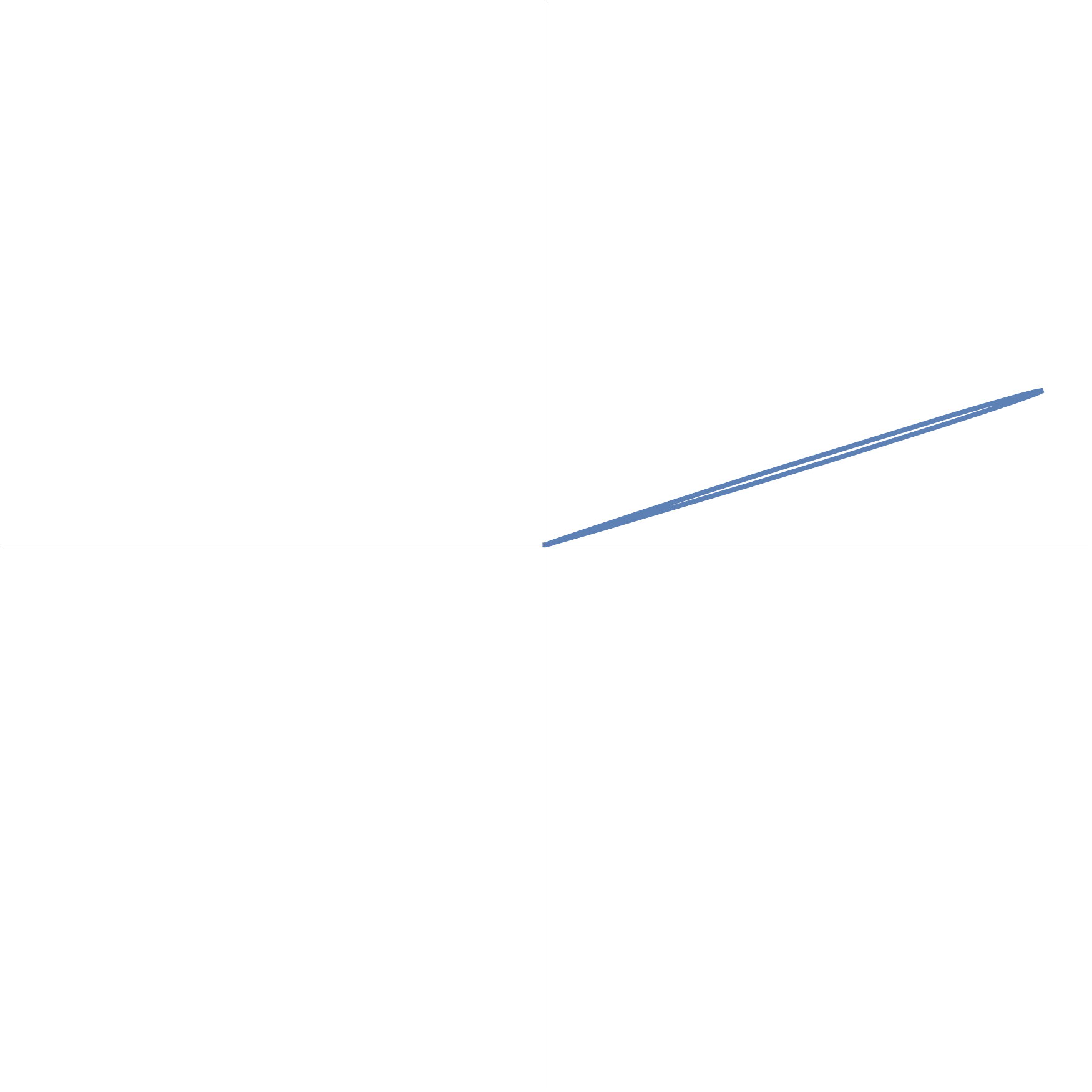} }}
\caption{\label{fig5a} Cusp Angular Distribution; A polar plot with speeds $\beta = 0.001, 0.333, 0.666, 0.999$. The vertical axes is $x$ and the horizontal axis is $z$. The electron moves in both dimensions.  Here there is a single parameter $\kappa = \tau$.  The limiting angle is zero despite appearances in the fourth frame, as the scaling for the maximum angle of radiation is slower than the other stationary worldlines.  That is, much higher speeds than $99.9\%$ are required to achieve beaming along the horizontal $z$ axis.     } 
\end{center}
\end{figure} 

\subsection*{Infrator}
\begin{figure}[ht]
\begin{center}
\mbox{
\subfigure{\includegraphics[width=1.3in]{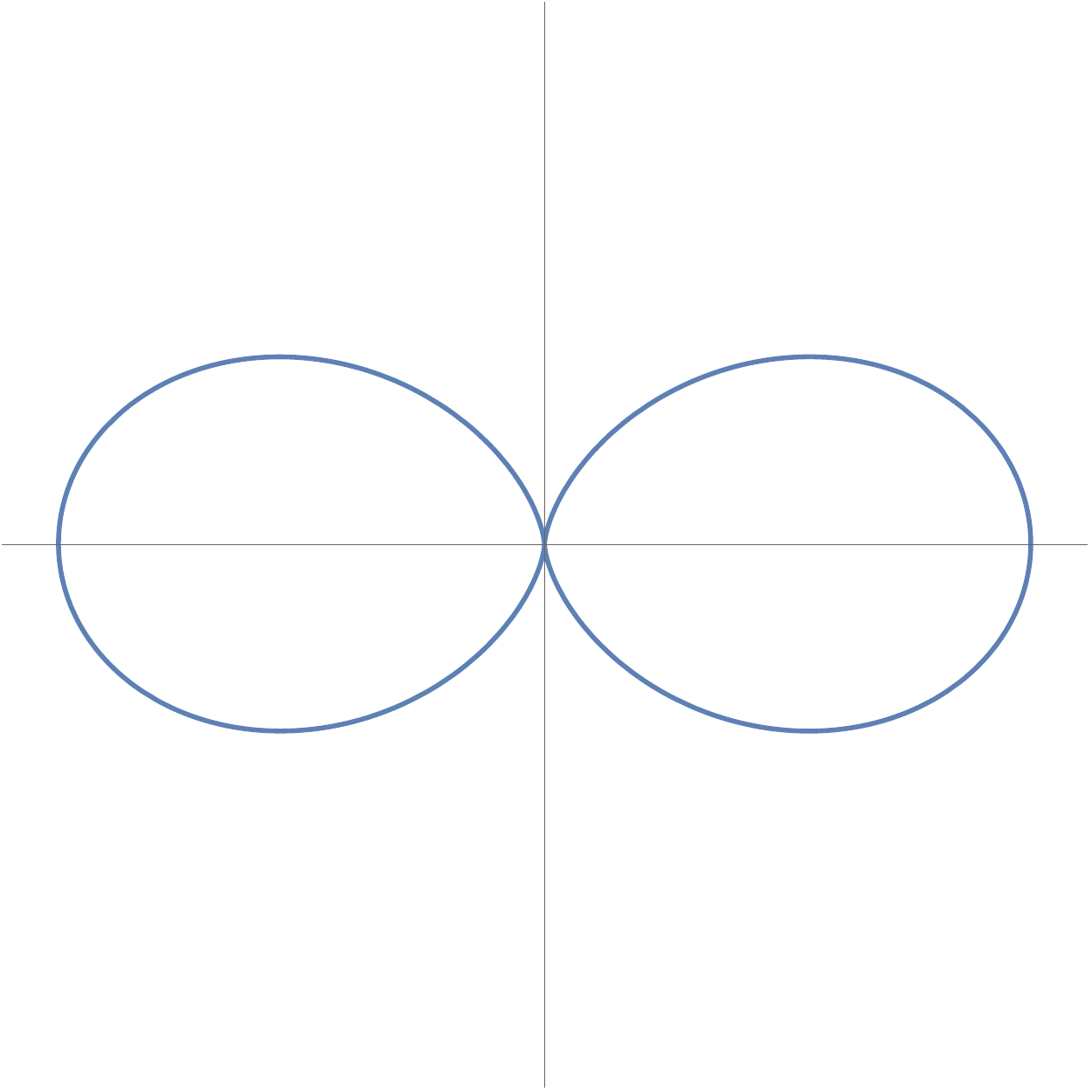}}\quad
\subfigure{\includegraphics[width=1.3in]{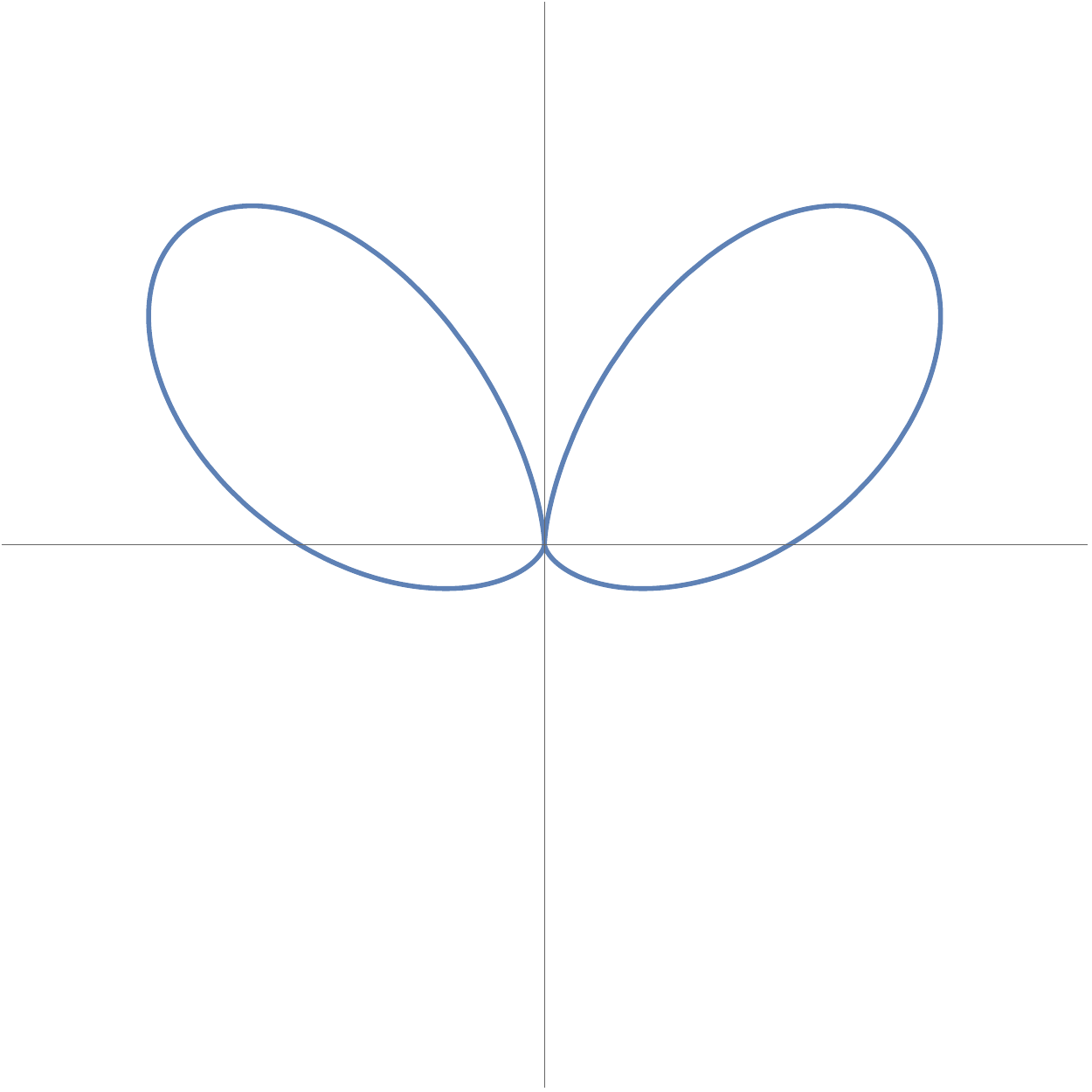}}\quad
\subfigure{\includegraphics[width=1.3in]{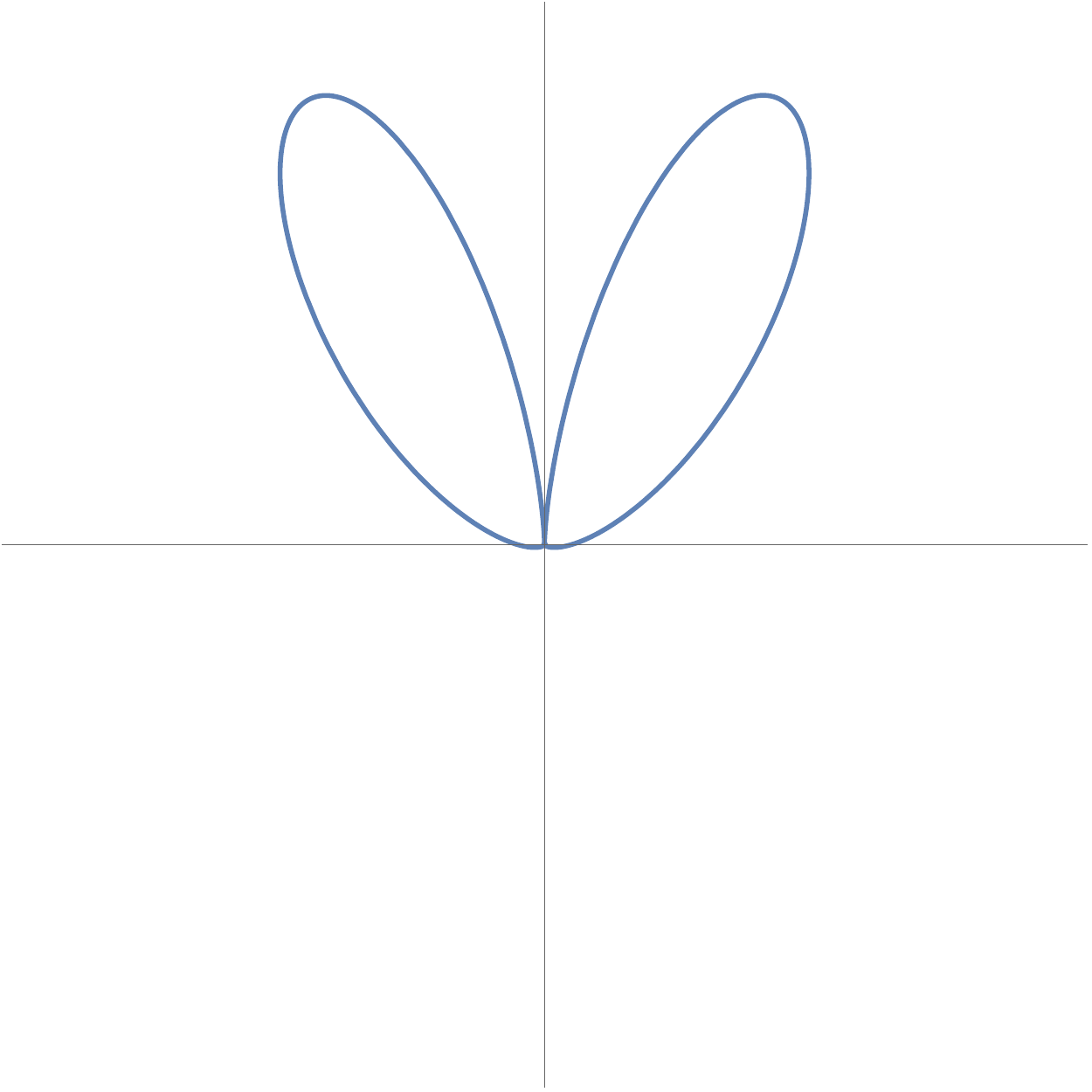}}\quad
\subfigure{\includegraphics[width=1.3in]{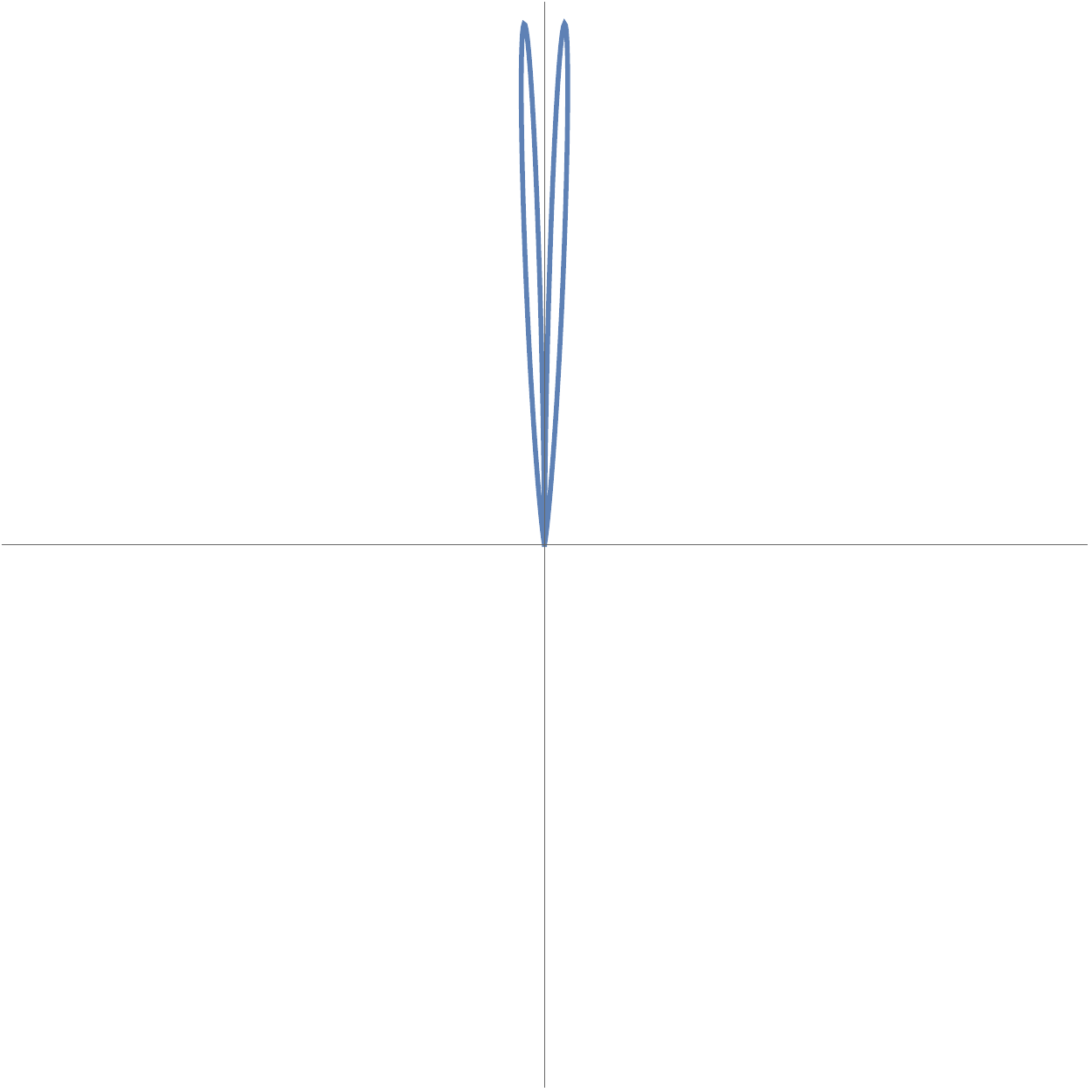} }}
\caption{\label{fig456} Catenary Angular Distribution; The catenary hangs as a chain would with the vertical axis $x$ and horizontal axis as $z$.  A polar plot with $v_R=0.0001$ and $\beta = 0.001, 0.333, 0.665, 0.997$. The electron moves in both dimensions but at high speeds moves mostly in the $x$-direction (thus the reason for the beaming in the $x$-direction).  Notice the symmetry to that of Nulltor, which is mostly coincidental due to the very low Rindler drift in Infrator's motion.  A very small drift means a very small torsion, similar to Nulltor, but Infrator is inherently a 2D motion (Nulltor is 1D), and so the beaming of Infrator is directed differently as can be seen in the following Figures.  } 
\end{center}
\end{figure} 

\begin{figure}[ht]
\begin{center}
\mbox{
\subfigure{\includegraphics[width=1.3in]{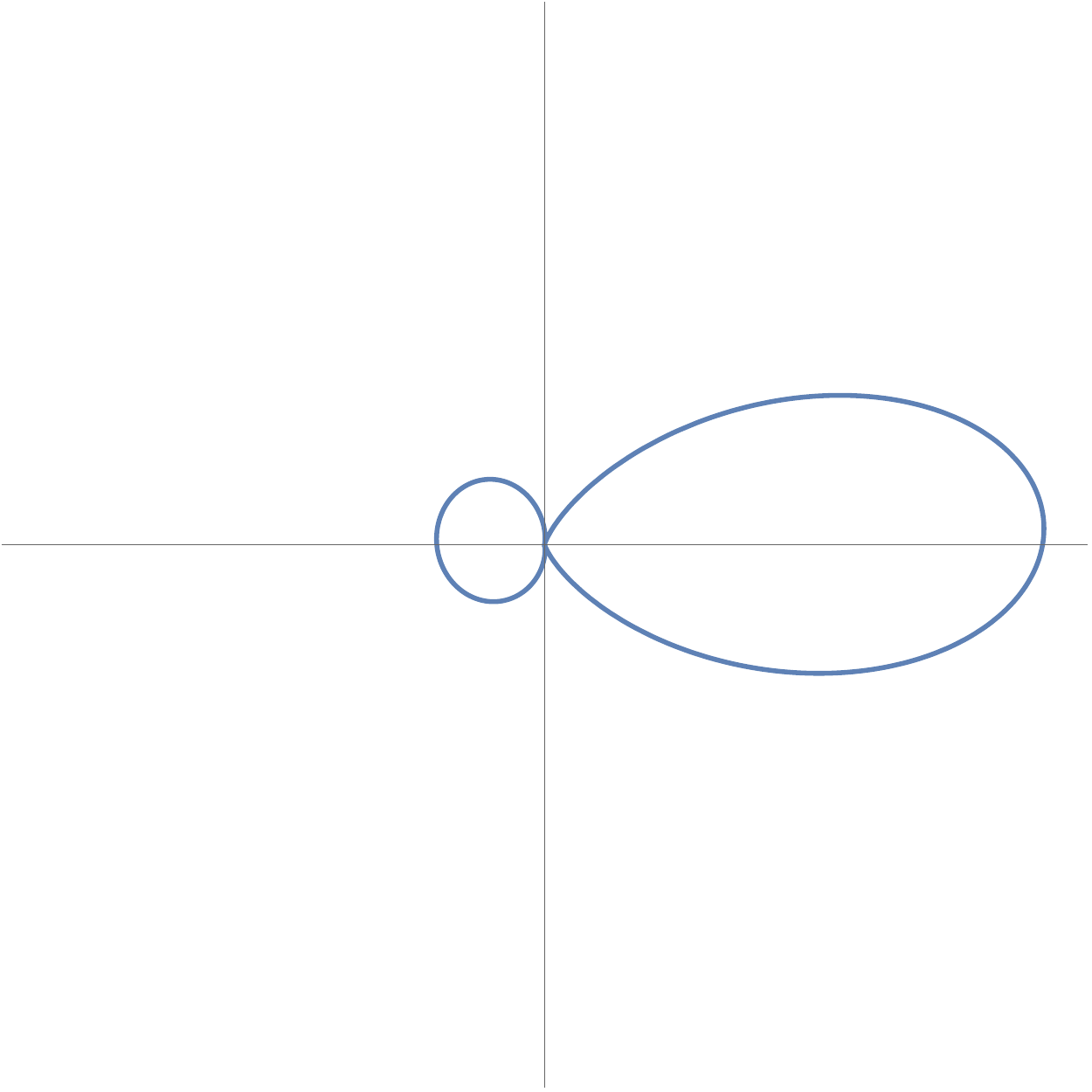}}\quad
\subfigure{\includegraphics[width=1.3in]{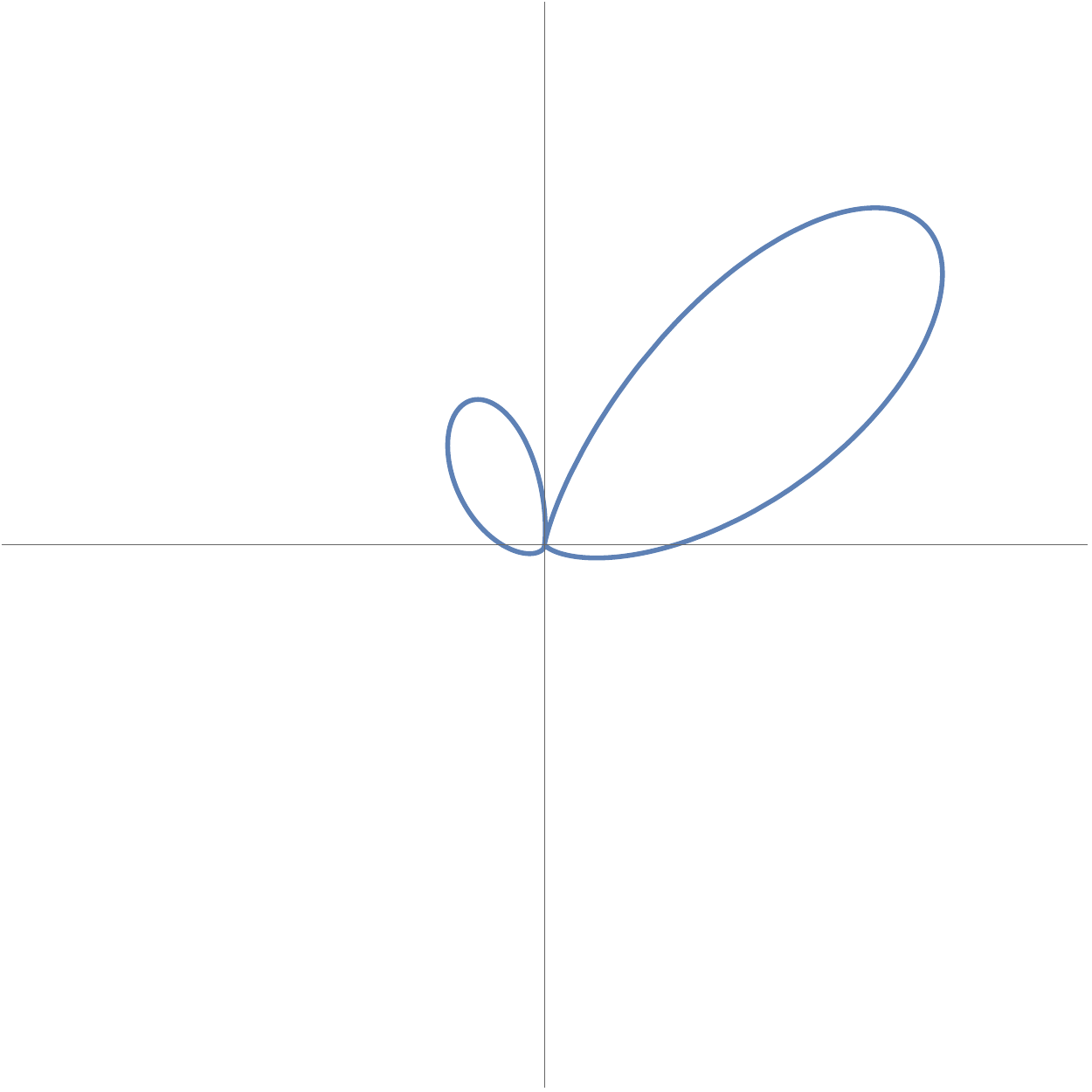}}\quad
\subfigure{\includegraphics[width=1.3in]{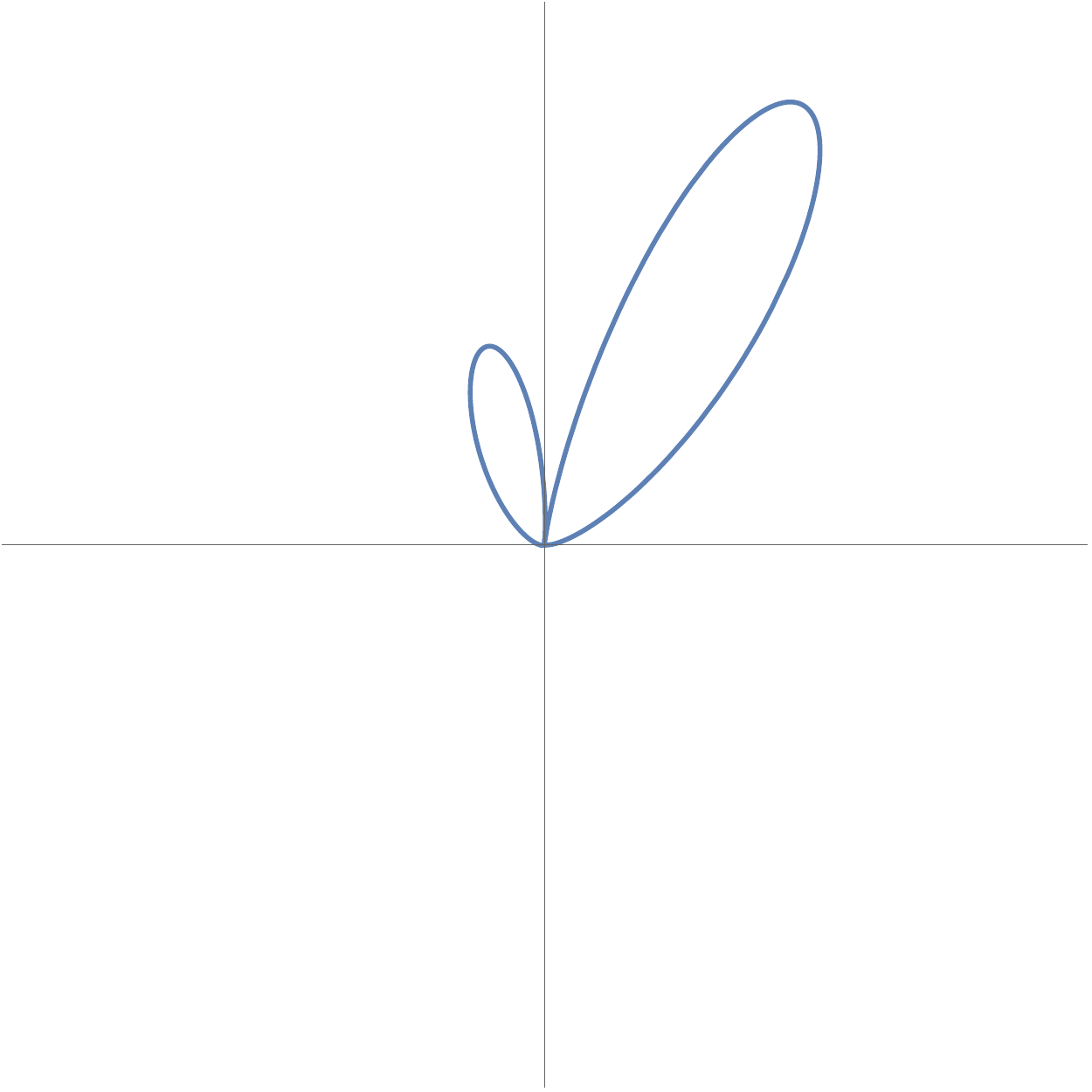}}\quad
\subfigure{\includegraphics[width=1.3in]{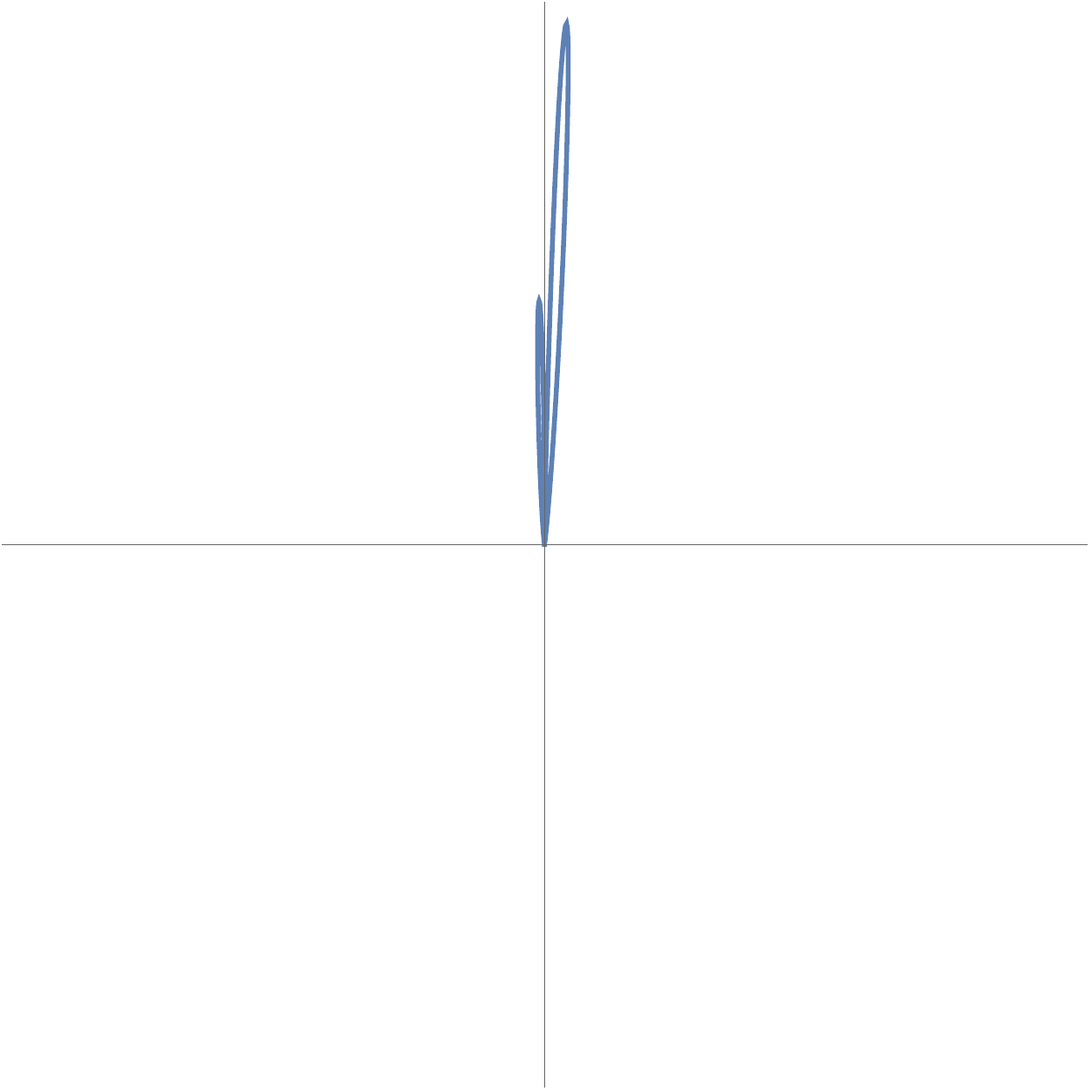} }}
\caption{\label{fig123} Catenary Angular Distribution; Additional torsion is added with substantial Rindler drift: $v_R=0.25$.  Here $\beta = 0.251, 0.500, 0.749, 0.998$ for the four plots respectively.  } 
\end{center}
\end{figure} 

\begin{figure}[ht]
\begin{center}
\mbox{
\subfigure{\includegraphics[width=1.3in]{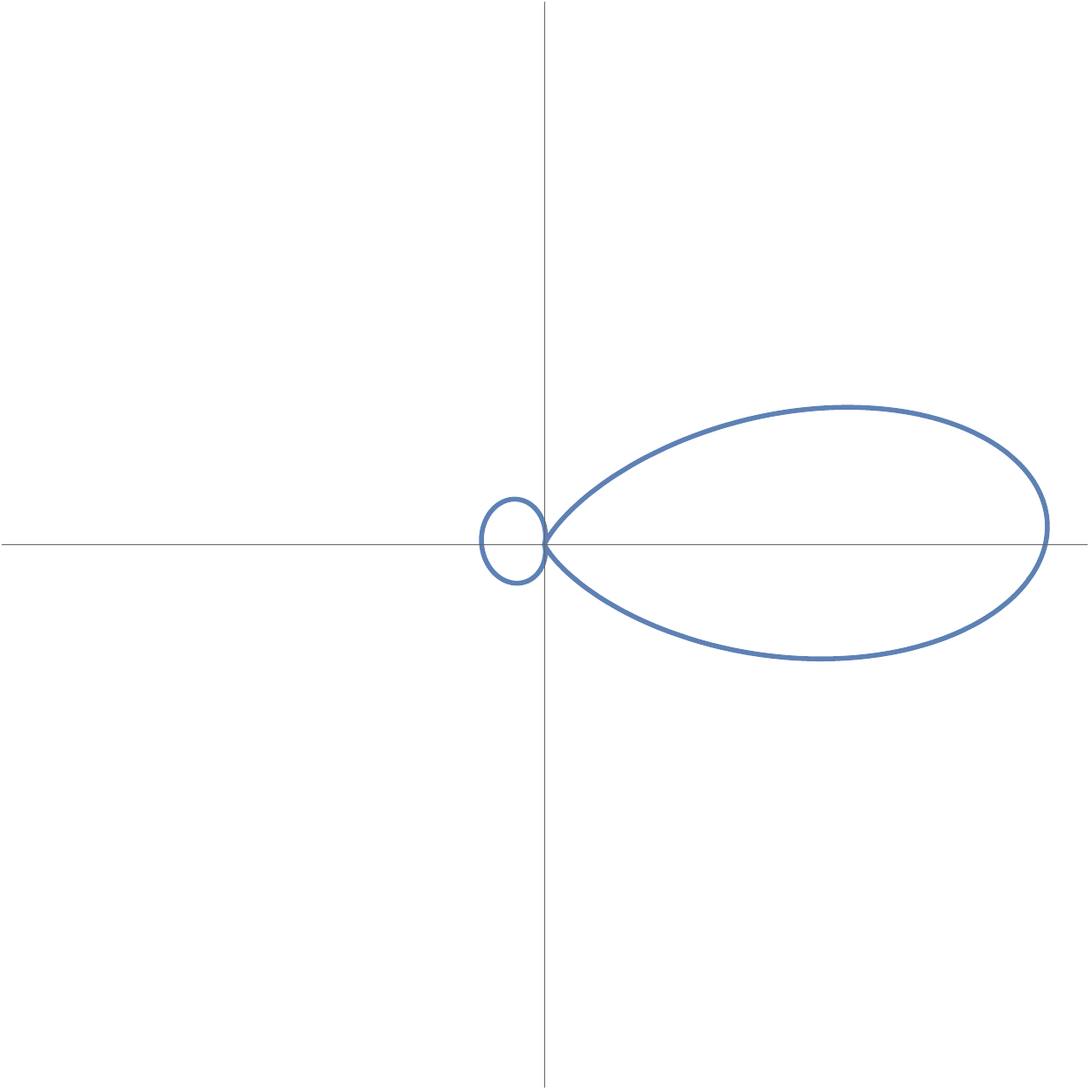}}\quad
\subfigure{\includegraphics[width=1.3in]{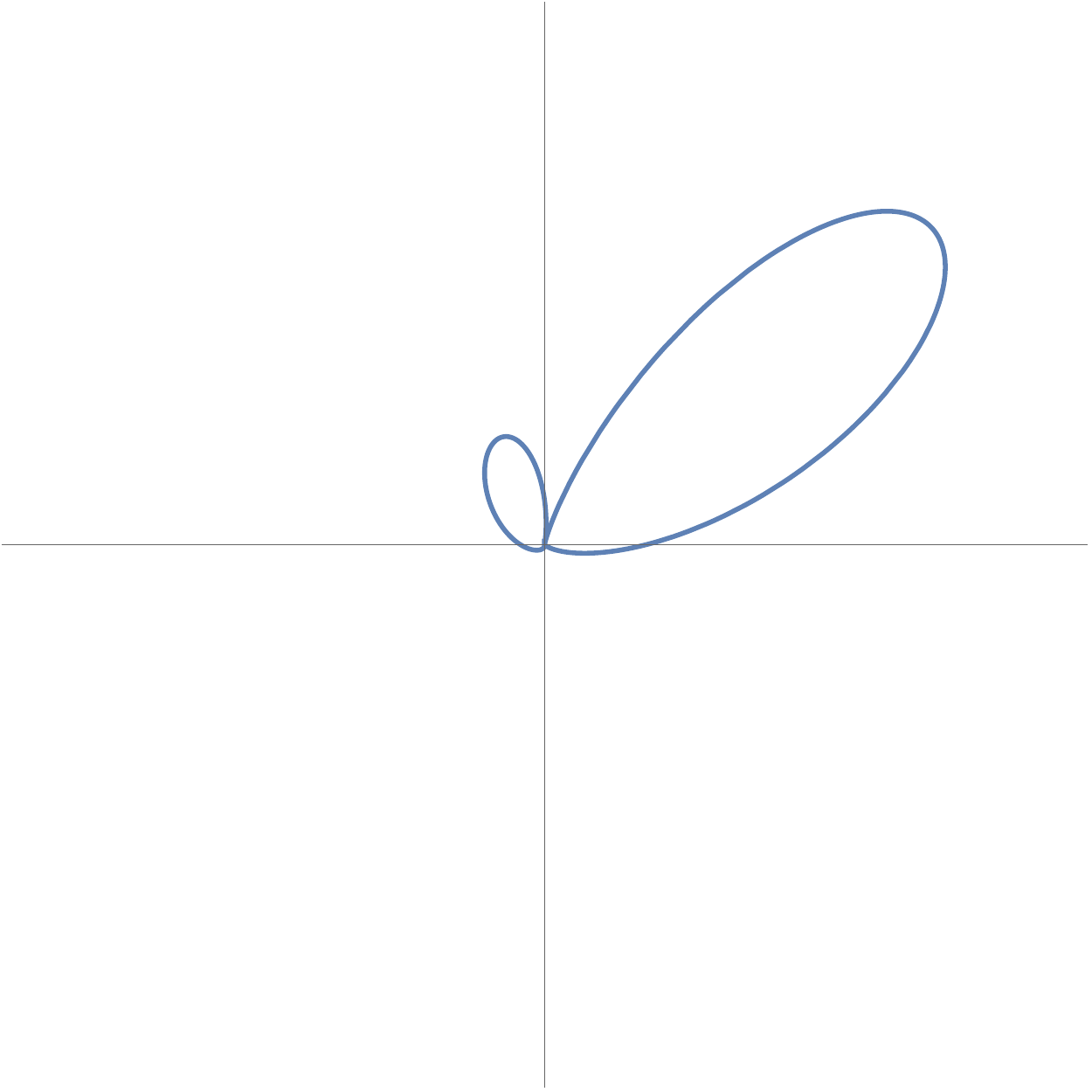}}\quad
\subfigure{\includegraphics[width=1.3in]{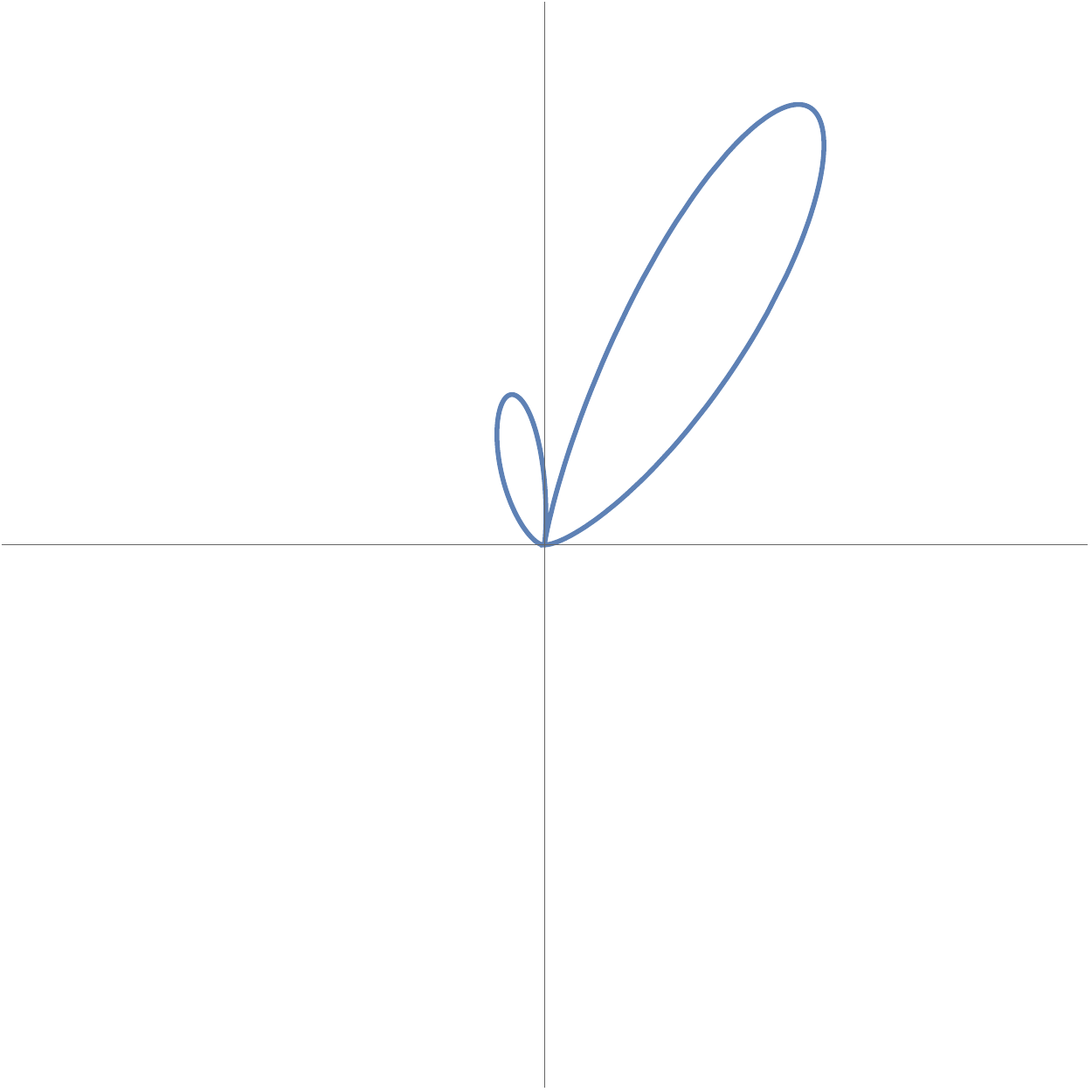}}\quad
\subfigure{\includegraphics[width=1.3in]{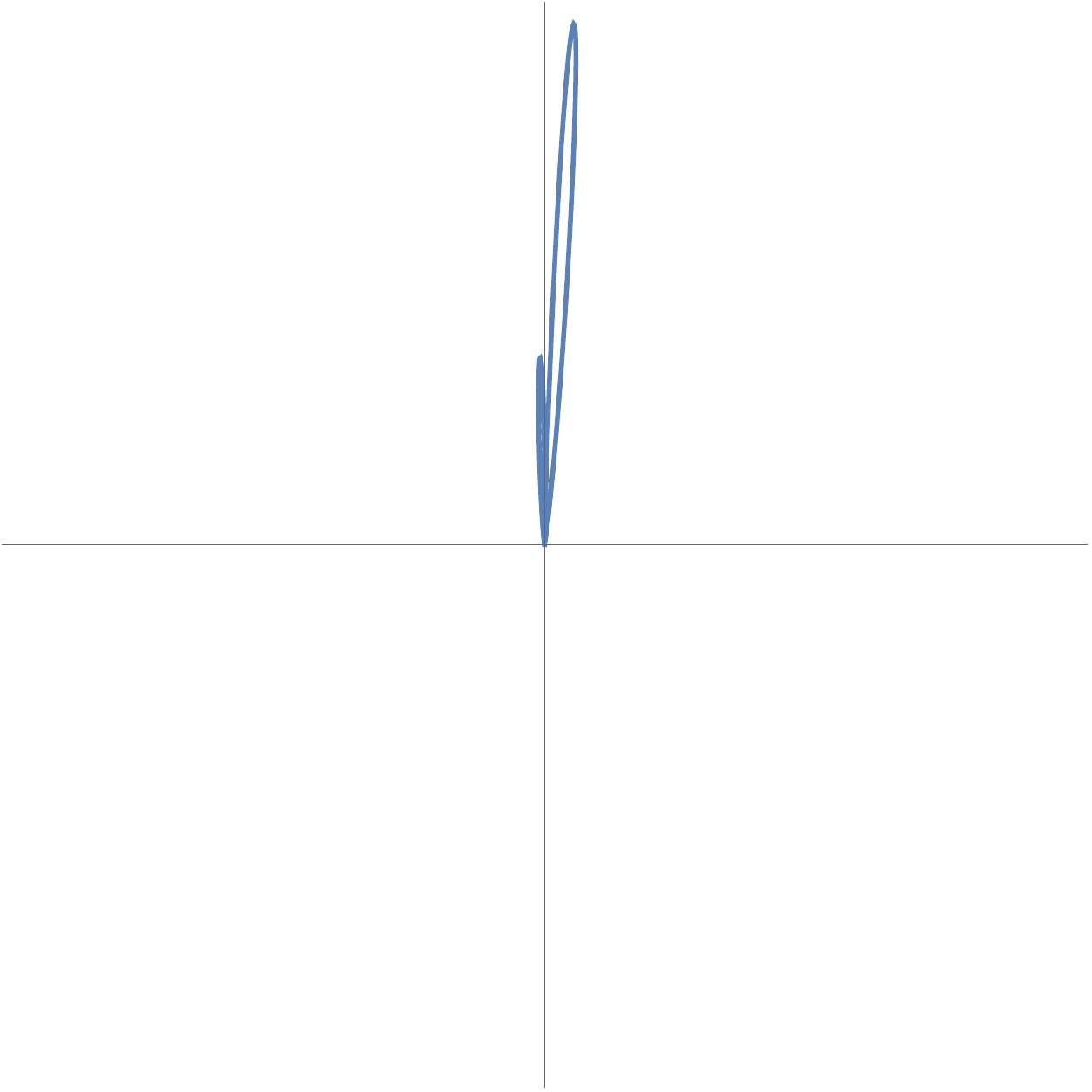} }}
\caption{\label{figabc} Catenary Angular Distribution; Even more Rindler drift requires an initial starting speed where $\beta > v_R$.  Here $v_R=0.333$, and $\beta = 0.334, 0.555, 0.776, 0.997$.  } 
\end{center}
\end{figure}

\newpage




\subsection*{Nulltor}
\begin{figure}[ht]
\begin{center}
\mbox{
\subfigure{\includegraphics[width=1.2in]{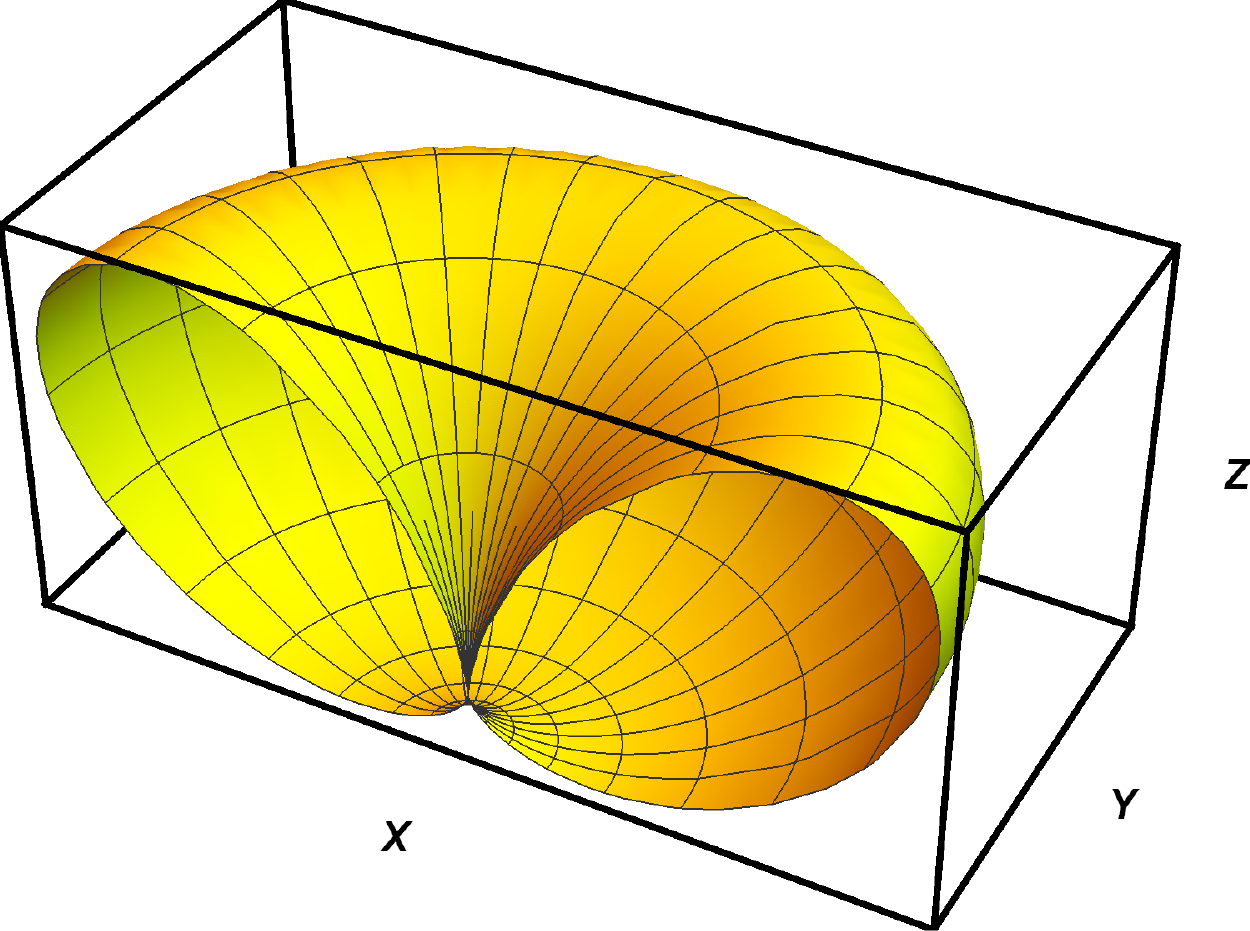}}\quad
\subfigure{\includegraphics[width=1.2in]{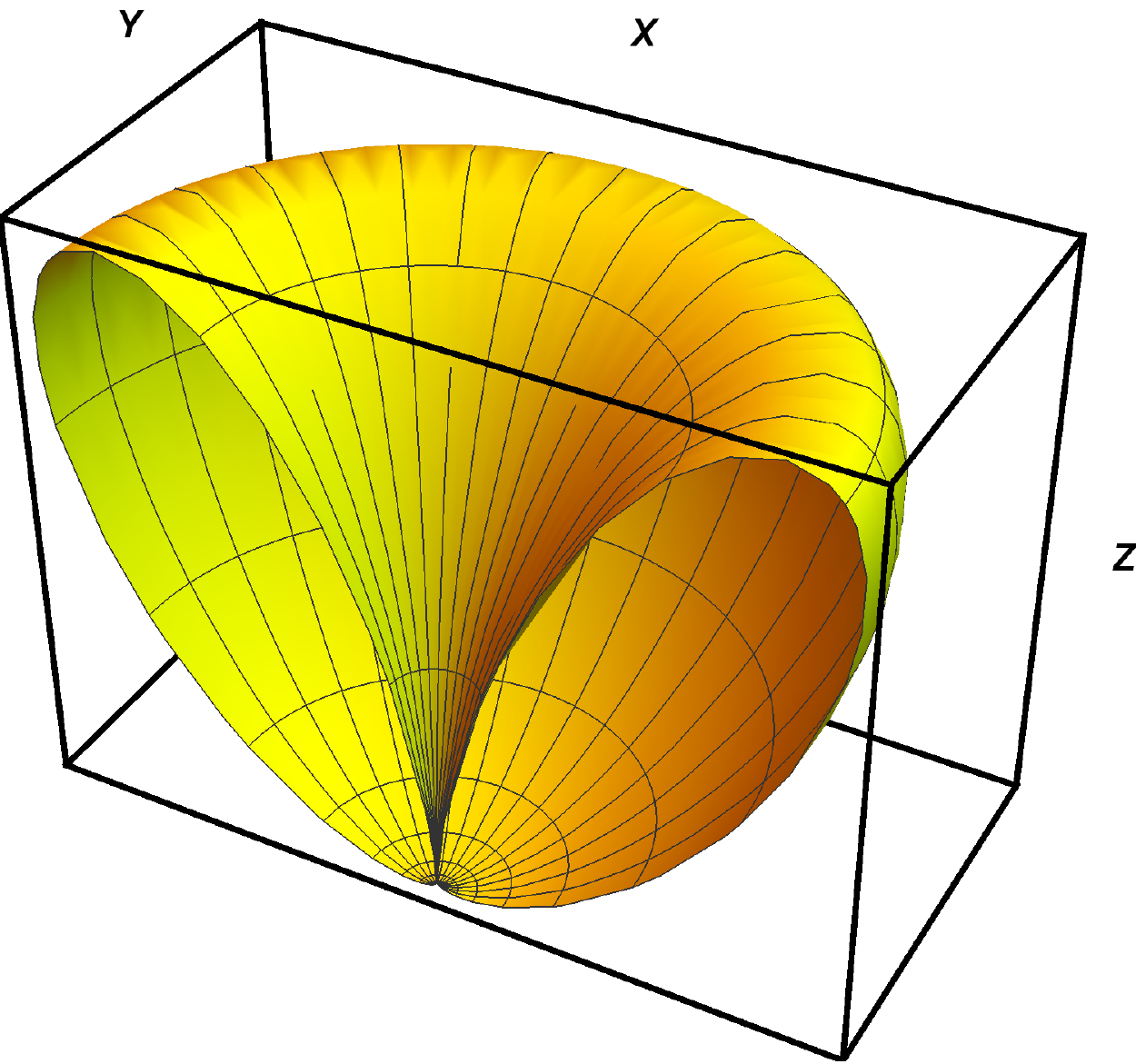}}\quad
\subfigure{\includegraphics[width=1.2in]{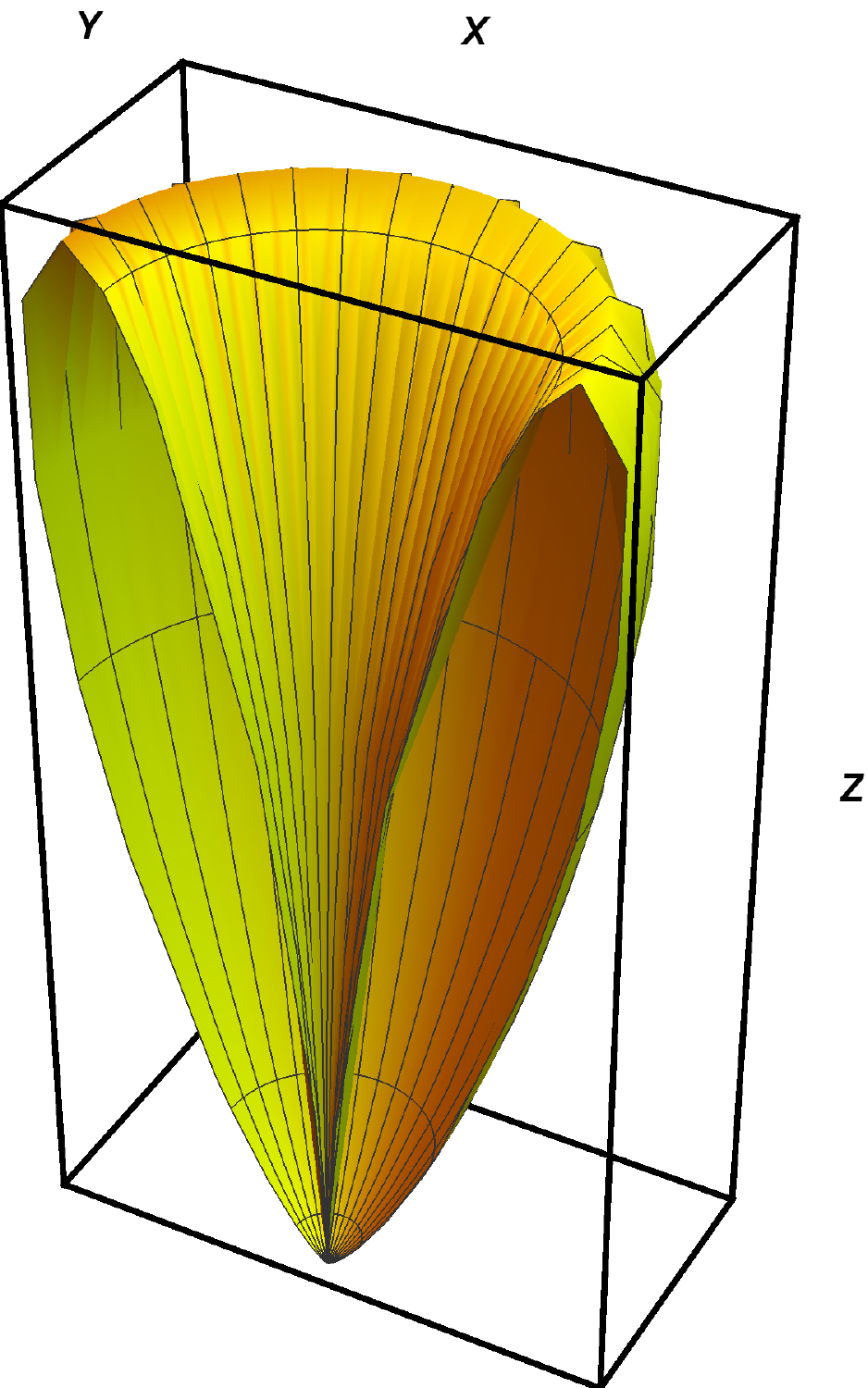} }}
\caption{\label{fig3a}3-Dimensional Rectilinear Angular Distribution also known as "Bremsstrahlung radiation"; A polar plot with different velocities for the particle, $\beta =  0.333, 0.666, 0.999$.  The electron moves forward along the $+z$  direction. The plot range is $\theta=[0,\pi]$ and $\phi=[0,\pi]$. }
\end{center}
\end{figure}  
\newpage
\subsection*{Ultrator}
\begin{figure}[ht]
\begin{center}
\mbox{
\subfigure{\includegraphics[width=1.0in]{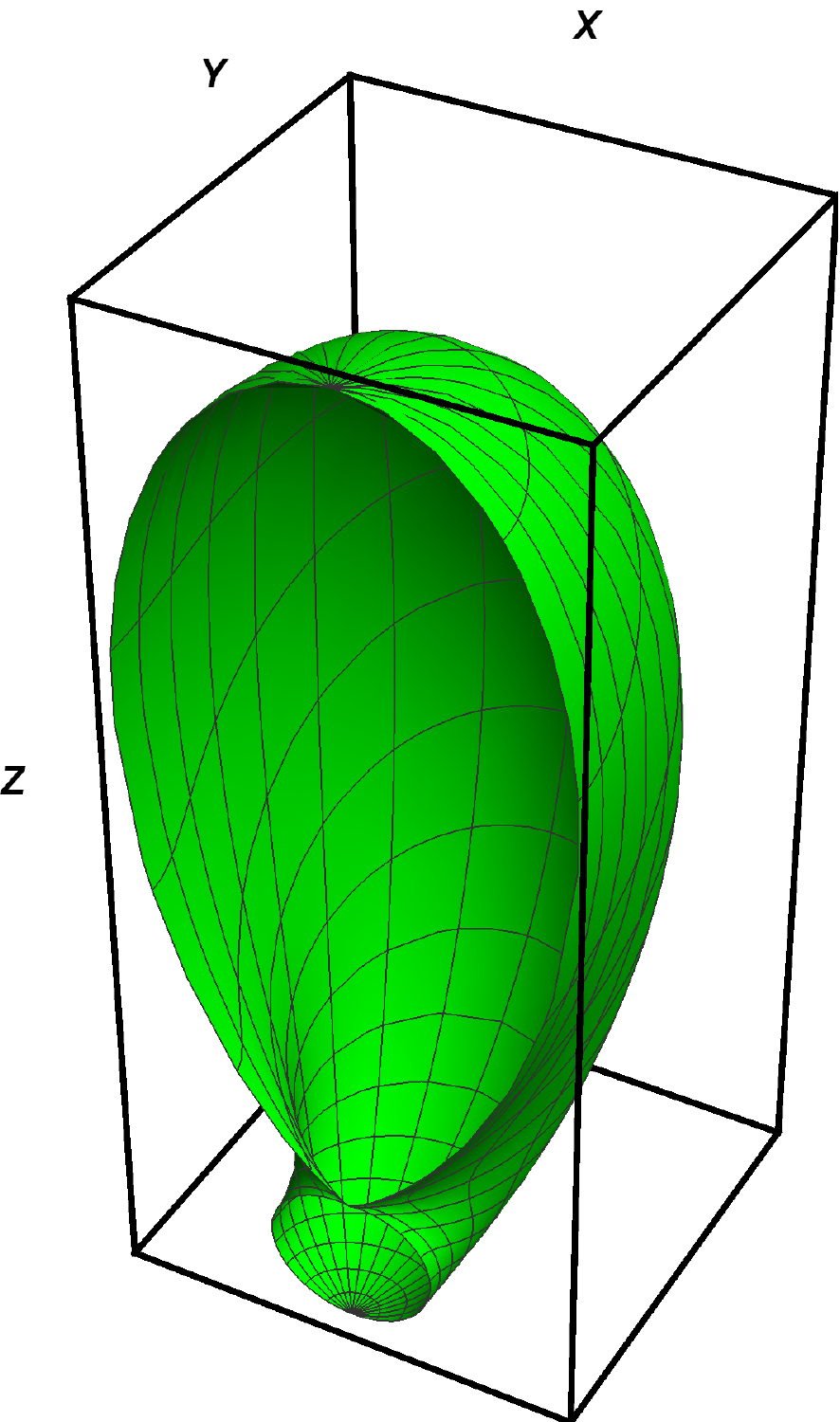}}\quad
\subfigure{\includegraphics[width=1.0in]{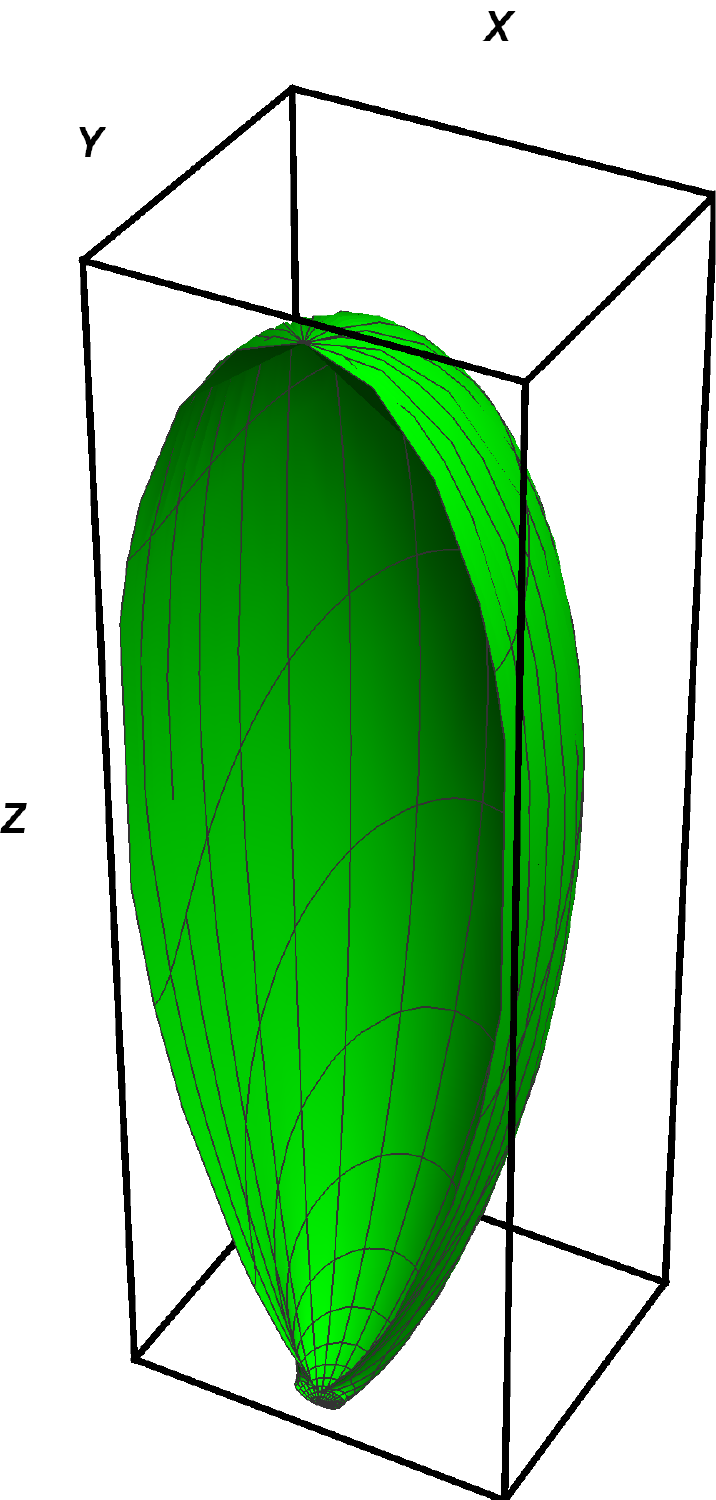}}\quad
\subfigure{\includegraphics[width=1.0in]{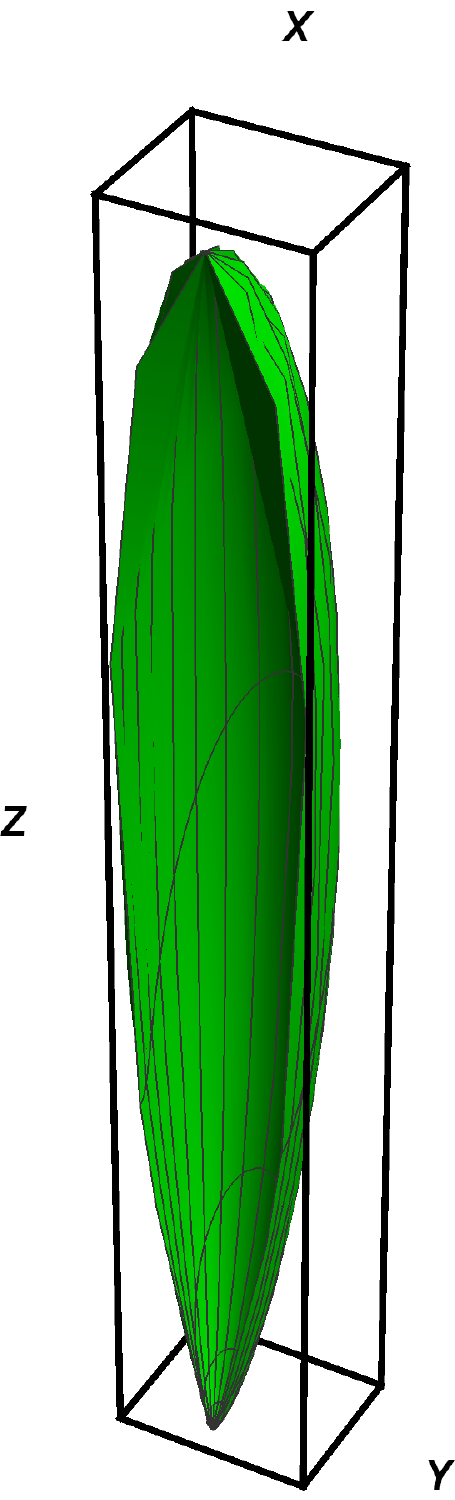}}}
\caption{\label{fig4} Synchrotron Angular Distribution in 3D; A spherical 3D plot.  In the first figure, the particle is moving with  $33\%$ speed of light, in the second and third figures, we have $66\%$ and $99\%$ speed of light respectively. The electron moves in the $z$ direction, but around the $x$ axis in circular motion.  The plot range is $\theta=[0,\pi]$ and $\phi=[0,\pi]$.} 
\end{center}
\end{figure} 
\newpage
\subsection*{Parator}
\begin{figure}[ht]
\begin{center}
\mbox{
\subfigure{\includegraphics[width=1.4in]{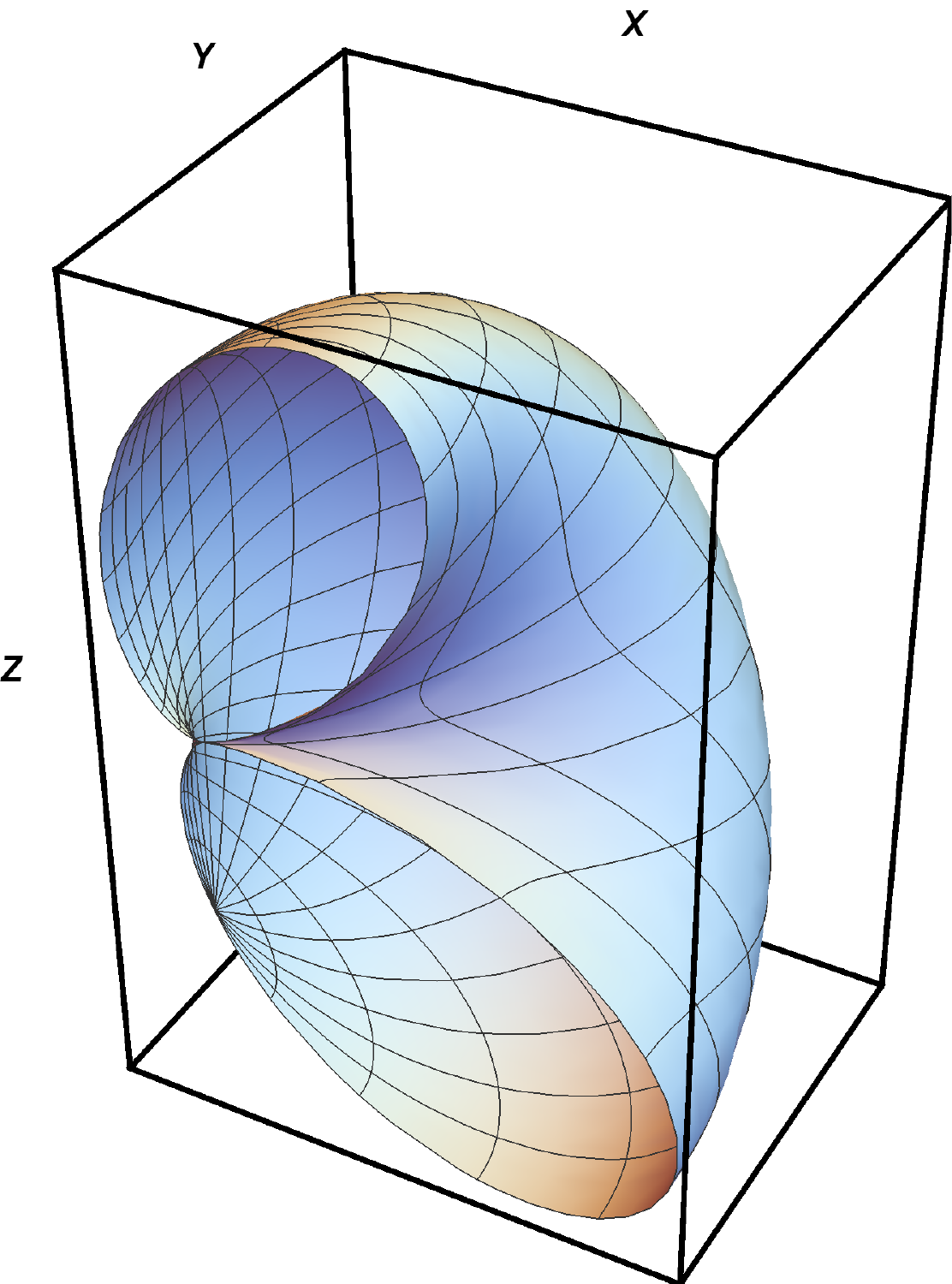}}\quad
\subfigure{\includegraphics[width=1.4in]{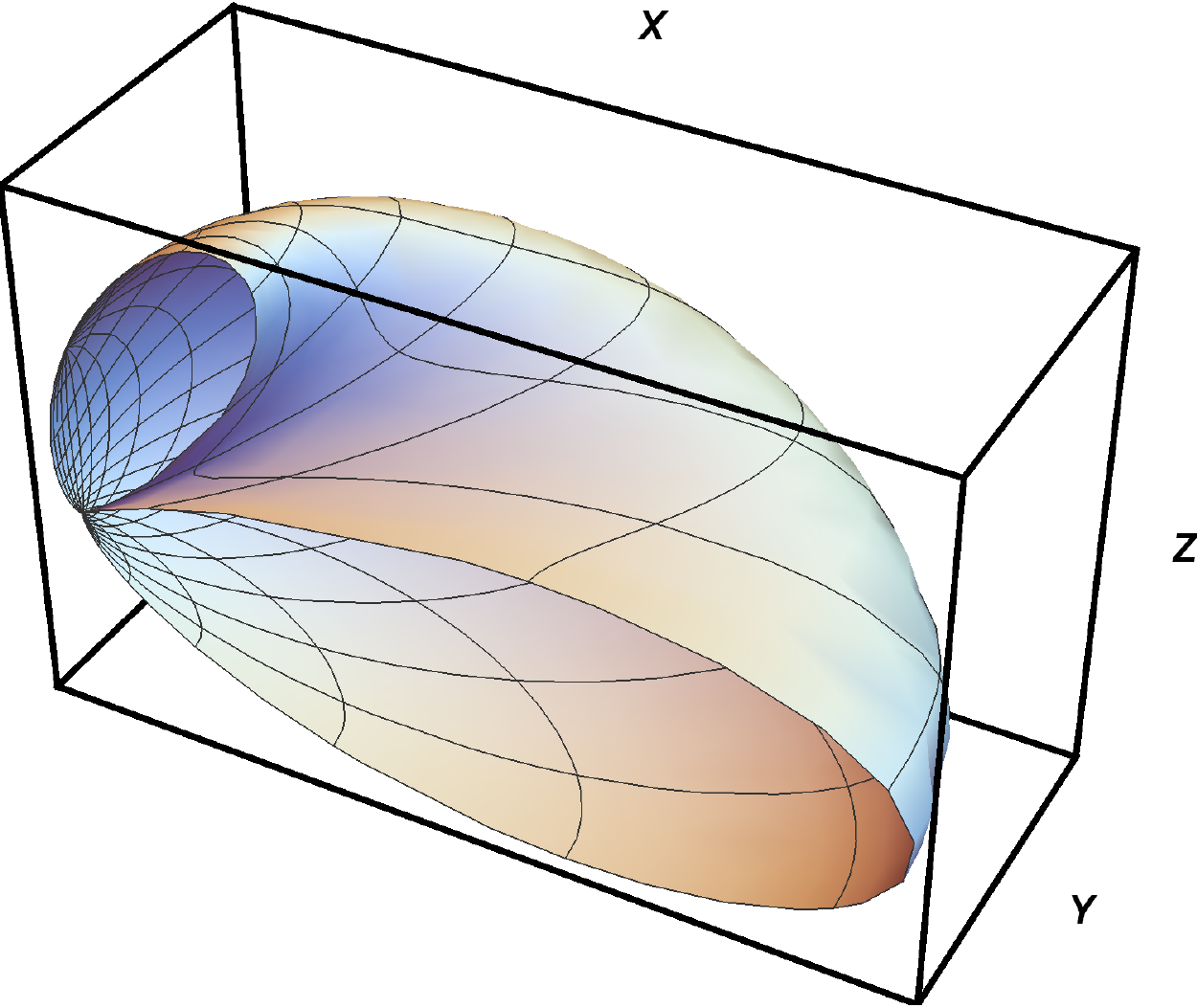}}\quad
\subfigure{\includegraphics[width=1.4in]{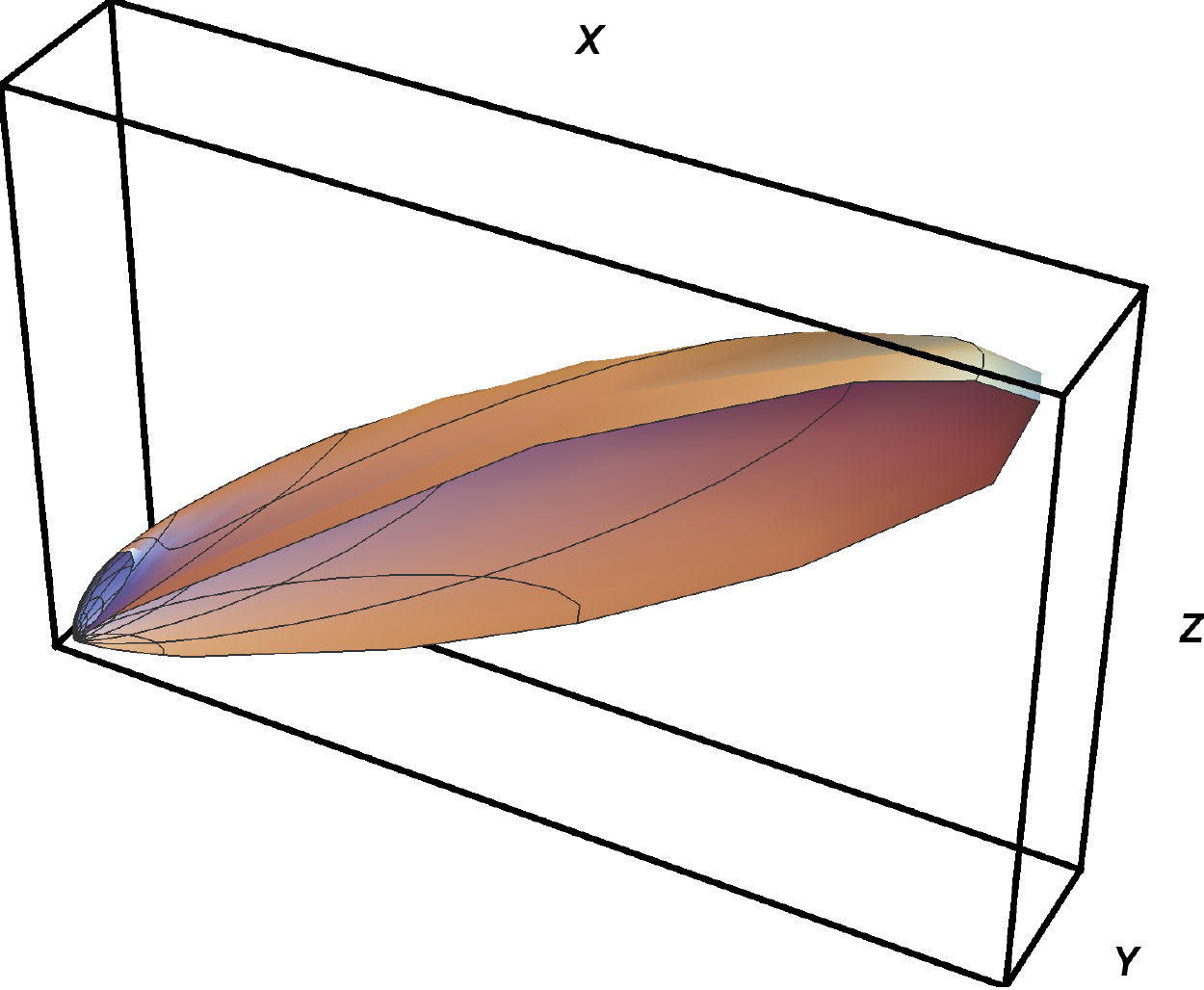} }}
\caption{\label{fig5} Cusp Angular Distribution in 3D pace; A polar plot with speeds $\beta =  0.333, 0.666, 0.999$. The electron moves between $x$ and $z$ dimensions.  The plot range is $\theta=[0,\pi]$ and $\phi=[0,\pi]$. Here curvature and torsion are equal,  $\kappa = \tau$. As it was mentioned before for Parator motion the limiting angle is zero despite appearances in the third frame, as the scaling for the maximum angle of radiation is slower than the other stationary worldlines.}
\end{center}
\end{figure}
\subsection*{Infrator}
\begin{figure}[ht]
\begin{center}
\mbox{
\subfigure{\includegraphics[width=1.4in]{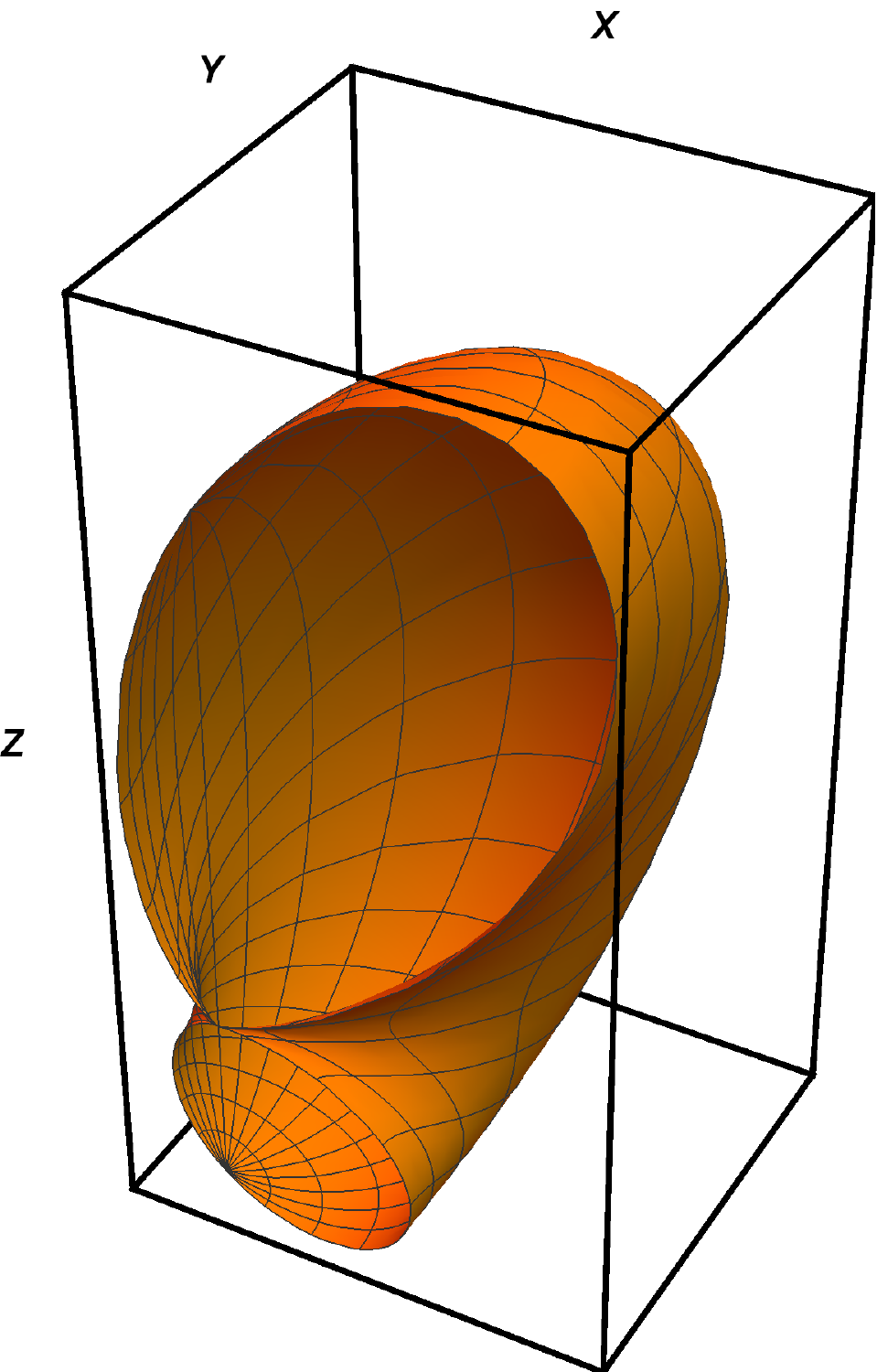}}\quad
\subfigure{\includegraphics[width=1.4in]{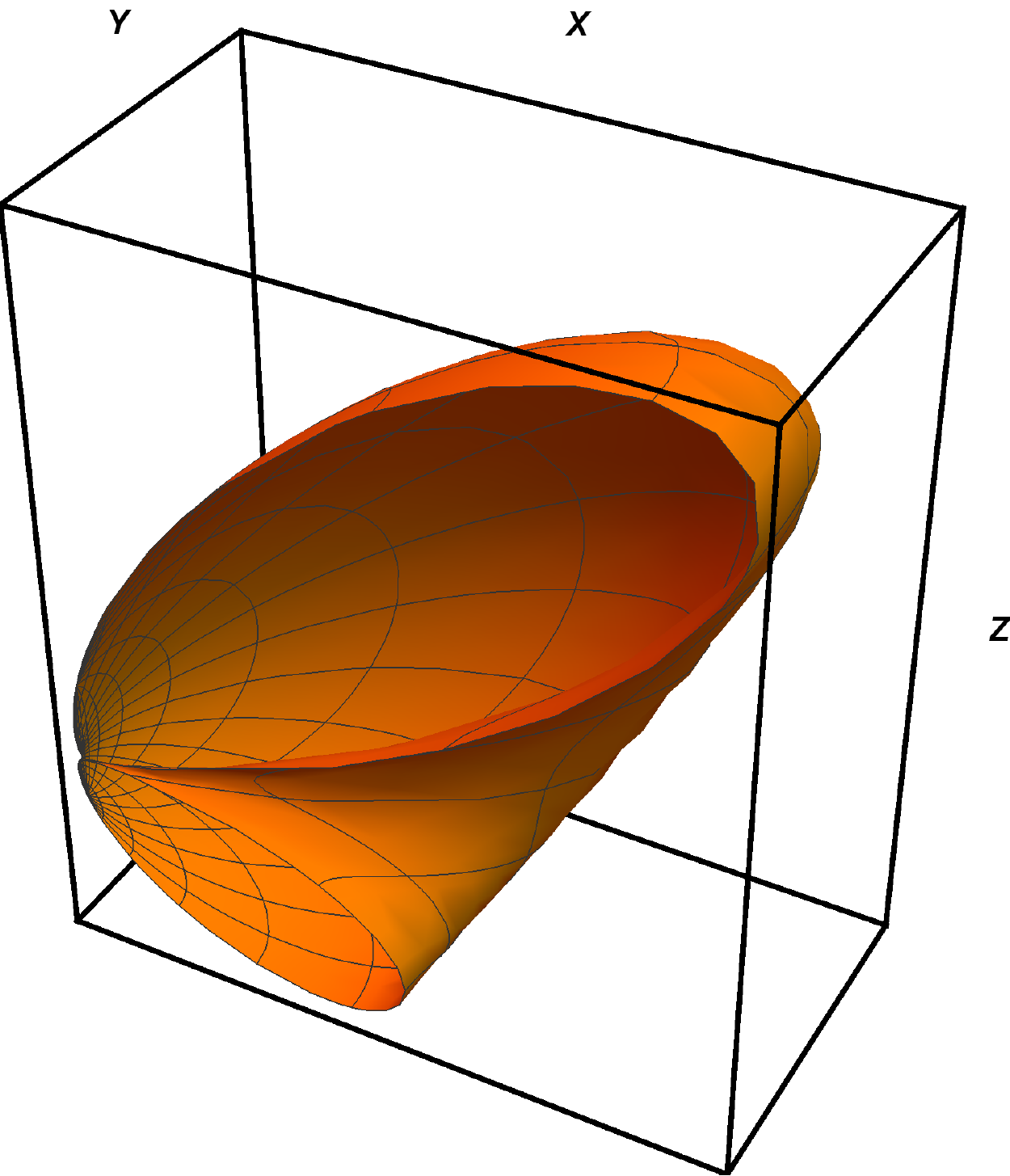}}\quad
\subfigure{\includegraphics[width=1.4in]{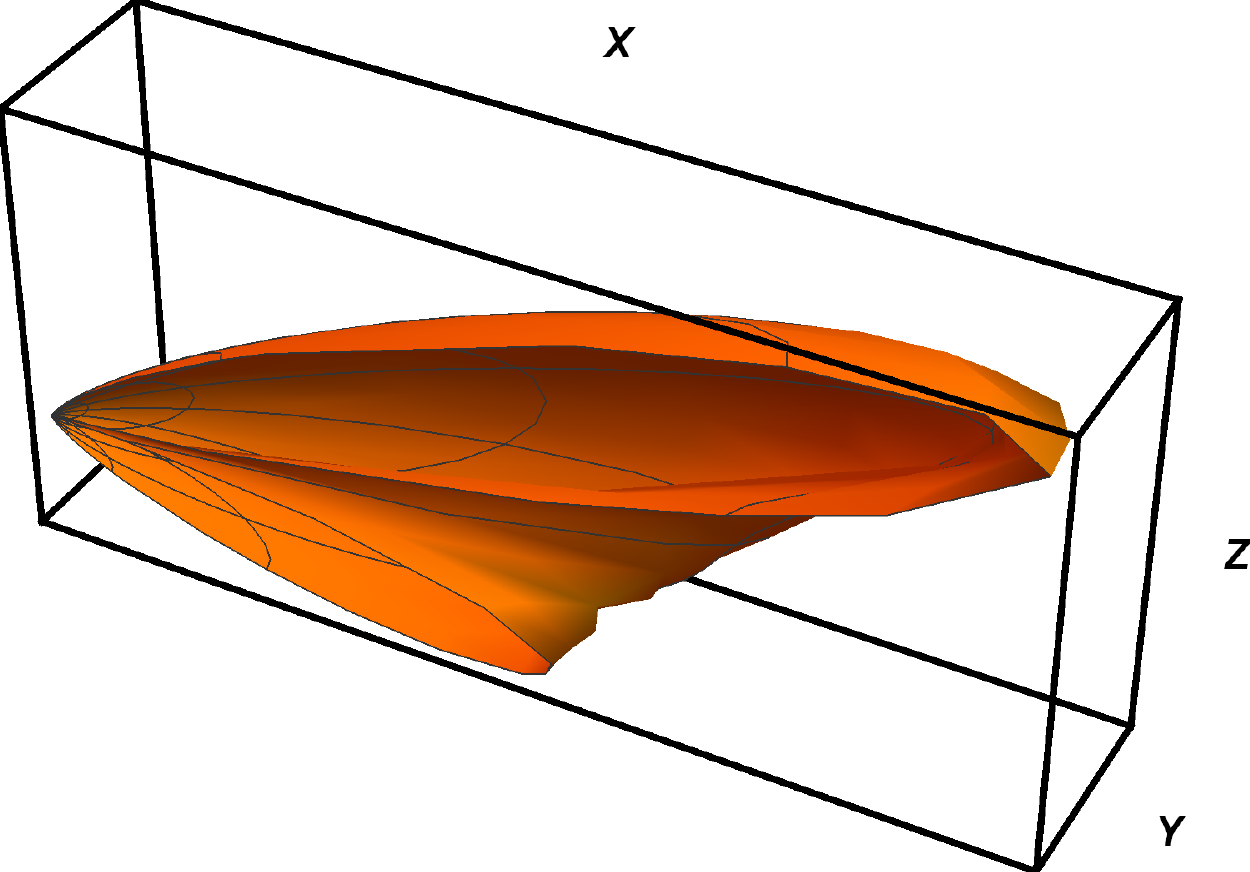} }}
\caption{\label{figdef} Catenary Angular Distribution in 3D space; Additional torsion is added with substantial Rindler drift: $v_R=0.2$.  Here $\beta = 0.332, 0.664, 0.996$ . The direction of the electron appears as a resulting vector from components along $x$ and $z$ dimensions.  The plot range is $\theta=[0,\pi]$ and $\phi=[0,\pi]$.} 
\end{center}
\end{figure} 

\newpage
\subsection*{Hypertor}
\begin{figure}[ht]
\begin{center}
\mbox{
\subfigure{\includegraphics[width=1.4in]{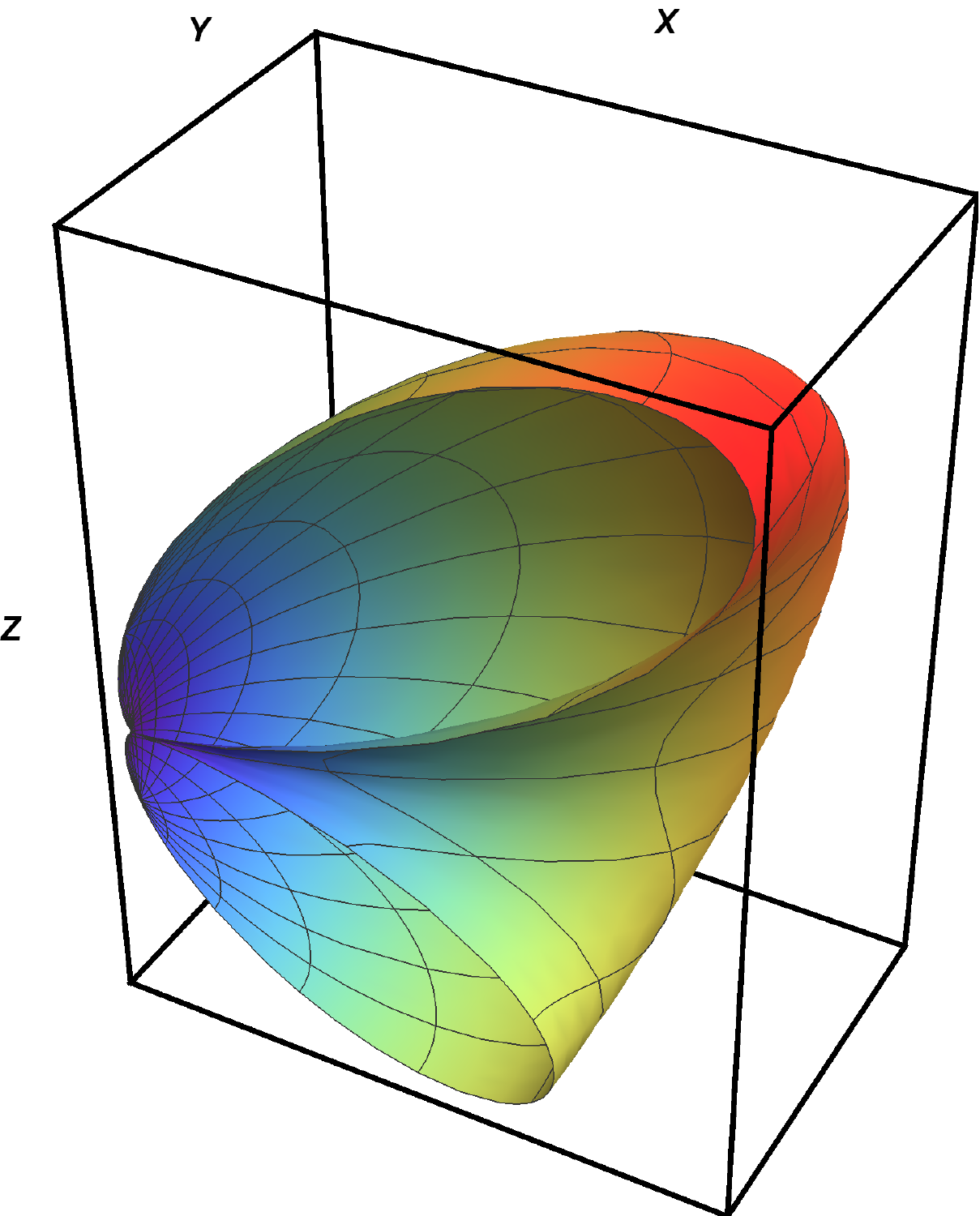}}\quad
\subfigure{\includegraphics[width=1.4in]{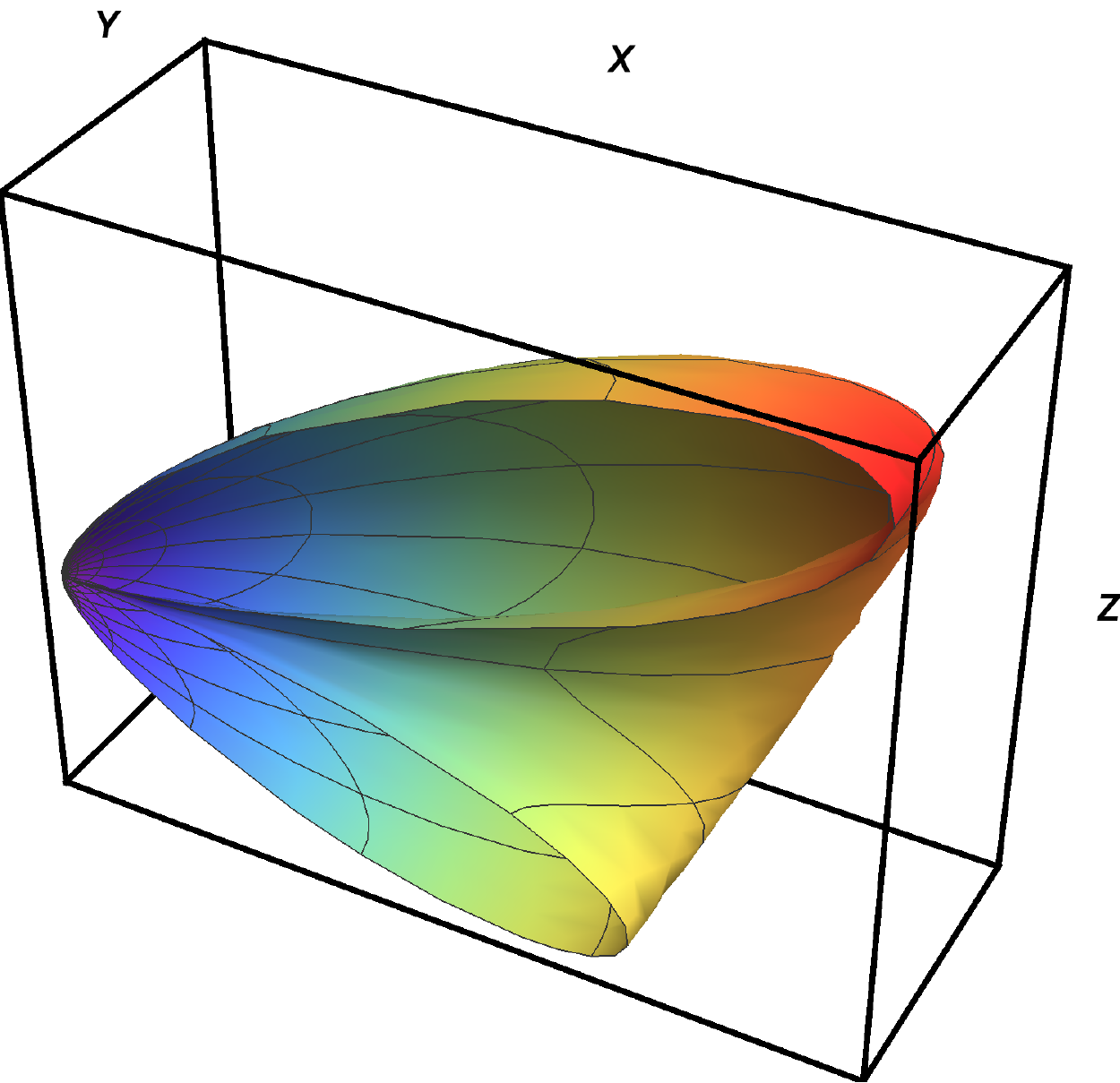}}\quad
\subfigure{\includegraphics[width=1.4in]{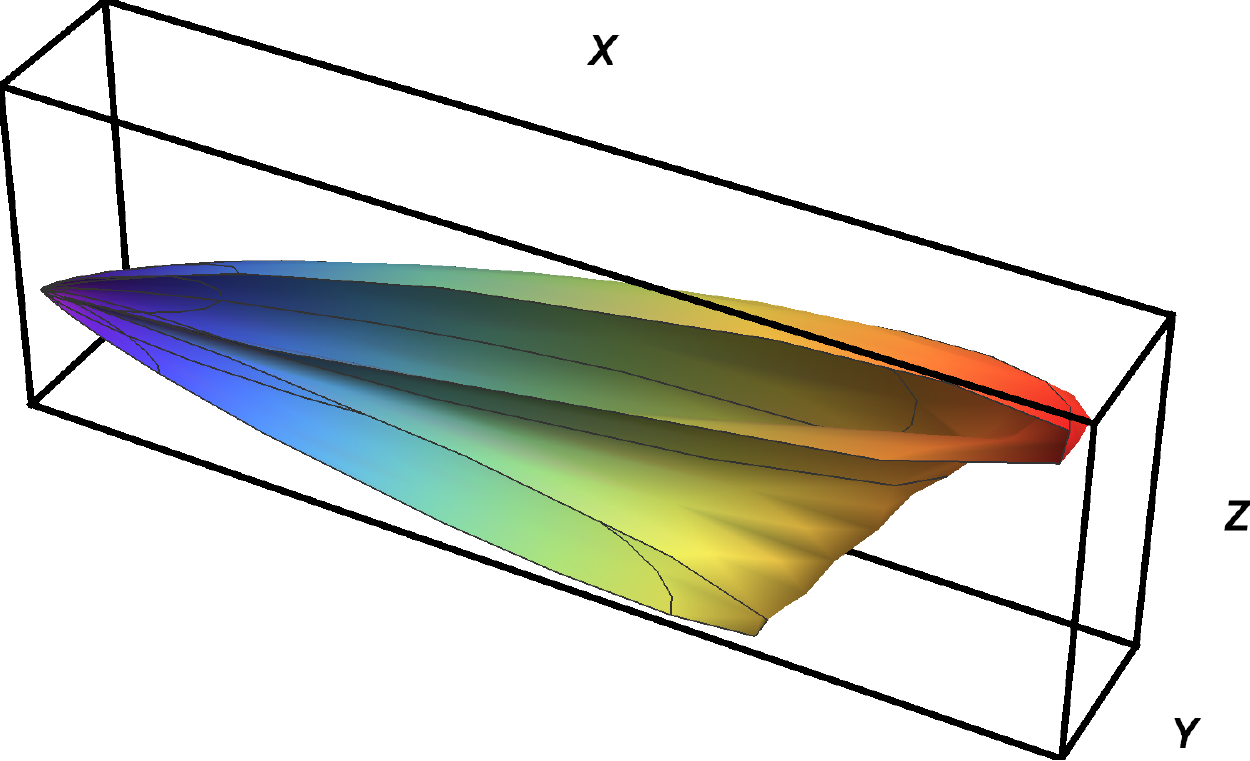} }}
\caption{\label{figh11} Hypertor Angular Distribution in 3D space; For the case when curvature invariants are  $\kappa = 1$, $\tau=\nu=0.1$ the radiation shape is displayed above. A polar plots with different velocities $\beta = 0.333, 0.665, 0.998$. In Hypertor motion the particle moves along a helix with a variable pitch trajectory and the radiation will be emitted in the prescribed directions. The plot range is $\theta=[0,\pi]$ and $\phi=[0,\pi]$.} 
\end{center}
\end{figure} 

\begin{figure}[ht]
\begin{center}
\mbox{
\subfigure{\includegraphics[width=1.4in]{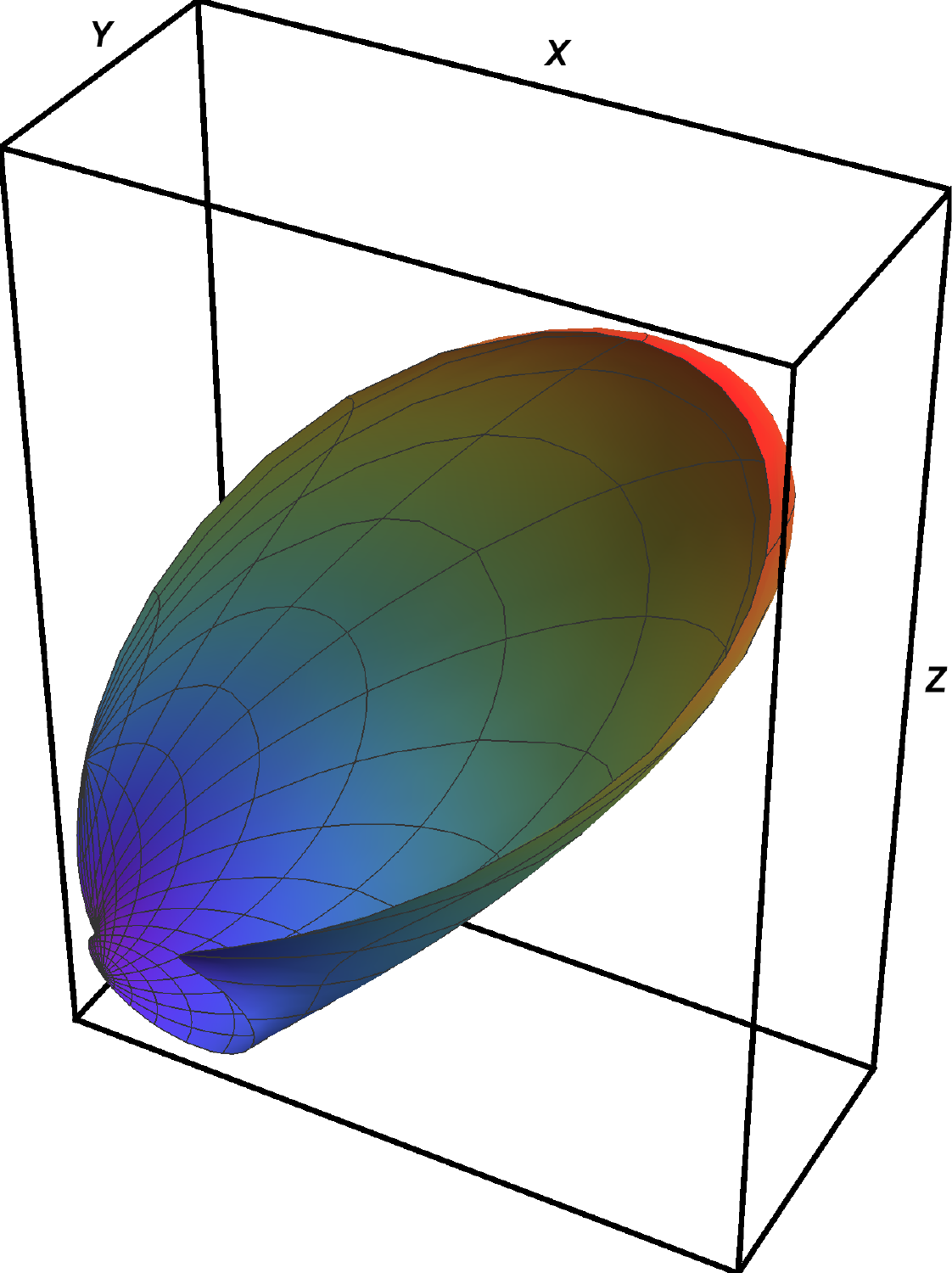}}\quad
\subfigure{\includegraphics[width=1.4in]{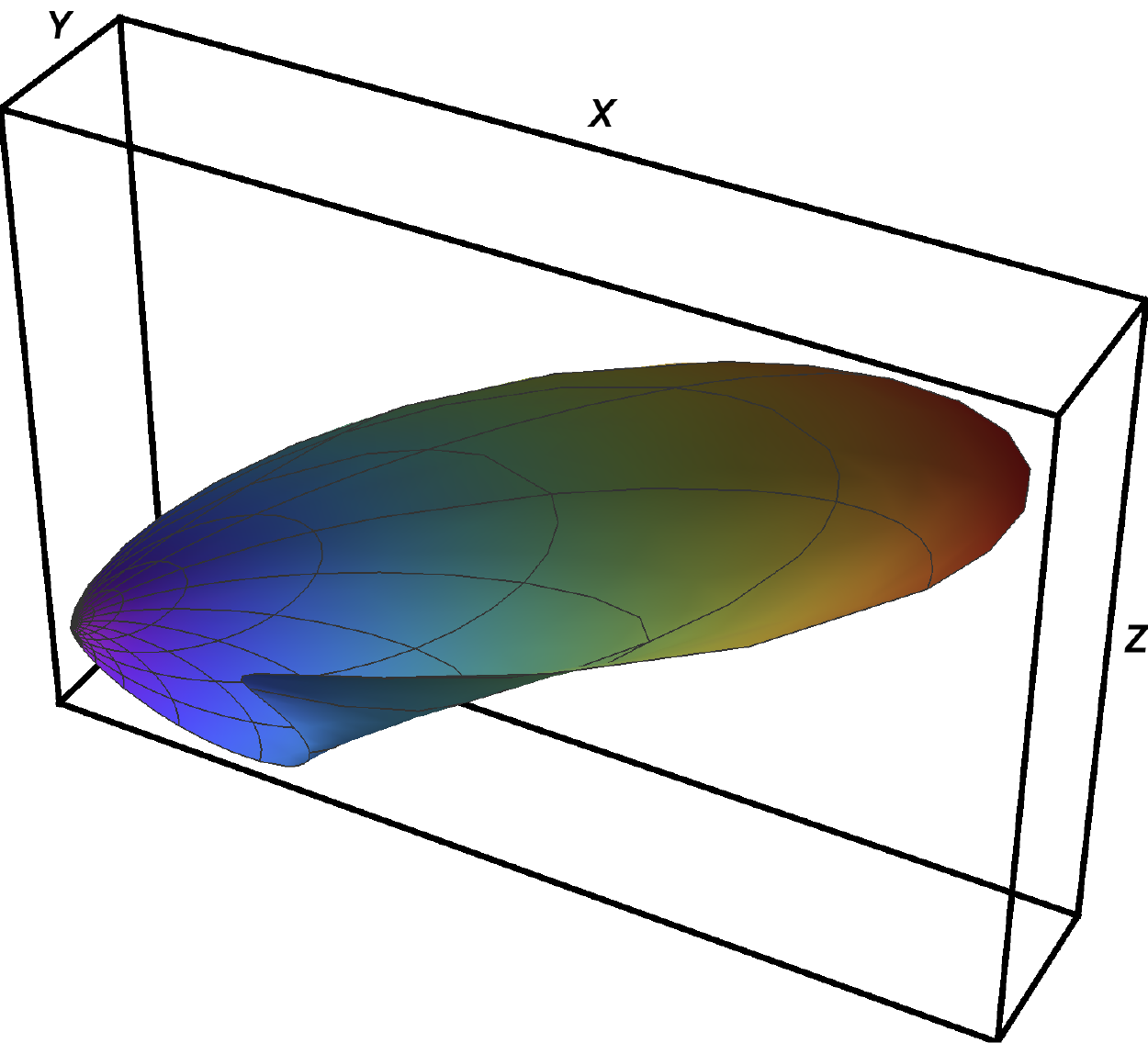}}\quad
\subfigure{\includegraphics[width=1.4in]{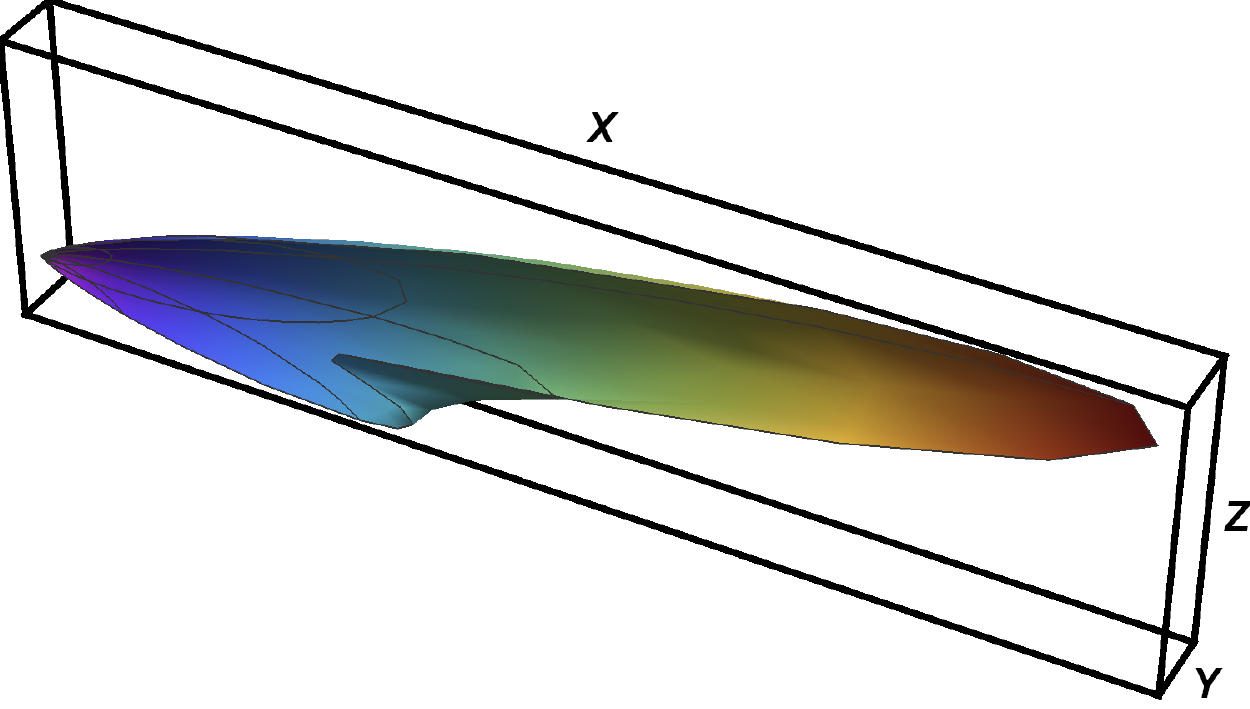}}}
\caption{\label{figh22}  Hypertor Angular Distribution in three dimensional space with  invariants  : $\kappa = 1,   \tau=\nu=0.5$, and speed  $\beta = 0.331 , 0.666 , 0.998$ . Compared to the previous example Fig. (\ref{figh11}) it can be seen how changes in invariants can affect to shape of radiation, i.e. quite a variety of change can happen to the ancillary lobe depending on the particular choice of torsion and hypertorsion. The plot range is $\theta=[0,\pi]$ and $\phi=[0,\pi]$.}
\end{center}
\end{figure}




\end{appendices}

\end{document}